\documentclass[a4paper,12pt,onecolumn,oneside]{book}

\usepackage{graphicx}
\usepackage{bm}
\usepackage{amsmath}
\usepackage{exscale}
\usepackage{amssymb}
\usepackage{dsfont}
\usepackage{txfonts}
\usepackage{german}
\usepackage{a4}
\usepackage{geometry}
\usepackage[latin1]{inputenc}

\newcommand*{\MRM}[3]{Magn.\ Reson.\ Med. #1;#2:#3.}
\newcommand*{\MRI}[3]{Magn.\ Reson.\ Imaging #1;#2:#3.}

\newcommand*{\JMR}[3]{J.\ Magn.\ Reson. #1;#2:#3.}
\newcommand*{\Circ}[3]{Circulation #1;#2:#3.}
\newcommand*{\PR}[3]{Phys.\ Rev.\ #1;#2:#3.}
\newcommand*{\PRL}[3]{Phys.\ Rev.\ Lett. #1;#2:#3.}
\newcommand*{\PRA}[3]{Phys.\ Rev.\ A #1;#2:#3.}

\newcommand*{\PRE}[3]{Phys.\ Rev.\ E #1;#2:#3.}

\newcommand*{\MAGMA}[3]{Magn.\ Reson.\ Mater.\ Phy. #1;#2:#3.}
\newcommand*{\JCP}[3]{J.\ Chem.\ Phys. #1;#2:#3.}

\newcommand*{\Jphysiol}[3]{J. Physiol. (Lond) #1;#2:#3.}
\newcommand*{\SAM}[3]{Stud. Appl. Math. #1;#2:#3.}
\newcommand*{\PNAS}[3]{Proc.\ Natl.\ Acad.\ Sci.\ U.\ S.\ A.\ #1;#2:#3.}
\newcommand*{\NMRBio}[3]{NMR Biomed.\ #1;#2:#3.}
\newcommand*{\Hyperfine}[3]{Hyperfine Interactions #1;#2:#3.}
\newcommand*{\JACC}[3]{J. Am. Coll. Cardiol. #1;#2:#3.}
\newcommand*{\ANYAS}[3]{Ann. N.Y. Acad. Sci. #1;#2:#3.}
\newcommand*{\AJP}[3]{Am. J. Phys. #1;#2:#3.}
\newcommand*{\JAMSSB}[3]{J. Austral. Math. Soc. Ser. B #1;#2:#3.}

\newcommand*{\MA}[3]{Math. Ann. #1;#2:#3.}
\newcommand*{\RMF}[3]{Revista Mexicana de Física #1;#2:#3.}
\newcommand*{\LAA}[3]{Linear Algebra and its Applications #1;#2:#3.}

\begin{document}
\renewcommand{\baselinestretch}{1.5}
\normalsize
\begin{titlepage}
\vfill
\begin{minipage}{\textwidth}
\vspace*{1cm}
\centering \Large
Aus der Medizinischen Klinik und Poliklinik I \\
der Universität Würzburg \\
Direktor: Professor Dr. med. Georg Ertl
\end{minipage}

\vspace*{2cm}
\begin{minipage}{\textwidth}
\centering \Huge \bf
Spindephasierung im Kroghschen\\Kapillarmodell des Myokards
\end{minipage}

\vspace{2cm}\begin{minipage}{\textwidth}
\centering{ \Large Inaugural - Dissertation\\
zur Erlangung der Doktorwürde der\\
Medizinischen Fakultät\\
der\\
Julius-Maximilians-Universität Würzburg\\
vorgelegt von\\
Christian H. Ziener\\
aus Weimar\\[8ex]
Würzburg, Juni 2011\\}
\end{minipage}

\vfill

\thispagestyle{empty}

\phantom{Leerzeile}

\flushleft
\begin{minipage}{\textwidth}
  \normalsize
  \begin{tabular}{ll}
    \multicolumn{2}{l}{}\\
                               & \\
    Referent:              &\hspace{-3,3cm} {Prof. Dr. med. Dr. rer. nat. Wolfgang R. Bauer} \\[4ex]
    Korreferent:           &\hspace{-3,3cm} {Prof. Dr. med. Thomas Dandekar } \\[4ex]
    Dekan:                 &\hspace{-3,3cm} {Prof. Dr. med. Matthias Frosch} \\ 
                                     & \\[40ex]
    Tag der mündlichen Prüfung: 09. 11. 2012\\
                                     & \\[40ex]
Der Promovend ist Arzt
  \end{tabular}
\end{minipage}

\newpage
\thispagestyle{empty}
\pagestyle{empty}

\tableofcontents

\pagestyle{empty}
\thispagestyle{empty}

\end{titlepage}

\pagestyle{empty}
\thispagestyle{empty}
\chapter*{\label{Kap.Einleitung}\vspace{-3cm} 1 Einleitung}
\addcontentsline{toc}{chapter}{1 Einleitung}
\pagestyle{plain}

\noindent
Die Magnetresonanzbildgebung hat in den letzten Jahren in der kardiologischen Diagnostik zunehmend an Bedeutung gewonnen. Durch dieses nichtinvasive Verfahren können pathologische Veränderungen diagnostiziert und auf Grundlage dieser Diagnose der entsprechenden Therapie zugeführt werden. Entscheidender Vorteil der kardialen Bildgebung ist der, dass nicht nur morphologische Veränderungen, sondern auch funktionelle Parameter des zu Grunde liegenden Pathomechanismus ermittelt werden können. Die quantitative Analyse der funktionellen Parameter des kardiovaskulären Systems ist entscheidend zum Verständnis der Pathogenese kardialer Erkrankungen. Insbesondere die Mikrozirkulation, von welcher der Stoffaustausch zwischen Blut und Herzmuskelgewebe abhängt, ist bei vielen Krankheitsbildern beeinträchtigt. Bei der Herzmuskelhypertrophie, wie sie zum Beispiel bei der chronischen Hypertonie vorkommt, verringert sich die Kapillardichte, was eine Verschlechterung der Gewebeversorgung nach sich zieht. Ein weiteres Beispiel für die diagnostische Bedeutung der Mikrozirkulation ist die Angina pectoris, die im Rahmen einer koronaren Herzkrankheit auftreten kann. Durch atherosklerotische Veränderungen an den Herzkranzgefäßen werden die nachgeschalteten Kapillaren nicht ausreichend durchblutet. Kompensatorisch werden vermehrt Kapillaren rekrutiert, die eine Durchblutung unter Ruhebedingungen sicherstellen. Bei körperlicher Belastung stehen dann keine weiteren Kapillaren zur Verfügung, d.h. die Perfusionsreserve des Myokards ist aufgebraucht, und die pektanginöse Symptomatik tritt auf. Mit der Kenntnis der Kapillardichte bzw. der Perfusionsreserve kann der Grad der Stenose der vorgeschalteten Koronarien abgeschätzt werden. 

Die quantitative Bestimmung der funktionellen Parameter der Mikrozirkulation ist also sowohl von Seiten der kardiologischen Grundlagenforschung als auch von der klinischen Seite her von großer Bedeutung. Die kardiale Magnetresonanztomographie bietet als diagnostisches Verfahren in vielerlei Hinsicht Vorteile gegenüber anderen Verfahren \cite{Hombach2005}. Zum einen können insbesondere bei hohen Feldstärken die Morphologie des Herzens oder sogar die Mikrostruktur in hoher Auflösung untersucht werden \cite{Koehler03}. Zum anderen können durch geeignete Wahl der angewandten Bildgebungssequenzen funktionelle Parameter ermittelt werden. Allerdings erfordert die Interpretation der gewonnenen Messwerte, dass der Zusammenhang zwischen den Mikrozirkulationsparametern und dem gemessenen MR-Signal bekannt ist. Ziel dieser Arbeit ist es, ein Gewebemodell für das Myokard zu entwickeln und einen quantitativen Zusammenhang zwischen den Mikrozirkulationsparametern und dem MR-Signal herzustellen. Dieser Zusammenhang kann dann genutzt werden, um aus den gemessenen MR-Signalen auf die Mikrozirkulationsparameter zu schließen. 

\begin{figure}
\begin{center}
\includegraphics[width=11cm]{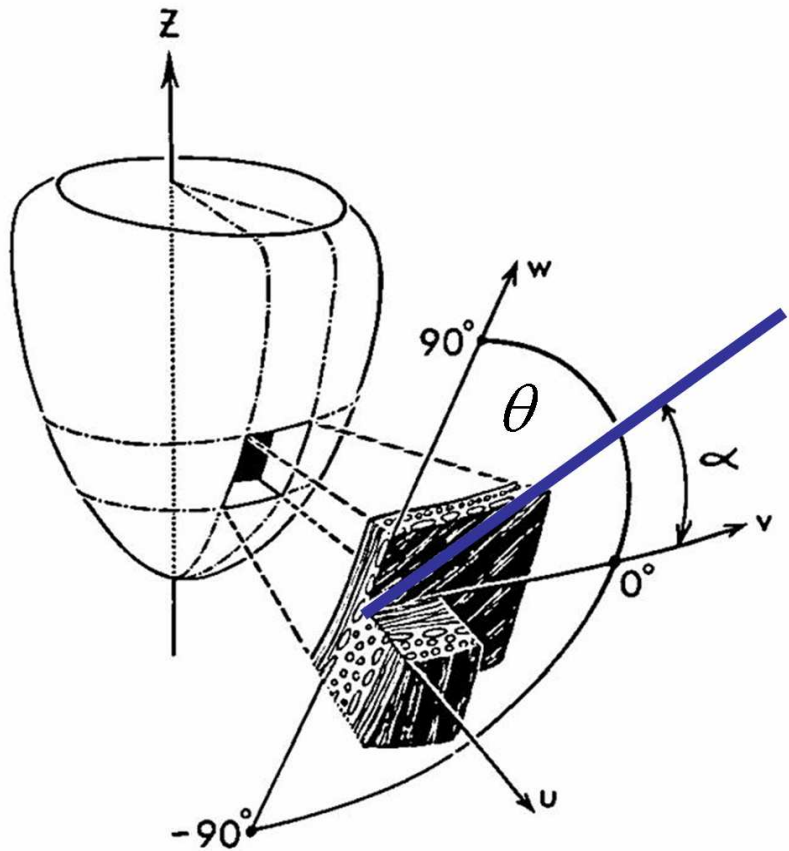}
\vspace{-0.5cm}
\caption[Faserarchitektur des Myokards]{\label{Fig:Orientierung}{\footnotesize Faserarchitektur des Myokards. Die blutgefüllten Kapillaren (blau) sind gegenüber dem äußeren Magnetfeld in $z$-Richtung um den Winkel $\theta$ geneigt (modifiziert nach \cite{Streeter69}).}}
\end{center}
\end{figure}

\begin{figure}
\begin{center}
\includegraphics[width=\textwidth]{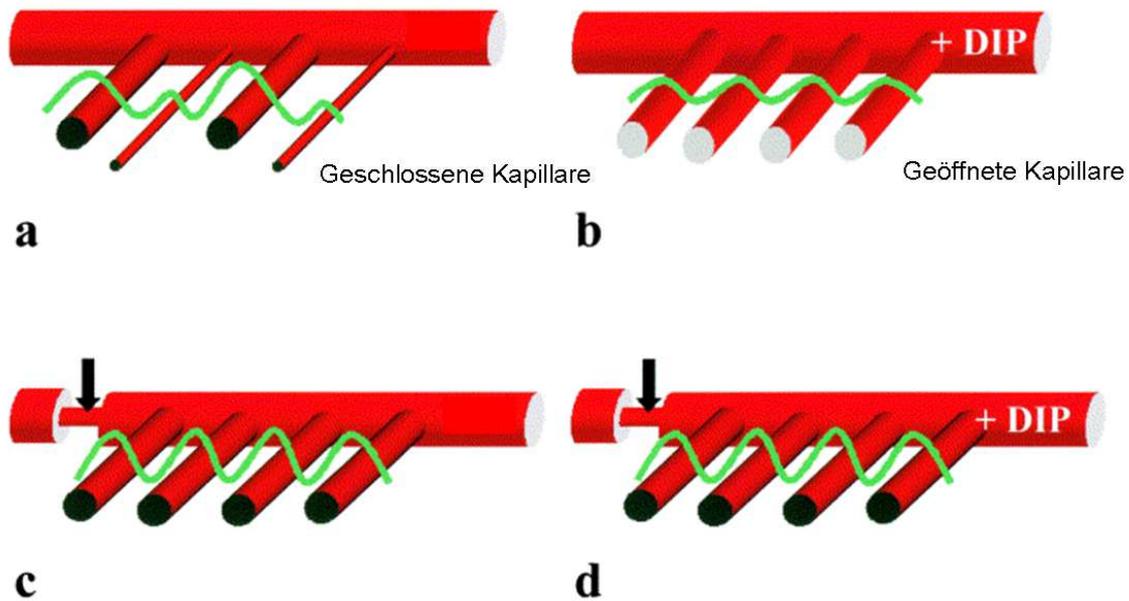}
\vspace{-1.5cm}
\caption[Funktionelle Kapillardichte]{\label{Fig:stress}{\footnotesize Funktionelle Kapillardichte bei gesundem (a und b) und atherosklerotisch verengtem (c und d) Kranzgefäß unter Ruhebedingungen (a und c) und bei pharmakologischem Stress (b und d) durch den Vasodilatator Dipyridamol (DIP). Das von den offenen Kapillaren erzeugte lokale Magnetfeld ist durch die grüne Welle veranschaulicht (modifiziert nach \cite{Wacker03}).}}
\end{center}
\end{figure}

Zur Gewinnung eines Einblicks in die Untersuchungstechnik der Magnetresonanztomographie ist es sinnvoll, die grundlegenden Mechanismen zu rekapitulieren. Während der magnetresonanztomographischen Untersuchung des Herzens befindet sich das Myokard in einem starken äußeren Magnetfeld mit der Flussdichte $B_0$. Durch Anlegen von Magnetfeldgradienten und Einstrahlen von Hochfrequenzpulsen können aus den MR-Signalen Bilder des Myokards gewonnen werden. Zur Entstehung des MR-Signals tragen die Protonen der Wassermoleküle bei, die sich in den Herzmuskelzellen und im Interstitium des Myokards befinden. Wie in der Abb. \ref{Fig:Orientierung} dargestellt ist, besteht das Myokard aus Muskelzellen, die von Kapillaren versorgt werden.
Auf Grund der magnetischen Eigenschaften des Blutes erzeugen diese blutgefüllten Kapillaren ein lokales inhomogenes Magnetfeld. Die Spins der Protonen diffundieren im lokalen inhomogenen Magnetfeld der Kapillaren und führen so zur Spindephasierung und damit zum Signalabfall. Das MR-Signal im Myokard entsteht also vollkommen analog zum BOLD-Effekt \cite{Ogawa90}, der in der funktionellen Bildgebung zur Charakterisierung metabolischer Prozesse im Gehirn genutzt wird. Im Myokard wird das lokale Magnetfeld hauptsächlich vom Kapillarradius und von der Kapillardichte bestimmt. Eine Veränderung dieser funktionellen Parameter beeinflusst das MR-Signal in charakteristischer Weise und kann diagnostisch genutzt werden. Dies soll am Beispiel eines Herzkranzgefäßes verdeutlicht werden. Im gesunden Zustand sind unter Ruhebedingungen nicht alle Kapillaren, die dem Herzkranzgefäß nachgeschaltet sind, geöffnet (Abb. \ref{Fig:stress}a).
Allerdings tragen auf Grund der magnetischen Eigenschaften des Blutes nur diese geöffneten Kapillaren zum lokalen Magnetfeld und damit zum Signalabfall bei. Durch die Gabe eines Vasodilatators (Dipyridamol) wird pharmakologischer Stress induziert, durch den sich sämtliche Kapillaren öffnen (Abb. \ref{Fig:stress}b). Infolge des vergrößerten Blutvolumens verstärkt sich das lokale Magnetfeld und das gemessene Signal fällt schneller ab. Liegt eine Engstelle im betreffenden Herzkranzgefäß vor, so sind bereits unter Ruhebedingungen sämtliche Kapillaren geöffnet (Abb. \ref{Fig:stress}c), um eine ausreichende Perfusion des Myokards zu sichern. Die Gabe eines Vasodilatators kann nun keine Rekrutierung weiterer Kapillaren bewirken (Abb. \ref{Fig:stress}d) und demzufolge ändern sich das lokale Magnetfeld und damit auch das gemessene MR-Signal nicht.

Zum Verständnis dieser Effekte und deren Auswirkung auf das MR-Signal ist es erforderlich, einen Zusammenhang zwischen den Mikrozirkulationsparametern und den Dephasierungsmechanismen herzustellen. Dabei müssen zwei entscheidende Effekte berücksichtigt werden: die Diffusion der Spins um die Kapillare und die Suszeptibilitätseffekte des Blutes innerhalb der Kapillare. Hierin liegt auch die Besonderheit des Myokards -- nämlich dass keiner der beiden Effekte vernachlässigt werden kann. In einer der ersten Arbeiten zur Beschreibung der Dephasierung im Myokard konnten Bauer et al. einen Zusammenhang zwischen den Gewebeparametern und der transversalen Relaxationszeit $T_2^*$ finden. Diese transversale Relaxationszeit ist eine Näherung für einen rein monoexponentiellen MR-Signalzerfall und eignet sich zur Interpretation von Gradientenechobildern. 

Die von Bauer et al. genutzte Strong-Collision-Näherung \cite{Bauer99} berücksichtigt sowohl Diffusionseffekte als auch Suszeptibilitätseffekte; durch die verwendete Mean-Relaxation-Time-Näherung wurde der Signal-Zeit-Verlauf durch eine monoexponentielle Funktion ersetzt, was schließlich zur Relaxationszeit $T_2^*$ führte. In einer weitergehenden Analyse konnten Ziener et al. den Formalismus der diffusionsabhängigen Frequenzverteilungen entwickeln \cite{Ziener07PRE}, und so die Mean-Relaxation-Time-Näherung umgehen. In der vorliegenden Arbeit wird gezeigt, dass es möglich ist, ohne den Umweg der diffusionsabhängigen Frequenzverteilungen das Dephasierungsverhalten auf Grundlage der Strong-Collision-Näherung zu beschreiben. Des Weiteren wird gezeigt, dass zur Beschreibung der Spindephasierung auch auf diese Näherung verzichtet werden kann. 

Als Gewebemodell für das Myokard wird das in Abschnitt 2.1 beschriebene Kroghsche Kapillarmodell genutzt. Allgemeine Gesichtspunkte zur Spindephasierung im Kroghschen Kapillarmodell werden in Abschnitt 2.2 erörtert. Die Anwendung der Strong-Collision-Näherung zur Beschreibung der Spindephasierung erfolgt in Abschnitt 2.3. Dass es auch möglich ist, die zu Grunde liegende Differenzialgleichung, welche den Magnetisierungs-Zeit-Verlauf beschreibt, analytisch zu lösen, wird in Abschnitt 2.4 gezeigt, und zwar indem ein Separationsansatz angewandt wird. Die Experimente zur Bestätigung der analytischen Ergebnisse werden in Abschnitt 2.5 beschrieben. Die Ergebnisse der Anwendung der Strong-Collision-Näherung sind in Abschnitt 3.1 dargestellt, während in Abschnitt 3.2 die analytische Lösung -- basierend auf einem Separationsansatz -- erläutert wird. Im Abschnitt 3.3 werden dann die theoretischen Resultate mit den experimentellen Ergebnissen verglichen. Die Erkenntnisse werden in Kapitel 4 diskutiert und in Kapitel 5 zusammengefasst. 

\chapter*{\label{Kap.Material}\vspace{-3cm} 2 Material und Methoden}
\addcontentsline{toc}{chapter}{2 Material und Methoden}

\section*{\normalsize{2.1 Kroghsches Kapillarmodell}}
\begin{figure}
\begin{center}
\includegraphics[width=\textwidth]{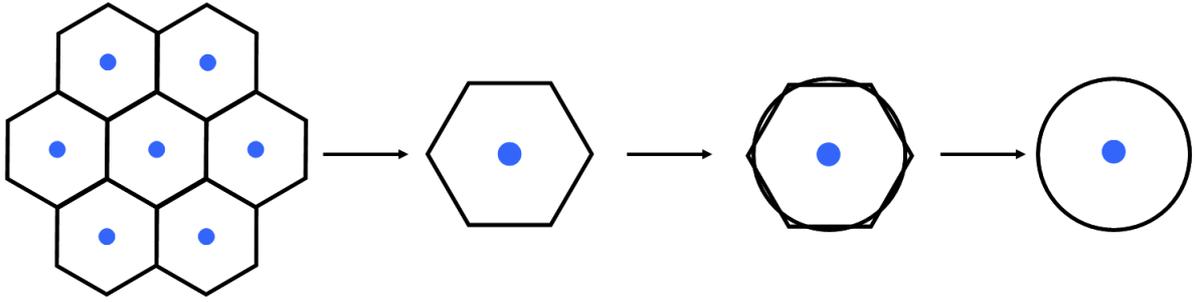}
\vspace{-1.0cm}
\caption[Kroghsches Kapillarmodell]{\label{Fig:Krogh}{\footnotesize Kroghsches Kapillarmodell. Die regelmäßige Anordnung der Kapillaren wird durch eine einzige Kapillare mit umgebenden Gewebszylinder ersetzt.}}
\end{center}
\end{figure}

\begin{figure}
\begin{center}
\includegraphics[width=\textwidth]{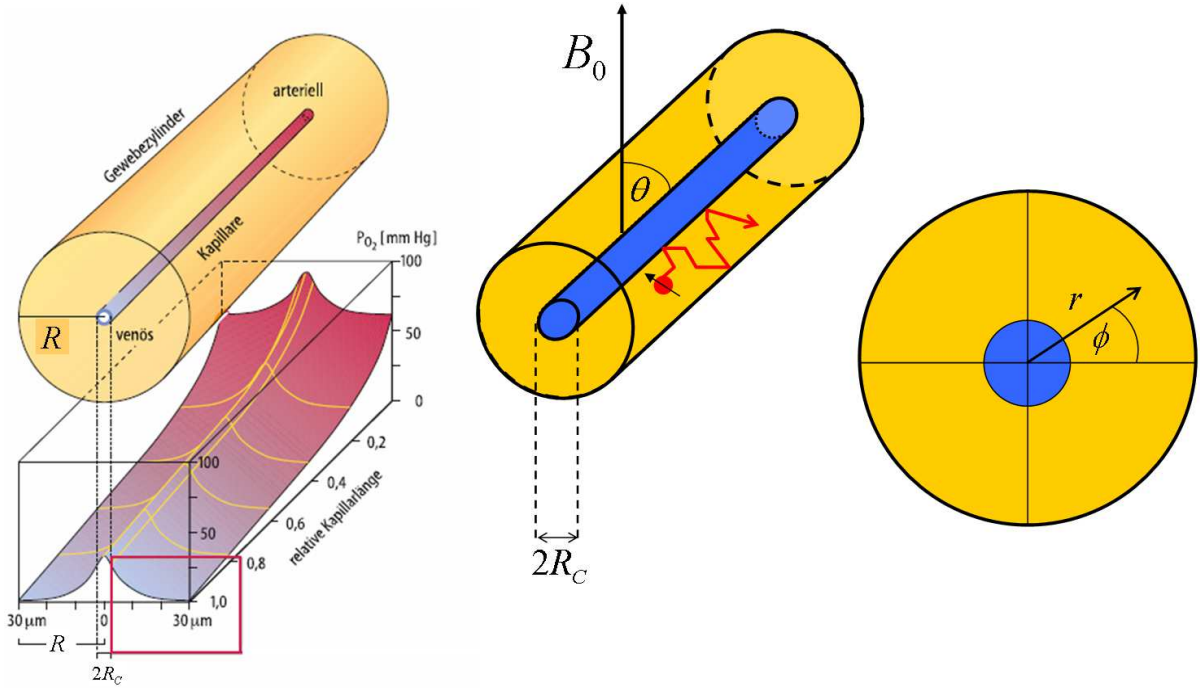}
\vspace{-1.5cm}
\caption[Versorgungsbereich einer Kapillare]{\label{Fig:zylinder}{\footnotesize Links: Versorgungsbereich einer Kapillare mit dem entsprechenden Sauerstoffpartialdruck (nach \cite{Schmidt05}). Mitte: Dephasierungsgebiet, in dem die Diffusion (rote Trajektorie) und die Dephasierung stattfinden. Rechts: Querschnitt durch Kapillare und Dephasierungsgebiet mit Polarkoordinaten.}}
\end{center}
\end{figure}

\addcontentsline{toc}{section}{2.1 Kroghsches Kapillarmodell}
Ein Gewebemodell zur Beschreibung von Muskelgewebe wurde von Krogh bereits 1919 angewandt, um Sauerstoffpartialdrücke im Muskelgewebe zu beschreiben \cite{Krogh19a,Krogh19b,Krogh70}. Voraussetzung für die Anwendbarkeit dieses Modells ist eine regelmäßige und parallele Anordnung von Kapillaren mit dem Radius $R_{\text{C}}$ in einem Gewebeareal. Innerhalb eines Bildgebungsvoxels ist diese Voraussetzung erfüllt. Der Grundgedanke des Kroghschen Kapillarmodells besteht darin, dass eine einzelne Kapillare von einem konzentrischen Gewebszylinder umgeben ist, der durch diese Kapillare versorgt wird (siehe Abb. 72 in \cite{Krogh70}). Anstatt nun alle Kapillaren zu berücksichtigen, beschränkt man sich -- wie in Abb. \ref{Fig:Krogh} dargestellt --  auf eine Kapillare, wobei der Radius $R$ des umgebenden Gewebszylinders so gewählt wird, dass das regionale Blutvolumenverhältnis (RBV$=\eta$) konstant bleibt:
\begin{equation}
\label{RBV}
\eta=\frac{R_{\text{C}}^2}{R^2} \,.
\end{equation}
Im menschlichen Myokard haben die Kapillaren einen Radius von $R_{\text{C}}=2,\!75 \,\mu \text{m}$ und das regionale Blutvolumenverhältnis beträgt etwa $\eta = 0,\!084$ \cite{Bauer99}. Daraus ergibt sich ein zylinderförmiges Versorgungsgebiet mit einem Radius von $R = 9,\!5 \, \mu \text{m}$. In Abb. \ref{Fig:zylinder} ist der aus der Physiologie bekannte Krogh-Zylinder zur Beschreibung des Atemgaswechsels zwischen dem Blut und dem Gewebe dargestellt. Der Sauerstoffpartialdruck sinkt sowohl senkrecht zur Kapillare mit zunehmender Entfernung von dieser als auch entlang der Kapillare in Richtung des venösen Endes. Kapillarferne Zellen am venösen Ende des Krogh-Zylinders werden am schlechtesten mit Sauerstoff versorgt und sind am ersten von Hypoxie bedroht (rotes Rechteck in Abb. \ref{Fig:zylinder}). Zur Beschreibung des MR-Signalverhaltens interessiert in dieser Arbeit allerdings nicht der Sauerstoffpartialdruck im Versorgungszylinder, sondern die Dephasierung der Spins, die um die Kapillare diffundieren. Deshalb wird dieser Versorgungszylinder bei der Untersuchung des MR-Signals auch als Dephasierungszylinder bezeichnet. Die Diffusion der signalgebenden Protonen findet im Raum zwischen der Kapillare und dem umgebenden Zylinder im Gewebe mit dem Diffusionskoeffizienten $D = 1 \, \mu \text{m}^2 \text{ms}^{-1}$ statt \cite{Bauer99} (siehe Abb. \ref{Fig:zylinder}). Der Diffusionsprozess wird durch die charakteristische Korrelationszeit
\begin{equation}
\label{eEq19}
\tau=\frac{R_C^2}{4D}\,\frac{\text{ln}(\eta)}{\eta -1}
\end{equation}
beschrieben. Diese Korrelationszeit erlaubt es, zwischen verschiedenen Diffusionsregimen zu unterscheiden. Abhängig vom Diffusionskoeffizienten des umgebenden Mediums lassen sich die beiden Grenzfälle Motional-Narrowing-Regime (überwiegende Diffusion) \cite{Kennan94} und Static-Dephasing-Regime (vernachlässigbare Diffusion) \cite{Yablonskiy94} unterscheiden. Mit den typischen Parametern für das Myokard ergibt sich die Korrelationszeit von $\tau=5,\!1\,\text{ms}$. Diese Korrelationszeit wird im Weiteren wichtig sein, da sie mit einem weiteren Parameter des Myokards (der charakteristischen Frequenz auf der Oberfläche der Kapillare) das zu Grunde liegende Dephasierungsregime festlegt.
Ein weiteres Charakteristikum des Modells besteht darin, dass die diffundierenden Wassermoleküle an der Oberfläche der Kapillare und am Rand des äußeren Zylinders reflektiert werden. Diese reflektierenden Randbedingungen sind auf Grund der ursprünglich periodischen Anordnung gerechtfertigt und kompensieren den Einfluss der lokalen Magnetfelder der anderen umgebenden Kapillaren. 

\section*{\normalsize{2.2 Spindephasierung}}
\begin{figure}
\begin{center}
\includegraphics[width=10cm]{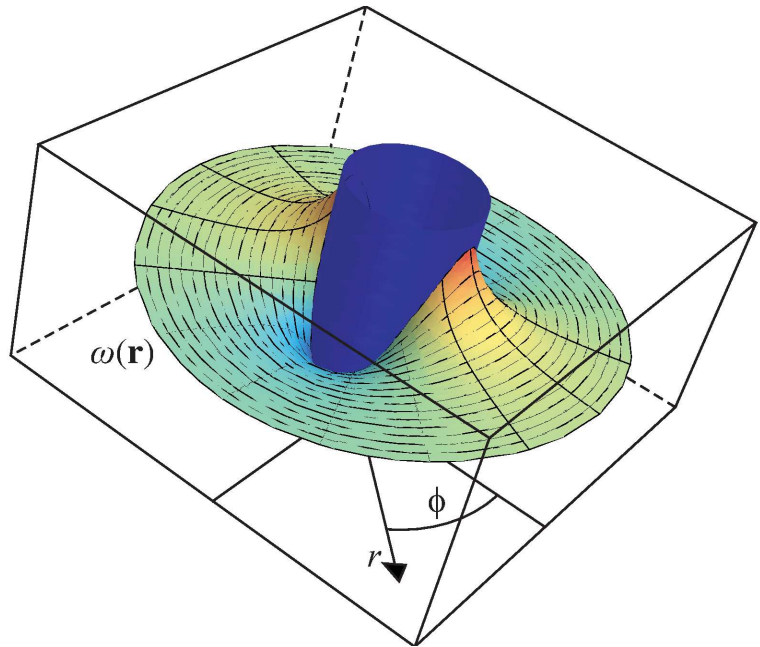}
\vspace{-0.5cm}
\caption[Kapillare im Voxel]{\label{Fig:Kapillare}{\footnotesize Kapillare im Voxel. Die lokale Resonanzfrequenz $\omega(\textbf{r})$ nach Gl. (\ref{Dipol}) ist in Polarkoordinaten schematisch dargestellt. Da ein zylinderförmiges Dephasierungsgebiet angenommen wird, ist die lokale Resonanzfrequenz nur bis zum Rand des Dephasierungszylinders mit dem Radius $R$ dargestellt. Die Kapillare hat den Radius $R_{\text{C}}$ und das regionale Blutvolumenverhältnis ist durch Gl. (\ref{RBV}) festgelegt.}}
\end{center}
\end{figure}

\addcontentsline{toc}{section}{2.2 Spindephasierung}
Während der magnetresonanztomographischen Untersuchung des Herzens befindet sich das Myokard in einem Magnetfeld mit der Flussdichte $B_0$. Entsprechend der Reduzierung des Problems auf eine zylinderförmige Kapillare wird nun ein einzelner Zylinder in einem äußeren Magnetfeld $B_0$ betrachtet (siehe Abb. \ref{Fig:zylinder}), wobei $\theta$ für den Neigungswinkel zwischen Zylinder und Magnetfeld steht. Dieser einzelne Zylinder wird von einem koaxialen Zylinder umgeben. Der Raum zwischen beiden Zylindern ist das Dephasierungsgebiet, in dem die Diffusion der Spins mit dem Diffusionskoeffizienten $D=1 \, \mu\text{m}^2 \text{ms}^{-1} $ stattfindet \cite{Bauer99}. In der zylinderförmigen Kapillare befindet sich Blut mit der magnetischen Suszeptibilität $\chi_i$. Die Suszeptibilität des Blutes ist unter anderem vom Hämatokrit $Hkt$ und vom Oxygenierungsgrad $Y$ des Hämoglobins abhängig \cite{Pauling36}. Desoxygeniertes Hämoglobin ($Y=0$) besitzt ungepaarte Elektronen und ist paramagnetisch (positive magnetische Suszeptibilität), während oxygeniertes Hämoglobin ($Y=1$) keine ungepaarten Elektronen enthält und deshalb diamagnetisch ist (negative magnetische Suszeptibilität). Das umgebenden Gewebe im Dephasierungszylinder hat die Suszeptibilität $\chi_e$. Die Suszeptibilitätsdifferenz zwischen der blutgefüllten Kapillare und dem umgebenden Gewebe berechnet sich nach der Gleichung
\begin{equation}
\Delta\chi = \chi_e - \chi_i = 2,26 \cdot 10^{-6} \cdot Hkt \cdot [1-Y] \,.
\end{equation}
Unter Ruhebedingungen beträgt diese Suszeptibilitätsdifferenz im menschlichen Myokard etwa $\Delta\chi = 6 \cdot 10^{-8}$ \cite{Bauer99}. Um den inneren Zylinder herum bewegen sich die Spins und präzedieren mit der Larmorfrequenz
\begin{align} \label{Dipol}
\omega(\textbf{r}) = \delta \omega R_{\text{C}}^2 \, \frac{\cos(2\phi)}{r^2} \quad \text{mit} \quad 
\delta\omega = 2 \pi \gamma \Delta\chi B_0 \sin^2 (\theta) \,,
\end{align}
wobei $\gamma=2,\!675 \cdot 10^8 \,\text{s}^{-1}\,\text{T}^{-1}$ das gyromagnetische Verhältnis und $\delta\omega=\omega(r=R_{\text{C}},\phi=0)$ die charakteristische Frequenz auf der Oberfläche der Kapillare sind \cite{Reichenbach01}. Die Polarkoordinaten $r$ und $\phi$ beschreiben die Position innerhalb des Dephasierungsgebietes (sie Abb. \ref{Fig:zylinder} rechts). In Abb. \ref{Fig:Kapillare} ist das lokale inhomogene Magnetfeld um eine Kapillare des Myokards schematisch dargestellt.
Unter der Annahme, dass die Kapillare senkrecht zum äußeren Magnetfeld orientiert sei ($\theta=90^{\circ}$), ergibt sich bei einer Feldstärke von $B_0 = 1,\!5 \, \text{T}$ eine charakteristische Frequenz von $\delta \omega = 151 \, \text{s}^{-1}$. Für spätere Betrachtungen ist es sinnvoll, die Periodizität dieser Resonanzfrequenz zu betrachten: $\omega(r,\phi) = \omega(r,\phi + \pi)$. Dies bedeutet, dass die Resonanzfrequenz punktsymmetrisch ist: $\omega(\textbf{r}) = \omega(-\textbf{r})$. 

\begin{figure}
\begin{center}
\includegraphics[width=\textwidth]{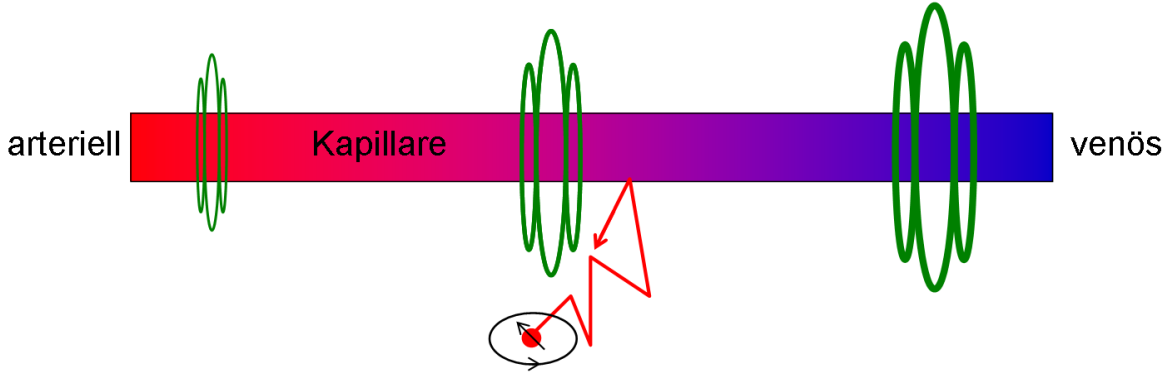}
\vspace{-1.5cm}
\caption[Induktion inhomogener perikapillärer Magnetfelder]{\label{Fig:Magnetfeldkap}{\footnotesize Induktion inhomogener perikapillärer Magnetfelder. Das sauerstoffreiche arterielle Blut enthält Oxyhämoglobin (diamagnetisch) und induziert ein schwaches perikapilläres Magnetfeld (grün). Das sauerstoffarme venöse Blut enthält Desoxyhämoglobin (paramagnetisch) und erzeugt ein starkes perikapilläres Magnetfeld. Transversal polarisierte Spins präzedieren (schwarze Kurve) und diffundieren (rote Trajektorie) im perikapillären Magnetfeld.}}
\end{center}
\end{figure}

In diesem lokalen perikapillären Magnetfeld, das durch die Kapillare erzeugt wird, präzedieren und dephasieren die Spins der Protonen der umgebenden Wassermoleküle im zylinderförmigen Dephasierungsvolumen. Dies ist schematisch in Abb. \ref{Fig:Magnetfeldkap} dargestellt. Am arteriellen Ende der Kapillare befindet sich oxygeniertes ($Y \approx 1$) Blut. Dieses diamagnetische Blut hat eine ähnlich große Suszeptibilität wie das umgebende diamagnetische Gewebe, weshalb auch nur ein schwaches perikapilläres Magnetfeld entsteht. Im venösen Ende der Kapillare erzeugt das desoxygenierte ($Y \approx 0,\!25$) paramagnetische Blut einen großen Suszeptibilitätsunterschied zum umgebenden Gewebe und damit auch ein starkes perikapilläres Magnetfeld. 

Durch das äußere Magnetfeld $B_0$ sind alle Spins, die sich im Dephasierungszylinder befinden, parallel zum Magnetfeld in $z$-Richtung ausgerichtet. Die transversale Magnetisierung $ m(\textbf{r},t) $ ist also gleich Null. Zum Zeitpunkt $t=0$ erfolgt die Einstrahlung eines Hochfrequenzpulses und die Magnetisierung besitzt nun die transversale Komponente $m_0$, wobei $\textbf{r}$ im Dephasierungsgebiet liegt. Dies ist die Anfangsbedingung der späteren Herleitung. Des Weiteren werden sowohl an der Oberfläche des inneren Zylinders mit dem Radius $R_{\text{C}}$ als auch an der Oberfläche des äußeren Zylinders mit dem Radius $R$ reflektierende Randbedingungen angenommen. Der Spin verlässt also nie das Dephasierungsvolumen, sondern wird immer wieder hineinreflektiert. Diese reflektierenden Randbedingungen sind ausführlich in \cite{Ziener08JCP} beschrieben.

Die transversale Magnetisierung $m(\textbf{r},t)=m_x(\textbf{r},t)+\text{i}m_y(\textbf{r},t)$ hat zum Zeitpunkt $t=0$ ihr Maximum und nimmt im Laufe der Zeit auf Grund der Spindephasierung ab. Zur Beschreibung der Zeitentwicklung dieser transversalen Magnetisierung wird die Bloch-Torrey-Gleichung \cite{Torrey56} genutzt:
\begin{align}
\label{BT}
&\partial_t m(\textbf{r},t) = [D \Delta - \text{i} \omega(\textbf{r}) - 1/T_2] m(\textbf{r},t)\,, \\
\label{Anfang}
&m(\textbf{r},0) = m_0 = \text{const} \quad \text{und} \\
\label{rand}
&\left. \partial_r m(\textbf{r},t) \right|_{r=R_{\text{C}}} =0 =\left. \partial_r m(\textbf{r},t) \right|_{r=R} \,,
\end{align}
wobei $T_2$ die intrinsische transversale Spin-Spin-Relaxationszeit des umgebenden Gewebes darstellt. Der Laplace-Operator $\Delta$ in Polarkoordinaten $\textbf{r}=(r,\phi)$, der in Gl. (\ref{BT}) benötigt wird, ist gegeben durch:
\begin{equation}
\Delta=\frac{\partial^2}{\partial r^2} + \frac{1}{r} \frac{\partial}{\partial r} + \frac{1}{r^2} \frac{\partial^2}{\partial \phi^2} \,.
\end{equation}
Dem Experiment zugänglich ist allerdings nicht die lokale Magnetisierung $m(\textbf{r},t)$, sondern nur das Signal aus dem gesamten Voxel:
\begin{equation}
S(t) = \int \text{d}^2 \textbf{r} \, m(\textbf{r},t) = M(t) \, \text{e}^{-\frac{t}{T_2}}\,.
\end{equation}
An dieser Stelle ist anzumerken, dass mit $S(t)$ ein experimentell messbares Signal bezeichnet wird, welches die intrinsische Relaxation auf Grund der Spin-Spin-Wechselwirkung mit der Relaxationszeit $T_2$ beinhaltet. Der Magnetisierungszerfall, der durch die Dephasierung im perikapillären Magnetfeld entsteht, wird mit $M(t)$ bezeichnet. Zwischen beiden Größen besteht, wie vorhergehend beschrieben, der Zusammenhang $S(t)=M(t)\exp{(-t/T_2)}$. 

Das oben eingeführte Kroghsche Kapillarmodell wird in der Physiologie oft genutzt, um Diffusionsprozesse in zylinderförmigen Objekten zu beschreiben. Thews analysierte solche Prozesse durch Betrachtung der Diffusionsgleichung zwischen zwei zylinderförmigen Objekten mit unterschiedlichen Randbedingungen \cite{Thews53}. Bei der Untersuchung der Spindephasierung tritt jedoch zusätzlich zum Diffusionsprozess zwischen den beiden Zylindern eine ortsabhängige Resonanzfrequenz $\omega(\textbf{r})$ auf, welches durch die magnetischen Eigenschaften der blutgefüllten Kapillare verursacht wird. Dadurch wird der Sachverhalt komplizierter, insbesondere da dieses Potenzial rein imaginär ist. Die zur Beschreibung der Dephasierung genutzte Bloch-Torrey-Gleichung spielt in der Magnetresonanzbildgebung eine große Rolle, vor allem zur Ermittlung des Einflusses der Diffusion auf die linearen Bildgebungsgradienten \cite{Haacke99}. Analytisch konnte die Bloch-Torrey-Gleichung bisher nur für den Fall eines linearen Gradienten gelöst werden \cite{Stoller91}, d.h. für den Fall, dass die lokale Larmorfrequenz linear vom Ort abhängt ($\omega(x)\propto x$). Zur Untersuchung der Dephasierung im Myokard muss man allerdings eine kompliziertere lokale Larmorfrequenz $\omega(\textbf{r})$ -- wie sie in Gl. (\ref{Dipol}) gegeben ist -- betrachten. 

Zwei entscheidende Mechanismen sind zum Verständnis der Signalentstehung zu unterscheiden. Zum einen diffundieren die Spins im Krogh-Zylinder um die Kapillare. Dieser Diffusionsprozess wird durch die Korrelationszeit in Gl. (\ref{eEq19}) beschrieben. Das Inverse der Korrelationszeit ($1/\tau \propto D/R_{\text{C}}^2$) wird als dynamische Frequenz bezeichnet, da sie mit der Diffusionsbewegung der Spins verbunden ist. Zum anderen wird durch die magnetischen Eigenschaften des Blutes auf der Oberfläche der Kapillare die charakteristische Frequenz $\delta\omega$ hervorgerufen (siehe Gl. (\ref{Dipol})), welcher auch als statische Frequenz bezeichnet wird. Diese beiden Frequenzskalen gilt es nun miteinander zu vergleichen. Wenn die Diffusion um die Kapillare sehr gering ist, ist auch die dynamische Frequenz sehr klein. Die Spins befinden sich fast immer an der gleichen Stelle und sind einer statischen Frequenz ausgesetzt. Deshalb spricht man auch vom Static-Dephasing-Regime ($1/\tau << \delta\omega$). Im entgegengesetzten Grenzfall ist die Diffusion sehr stark und die Dephasierung wird hauptsächlich von der Diffusionsbewegung beeinflusst. In diesem Fall befindet man sich im Motional-Narrowing-Regime ($1/\tau >> \delta\omega$). Beide Grenzfälle sind bisher gut untersucht. Das Static-Dephasing-Regime wurde von Yablonskiy und Haacke untersucht \cite{Yablonskiy94} und später von Kiselev und Posse erweitert \cite{Kiselev98PRL,Kiselev99MRM}. Im Motional-Narrowing-Grenzfall kann man die Gaußsche Näherung zur Beschreibung der Dephasierung anwenden, wie von Sukstanskii und Yablonskiy \cite{Sukstanskii03,Sukstanskii04} bzw. von Jensen und Chandra \cite{Jensen2000a,Jensen2000b} beschrieben. Im Myokard beträgt nach Gl.(\ref{eEq19}) die Korrelationszeit $\tau=5,\!1\,\text{ms}$ und damit folgt die dynamische Frequenz zu $1/\tau=196\,\text{s}^{-1}$. Die dynamische Frequenz liegt also in der gleichen Größenordnung wie die statische Frequenz $\delta \omega = 151 \, \text{s}^{-1}$ nach Gl. (\ref{Dipol}). Im Myokard stellt also keiner der beiden Grenzfälle den zu Grunde liegenden Dephasierungsmechanismus dar und somit lassen sich die bisher entwickelten Verfahren zur Beschreibung der Dephasierung nicht anwenden. Zur Beschreibung der Dephasierung im gesamten Dynamikbereich, d.h. vom Motional-Narrowing-Regime bis zum Static-Dephasing-Regime wurde von Bauer et al. die Strong-Collision-Näherung \cite{Bauer99T2,Bauer02} auf die Dephasierung im Myokard angewandt, um Aussagen über die transversale Relaxationszeit zu erhalten. Im Verlauf der vorliegenden Arbeit wird die Strong-Collision-Näherung genutzt, um den Signal-Zeit-Verlauf der Magnetisierung im Myokard zu untersuchen. Des Weiteren wird in dieser Arbeit erstmals eine exakte analytische Lösung der Bloch-Torrey-Gleichung für die Dephasierung im perikapillären Feld vorgestellt, die nicht auf Näherungen basiert sondern auf einem Separationsansatz. 

\section*{\normalsize{2.3 Strong-Collision-Näherung}}
\addcontentsline{toc}{section}{2.3 Strong-Collision-Näherung}
Ein Näherungsverfahren zur Lösung der Bloch-Torrey-Gleichung ist die Strong-Collision-Näherung \cite{Bauer99T2,Bauer02}. Hauptgedanke dieser Näherung ist es, den ursprünglichen Diffusionsprozess durch einen anderen stochastischen Prozess zu ersetzen. Zweckmäßigerweise betrachtet man aber nicht die Magnetisierung $M(t)$, sondern deren Laplace-Transformierte
\begin{equation}
\label{Laplacetransform}
\hat{M}(s) = \int_0^\infty \text{d}t \, \text{e}^{-st}M(t)\,.
\end{equation}
In früheren Arbeiten \cite{Bauer99,Bauer99PRL} gelang es, eine Ausdruck für die Laplace-Transformierte der Magnetisierung zu finden:
\begin{equation}
\label{Mdachallgemein}
\hat{M}(s) = \frac{1+\eta}{\sqrt{[s+\tau^{-1}]^2 + \eta^2\, \delta\omega^2} + \eta \sqrt{[s+\tau^{-1}]^2 + \delta\omega^2} - \tau^{-1}[1+\eta]} \,.
\end{equation}
Unter der Annahme eines monoexponentiellen Abfalls der Magnetisierung in der Form $M(t)=M(0) \text{e}^{-t/T_2^{'}}$ folgt mit Gl. (\ref{Laplacetransform}) der Zusammenhang $T_2^{'}=\hat{M}(0)$, bzw. mit $s=0$ in Gl. (\ref{Mdachallgemein}):
\begin{equation}
\label{T2strich}
T_2^{'} = \tau \frac{1+\eta}{\sqrt{1 + \eta^2\, \tau^2\delta\omega^2} + \eta \sqrt{1 + \tau^2\delta\omega^2} - 1 - \eta} \,.
\end{equation}
Der Diffusionsprozess um die Kapillare wird durch die Korrelationszeit $\tau$ beschrieben. Diese Korrelationszeit erlaubt es, zwischen verschiedenen Diffusionsregimen zu unterscheiden \cite{Ziener06MRI}. Im Motional-Narrowing-Regime (hohe Diffusion) strebt die Korrelationszeit gegen Null und der Einfluss der Suszeptibilitätseffekte mittelt sich durch die schnelle Bewegung der umgebenden Protonen heraus. In diesem Grenzfall $\tau \delta\omega = 0$ ergibt sich aus Gl. (\ref{Mdachallgemein}) für die Magnetisierung der Ausdruck
\begin{equation}
\label{MdachMN}
\hat{M}(s)=\frac{1}{s} \,.
\end{equation}
Im entgegengesetzten Grenzfall werden die Diffusion und damit die Bewegung der umgebenden Protonen vernachlässigt, d.h. sie befinden sich immer an der gleichen Position. Dementsprechend wird der Dephasierungsprozess nur durch die Suszeptibilitätseffekte beeinflusst. In diesem Grenzfall strebt die Korrelationszeit gegen unendlich ($\tau \delta\omega \to \infty$) und die Laplace-Transformierte der Magnetisierung ergibt sich nach Gl. (\ref{Mdachallgemein}) zu
\begin{equation}
\label{MdachSD}
\hat{M}(s) = \frac{1}{1-\eta}\frac{1}{s}\left[\sqrt{1+\frac{\eta^2\delta\omega^2}{s^2}} - \eta \sqrt{1+\frac{\delta\omega^2}{s^2}}\right] \,.
\end{equation}
Die inverse Laplace-Transformation lässt sich durchführen und für den Magnetisierungs-Zeit-Verlauf im Static-Dephasing-Regime erhält man den Ausdruck
\begin{align}
\label{FIDcyl}
\hat{M}(t) & =  \frac{\eta}{1-\eta} \int\limits_\eta^1 \frac{\mathrm{d}x}{x^2} J_0(x\delta\omega t) \\
& = \frac{1}{1-\eta}\left[ \, _1F_2\left(-\frac{1}{2};\frac{1}{2},1;-\left[\frac{\eta \delta \omega t}{2}\right]^2\right) - \eta \, _1F_2\left(-\frac{1}{2};\frac{1}{2},1;-\left[\frac{\delta \omega t}{2}\right]^2\right) \right]\,,
\end{align}
wobei die verallgemeinerte hypergeometrische Funktion oder auch Barnes erweiterte hypergeometrische Funktion durch
\begin{equation} \label{BEHF}
 _1F_2 \left(a;b,c;x \right) = \sum_{k=0}^{\infty}\frac{(a)_{k}}{(b)_{k} (c)_{k}} \, \frac{x^k}{k!}
\end{equation}
gegeben ist \cite{Oberhettinger72}. Das Pochhammer-Symbol ist definiert durch
\begin{equation} 
\label{PS}
(x)_k = \frac{\Gamma(x+k)}{\Gamma(x)} \,.
\end{equation}

Um nun den Magnetisierungs-Zeit-Verlauf $M(t)$ für jedes Diffusionsregime zu erhalten, kann man den Formalismus der diffusionsabhängigen Frequenzverteilungen nutzen \cite{Ziener07PRE}. Dabei lässt sich feststellen, dass der Signal-Zeit-Verlauf nicht immer monoexponentiell ist, sondern unter bestimmten Bedingungen ein oszillierendes Verhalten zeigt. Um besser zu verstehen, unter welchen Voraussetzungen ein bestimmtes Verhalten des Magnetisierungszerfalls zu beobachten ist, wird die inverse Laplace-Transformation von Gl. (\ref{Mdachallgemein}) analysiert. Im Zuge nachfolgender Analyse wird es möglich sein, die Singularitäten der Laplace-Transformierten der Magnetisierung $\hat{M}(s)$ mit dem zeitlichen Verhalten der Magnetisierung $M(t)$ zu verknüpfen.

\begin{figure}
\begin{center}
\includegraphics[width=\textwidth]{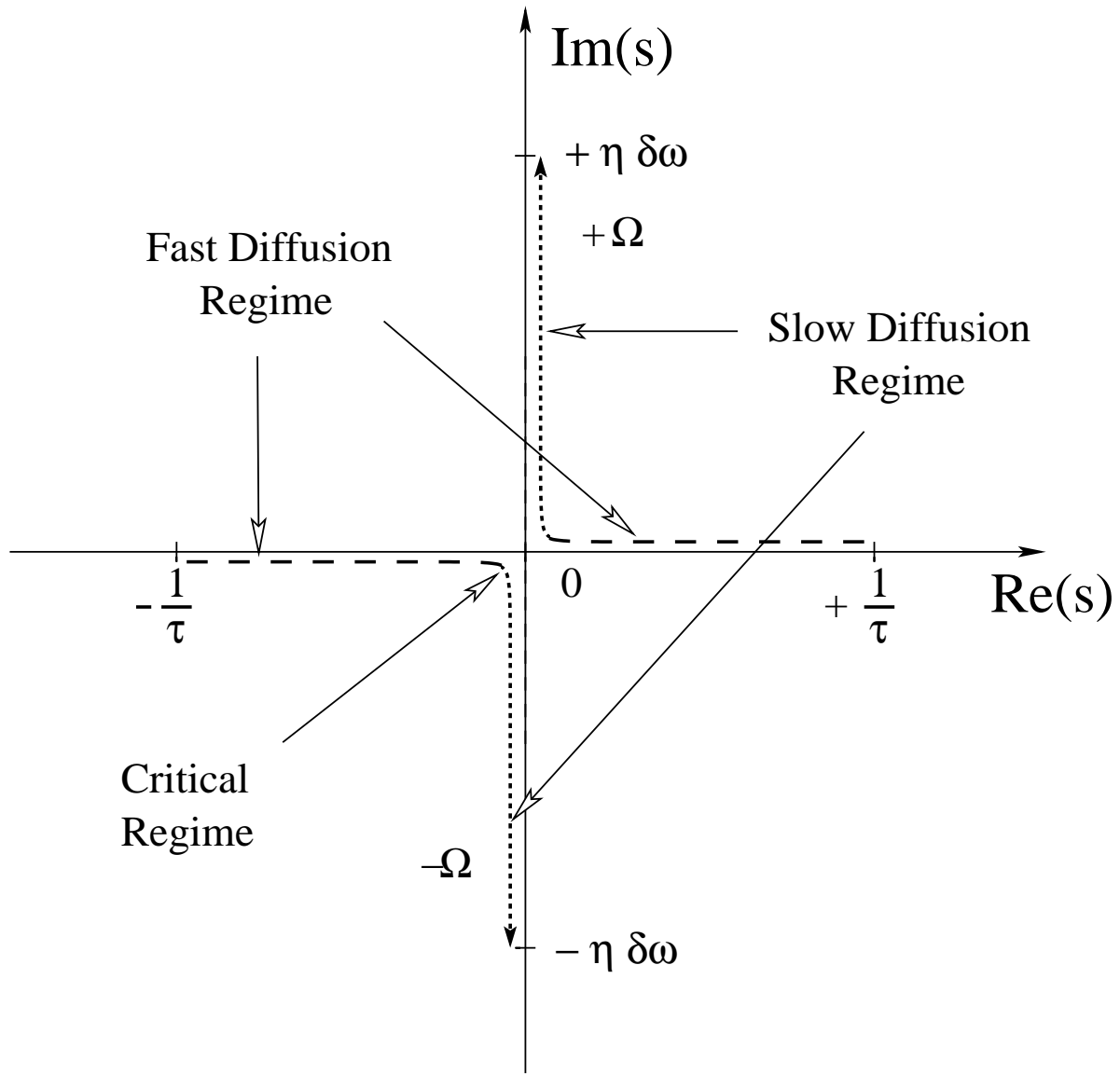}
\vspace{-0.5cm}
\caption[Klassifikation der Diffusionsregime]{\footnotesize Klassifikation der Diffusionsregime. Die Bewegungen der Singularitäten $+\Omega$ und $-\Omega$ bei Veränderung des Diffusionsregimes sind dargestellt. Startpunkt der Bewegungen ist das Motional-Narrowing-Regime ($+\Omega=+1/\tau$ und $-\Omega=-1/\tau$). Mit abnehmenden Diffusionskoeffizienten bewegen sich die Singularitäten auf der gestrichelten Linie (Fast Diffusion Regime) entlang der reellen Achse bis zum Koordinatenursprung (Critical Regime). Danach bewegen sich die Singularitäten auf der gepunkteten Linie (Slow Diffusion Regime) parallel zur imaginären Achse. Im Static-Dephasing-Grenzfall ($D=0$) streben die Singularitäten zu den Werten $+\Omega \to + \mathrm{i}\infty$ und $-\Omega \to - \mathrm{i}\infty$. Bei der Dephasierung im Myokard liegen die Singularitäten nach Gl. (\ref{Omega}) auf der reellen Achse an den Positionen $+\Omega=+191\, \text{s}^{-1}$ und $-\Omega=-191\, \text{s}^{-1}$.}
\label{fig:4_neu}
\end{center}
\end{figure}

Die Berechnung der inversen Laplace-Transformation von $\hat{M}(s)$ in Gl. (\ref{Mdachallgemein}) kann vereinfacht werden, da die Variable $s$ nur in der Kombination $s+\tau^{-1}$ in  Gl. (\ref{Mdachallgemein}) eingeht. Deshalb kann entsprechend dem Verschiebungstheorem der Laplace-Transformation die Magnetisierung in der Form
\begin{equation}
\label{mf}
M(t) = \text{e}^{-\frac{t}{\tau}}F(t)
\end{equation}
geschrieben werden, wobei die Laplace-Transformierte von $F(t)$ aus Gl. (\ref{Mdachallgemein}) mit der Substitution $s+\tau^{-1} \rightarrow s$ folgt: 
\begin{equation}
\label{Fdachallgemein}
\hat{F}(s) = \frac{1+\eta}{\sqrt{s^2 + \eta^2\, \delta\omega^2} + \eta \sqrt{s^2 + \delta\omega^2} - \frac{1+\eta}{\tau}} \,.
\end{equation}
Um nun die inverse Laplace-Transformation durchzuführen, ist es vorteilhaft, $\hat{F}(s)$ aus Gl. (\ref{Fdachallgemein}) als Quotienten zweier Polynome in der Form
\begin{equation}
\label{Fdachfraq}
\hat{F}(s) = \frac{1}{[1-\eta]^2\tau^3} \frac{g(s)}{[s^2-\Omega^2][s^2-\Phi^2]}
\end{equation}
darzustellen, mit dem Zähler
\begin{align}
\label{g}
g(s) = \left[ s^2 \tau^2 [\eta -1] - \eta -1 - 2\tau\sqrt{s^2 + \eta^2 \, \delta\omega^2} \right]\!\left[ 1 + \eta + \eta \tau \sqrt{s^2 + \delta\omega^2} - \tau \sqrt{s^2 + \eta^2 \,\delta\omega^2} \right]
\end{align}
und den Nullstellen des Nenners
\begin{equation}
\label{Omega}
\Omega \!=\! \frac{\sqrt{\!1 \!+\! \eta^2 \!-\!2\eta\sqrt{1 + \tau^2 \delta\omega^2[1-\eta]^2}}}{\tau[1-\eta]} \,\,\,\,\,\text{und}\,\,\,\,\, \Phi \!= \! \frac{\sqrt{\!1 \!+\! \eta^2 \!+\! 2\eta\sqrt{1 + \tau^2\delta\omega^2[1-\eta]^2}}}{\tau[1-\eta]} \,.
\end{equation}
Die Nullstellen des Nenners haben die Einheit einer Frequenz. Die Größe $\Phi$ nimmt für alle Parameter immer reelle Werte an. Die Frequenz $\Omega$ kann komplexe Werte annehmen und es lassen sich drei verschiedene Fälle unterscheiden. Diese Fallunterscheidung, ob $\Omega$ rein reell, gleich Null oder rein imaginär ist, wird zur Einteilung der Diffusionsregime in folgender Weise genutzt:
\begin{alignat}{4}
\label{1}
\tau\,\delta\omega & < \frac{1+\eta}{2\eta} &&: \Omega \in \mathds{R} \quad\quad \Rightarrow \text{Fast-Diffusion-Regime}\,, \\
\label{2}
\tau\,\delta\omega & = \frac{1+\eta}{2\eta} &&: \Omega =0 \quad\quad\Rightarrow \text{Critical-Regime}\,\,\,\,\text{und} \\
\label{3}
\tau\,\delta\omega & > \frac{1+\eta}{2\eta} &&: \Omega = \mathrm{i} |\Omega| \quad\Rightarrow \text{Slow-Diffusion-Regime}\,.
\end{alignat}
Demzufolge ist $\Omega$ entweder rein reell oder rein imaginär. In Abb. \ref{fig:4_neu} ist die Lage der Singularitäten $+\Omega$ und $-\Omega$ in der komplexen Ebenen dargestellt. 
Wie in Abschnitt 2.1 beschrieben, beträgt die Korrelationszeit im menschlichen Myokard $\tau=5,\!1\,\text{ms}$. Die charakteristische Frequenz auf der Oberfläche der Kapillare nimmt bei $B_0=1,\!5\,\text{T}$ den Wert $\delta\omega=151 \, \text{s}^{-1}$ an. Mit dem regionalen Blutvolumenverhältnis $\eta=0,\!084$ ergibt sich für die Frequenz $\Omega$ nach Gl. (\ref{Omega}) der Wert $\Omega=191 \,\text{s}^{-1}$. Demnach ist das Fast-Diffusion-Regime das zu Grunde liegende Diffusionsregime im Myokard. 

Die Funktion $g(s)$ besitzt die Nullstellen $g(+\Phi)=0$ und $g(-\Phi)=0$. Es lässt sich zeigen, dass in Gl. (\ref{Fdachfraq}) der Ausdruck $g(s)/[s^2-\Phi^2]$ keine Singularitäten besitzt. Deshalb hat die Laplace-Transformierte $\hat{F}(s)$ nur die zwei Singularitäten $+\Omega$ und $-\Omega$, die entweder rein reelle oder rein imaginäre Werte annehmen können.

\begin{figure}
\begin{center}
\includegraphics[width=\textwidth]{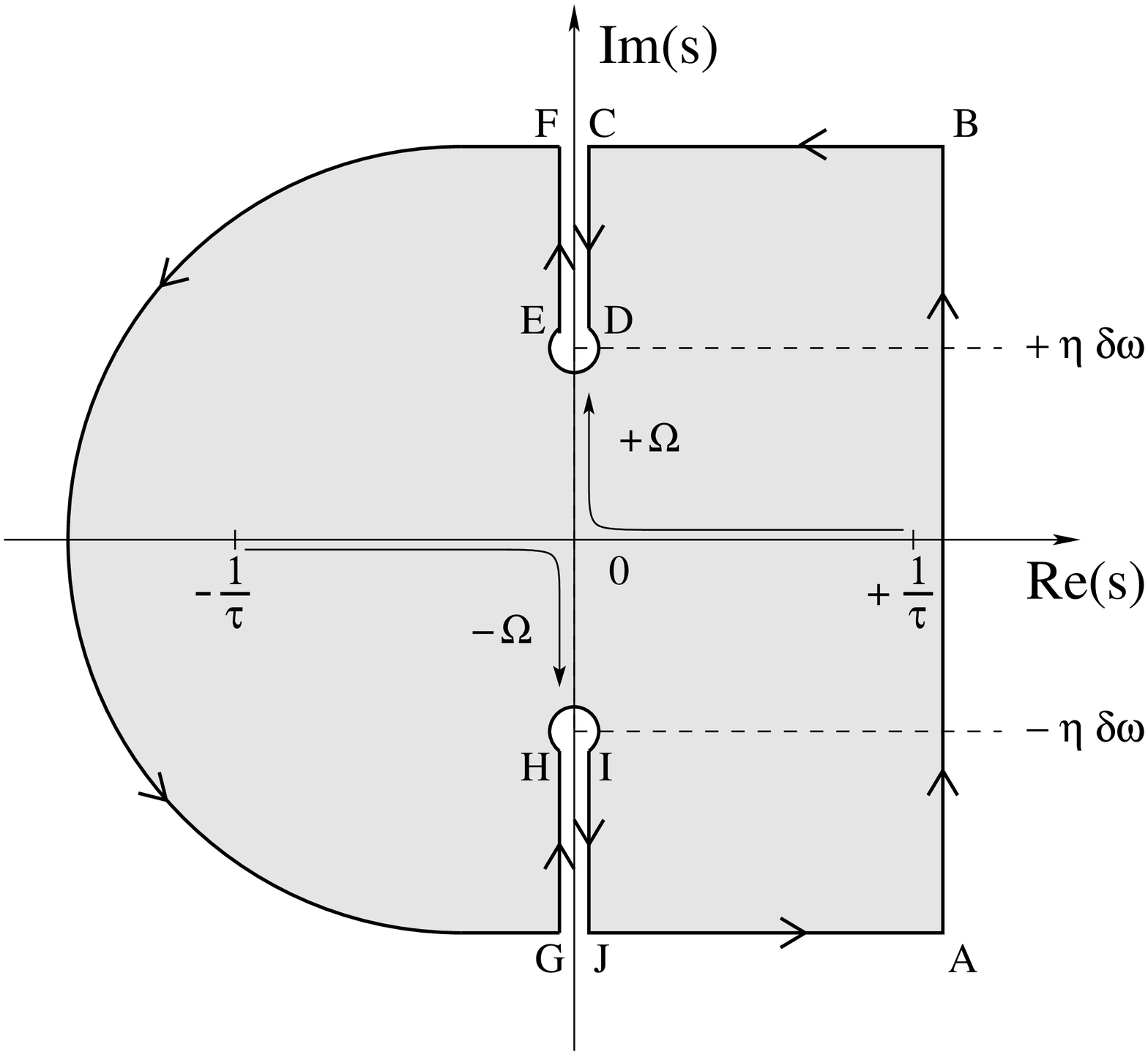}
\vspace{-0.5cm}
\caption[Bromwich-Linie zur Auswertung des Integrals (\ref{BromwichIntegral})]{\footnotesize Bromwich-Linie zur Auswertung des Integrals (\ref{BromwichIntegral}). Wie in Abb. \ref{fig:4_neu} sind die Bewegungen der Singularitäten $+\Omega$ und $-\Omega$ bei Veränderung des Diffusionsregimes dargestellt. Startpunkt der Bewegungen ist das Motional-Narrowing-Regime ($+\Omega=+1/\tau$ und $-\Omega=-1/\tau$). Im Fast-Diffusion-Regime bewegen sich die Singularitäten entlang der reellen Achse bis zum Koordinatenursprung, wo sie sich im Critical-Regime treffen. Im Slow-Diffusion-Regime bewegen sich die Singularitäten parallel zur imaginären Achse. Sobald die Frequenz $\Omega$ den Wert $\text{i}\eta\delta\omega$ auf der imaginären Achse erreicht, wandern die Singularitäten aus dem Integrationsbereich heraus, und man befindet sich definitionsgemäß im Strong-Dephasing-Regime. Im Static-Dephasing-Grenzfall ($D=0$) streben die Singularitäten zu den Werten $+\Omega \to + \mathrm{i}\infty$ und $-\Omega \to - \mathrm{i}\infty$.}
\label{fig:2}
\end{center}
\end{figure}

Um die Funktion $F(t)$ zu erhalten, wird die Mellin-Formel
\begin{align}
\label{Mellin}
F(t) = \frac{1}{2\pi\text{i}} \int\limits_{1/\tau-\text{i}\infty}^{1/\tau+\text{i}\infty} \mathrm{d}s\ \mathrm{e}^{st} \hat{F}(s)
\end{align}
herangezogen, wobei die reelle Zahl $1/\tau$ rechts aller Singularitäten liegt. Das Integral in Gl. (\ref{Mellin}) kann ausgewertet werden, indem man die Differenz
\begin{align}
\label{BromwichIntegral}
F(t) = \frac{1}{2\pi\text{i}} \ointctrclockwise \mathrm{d}s\ \mathrm{e}^{st} \hat{F}(s) - \frac{1}{2\pi\text{i}} \int\limits_{\Gamma} \mathrm{d}s\ \mathrm{e}^{st} \hat{F}(s) \,,
\end{align}
betrachtet, wobei der Integrationsweg, die sog. Bromwich-Linie, in Abb. \ref{fig:2} dargestellt ist und der Integrationsweg $\Gamma$ durch die Kontur $BCDEFGHIJA$ beschrieben wird.
Der Integrationsweg zur Berechnung des Bromwich-Integrals muss so gewählt werden, dass alle Singularitäten der zu transformierenden Funktion $\hat{F}(s)$ kleinere Realteile als die Linie $AB$ des Integrationswegs in Abb. \ref{fig:2} haben \cite{Spiegel77}. Der Integrationsweg wird über der linken Halbebene geschlossen \cite{Spiegel76}. Demzufolge liegen die Residuen der beiden Singularitäten $+\Omega$ und $-\Omega$ im Integrationsbereich und tragen zum Integral bei.

\begin{figure}
\begin{center}
\includegraphics[width=13cm]{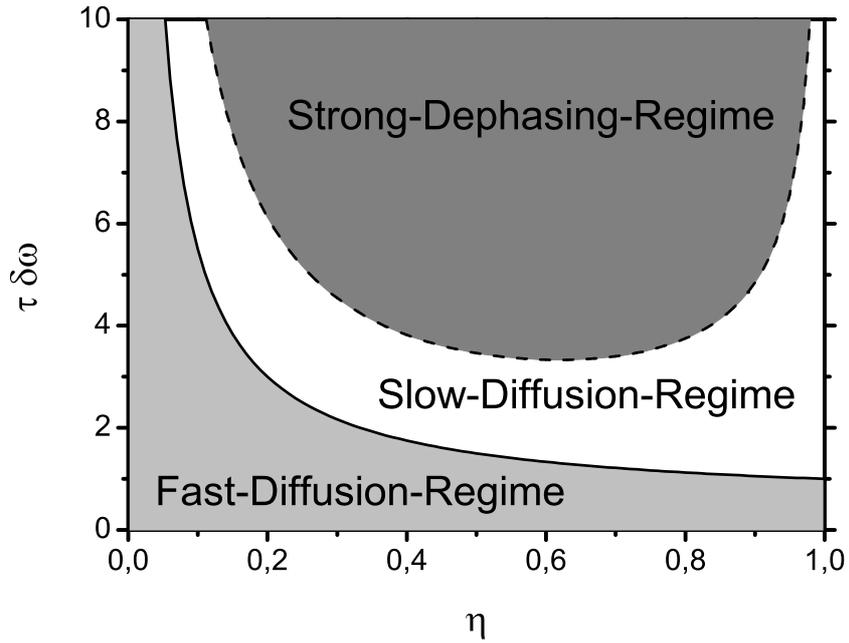}
\vspace{-1.0cm}
\caption[Einteilung der Diffusionsregime]{\footnotesize Einteilung der Diffusionsregime. Die durchgezogene Kurve ($\tau\delta\omega=[1+\eta]/[2\eta]$) charakterisiert das Critical-Regime entsprechend Gl. (\ref{2neu}). Die gestrichelte Kurve ($\tau\delta\omega=\sqrt{[1+\eta]/[1-\eta]}/\eta$) markiert die Grenze zwischen dem Slow-Diffusion-Regime (siehe Bedingung (\ref{3neu})) und dem Strong-Dephasing-Regime (siehe Bedingung (\ref{4neu})).} 
\label{fig:Bereiche}
\end{center}
\end{figure}

Des Weiteren muss berücksichtigt werden, dass die Funktion $\hat{F}(s)$ in Gl. (\ref{Fdachfraq}) die beiden Verzweigungslinien $CDEF$ und $GHIJ$ auf der imaginäre Achse aufweist. Diese Verzweigungslinien ergeben sich aus den Quadratwurzeln, die in der Funktion $g(s)$ in Gl. (\ref{g}) enthalten sind und beginnen auf der imaginären Achse bei den Werten $\pm \text{i}\eta\,\delta\omega$. Das Ringintegral in Gl. (\ref{BromwichIntegral}) kann mit Hilfe des Residuensatzes als Summe aller Residuen berechnet werden, die innerhalb des geschlossenen Integrationsweges liegen. Sobald sich jedoch die Punkte $\pm \Omega$ auf der imaginären Achse in die Verzweigungslinien hineinbewegen, also $+ |\Omega| > + \eta \delta\omega$ und $- |\Omega| < - \eta \delta\omega$, liegen die Singularitäten $\pm \Omega$ außerhalb des Integrationsgebietes und deren Residuen tragen nicht zur Berechnung des Integrals bei. Dieser Fall ($|\Omega|>\eta\delta\omega$) wird als das Strong-Dephasing-Regime bezeichnet. Zur Magnetisierung trägt also nur die Integration um die Verzweigungslinien bei. Mit Hilfe der Definition von $\Omega$ aus Gl. (\ref{Omega}) lässt sich ermitteln, dass das Strong-Dephasing-Regime vorliegt, wenn $\tau\delta\omega \geq \sqrt{[1+\eta]/[1-\eta]}/\eta$. Damit lässt sich die Einteilung der Diffusionsregime nach den Beziehungen (\ref{1}) - (\ref{3}) erweitern:
\begin{alignat}{4}
\label{1neu}
0 < \tau\,\delta\omega & < \frac{1+\eta}{2\eta} &&\quad\quad\Rightarrow \text{Fast-Diffusion-Regime}\,, \\[1ex]
\label{2neu}
\tau\,\delta\omega & = \frac{1+\eta}{2\eta} &&\quad\quad\Rightarrow \text{Critical-Regime}\,, \\
\label{3neu}
\frac{1+\eta}{2\eta} < \tau\,\delta\omega & < \frac{1}{\eta}\sqrt{\frac{1+\eta}{1-\eta}} &&\quad\quad\Rightarrow \text{Slow-Diffusion-Regime} \,\,\,\,\text{und} \\[0ex]
\label{4neu}
\frac{1}{\eta}\sqrt{\frac{1+\eta}{1-\eta}} \leq \tau\,\delta\omega & < \infty &&\quad\quad\Rightarrow \text{Strong-Dephasing-Regime}\,.
\end{alignat}
In Abb. \ref{fig:Bereiche} sind die so definierten Diffusionsregime dargestellt.

Letztendlich kann die inverse Laplace-Transformation in der Form
\begin{align}
F(t) = \sum \text{Residuen von } \mathrm{e}^{st} \hat{F}(s) - \frac{1}{2\pi\text{i}} \left[ \int\limits_{CDEF} \mathrm{d}s\ \mathrm{e}^{st}\hat{F}(s) + \int\limits_{GHIJ} \mathrm{d}s\ \mathrm{e}^{st}\hat{F}(s) \right]
\end{align}
geschrieben werden. Demzufolge kann im Fall $\Omega \neq 0$ die Funktion $F(t)$ in der Form
\begin{align}
\label{Mres}
F(t) = \mathrm{res}\left(\mathrm{e}^{st} \hat{F}(s);+\Omega\right)+\mathrm{res}\left(\mathrm{e}^{st} \hat{F}(s);-\Omega\right) +h(t)
\end{align}
geschrieben werden, wobei $\mathrm{res}\left(\mathrm{e}^{st} \hat{F}(s);+\Omega\right)$ und $\mathrm{res}\left(\mathrm{e}^{st} \hat{F}(s);-\Omega\right)$ die Residuen der Singularitäten sind und
\begin{align}
\label{hint}
-2\pi\text{i}h(t) = \int\limits_{CDEF} \mathrm{d}s\ \mathrm{e}^{st}\hat{F}(s) + \int\limits_{GHIJ} \mathrm{d}s\ \mathrm{e}^{st}\hat{F}(s)
\end{align}
der Beitrag der Integration um die Verzweigungslinien ist.

Wenn beide Singularitäten übereinstimmen $(\Omega = 0)$, dann werden die beiden Pole erster Ordnung ein einziger Pol zweiter Ordnung und die Funktion $F(t)$ kann in der Form
\begin{align}
\label{Mnureinres}
F(t) = \mathrm{res}\left(\mathrm{e}^{st} \hat{F}(s);0\right) +h(t)
\end{align}
geschrieben werden. Die Bewegungen der Singularitäten sind eng verknüpft mit dem zu Grunde liegenden Diffusionsregime und stehen in direktem Zusammenhang mit dem vorherrschenden Dephasierungsmechanismus.

Die Berechnung der Residuen an den Stellen $+\Omega$ und $-\Omega$ ist unkompliziert:
\begin{align}
\nonumber
\mathrm{res}\left(\mathrm{e}^{st} \hat{F}(s);+\Omega\right) & = \mathrm{res}\left(\mathrm{e}^{st} \frac{1}{[1-\eta]^2 \tau^3} \frac{g(s)}{[s^2-\Omega^2][s^2-\Phi^2]} ;+\Omega\right)\\
\label{alpha}
& = +\frac{\mathrm{e}^{+\Omega t}}{2\tau\Omega} \underbrace{\frac{1}{[1-\eta]^2 \tau^2}\frac{g(\Omega)}{\Omega^2-\Phi^2}}_{= +\alpha}\\[-2ex]
\label{res1}
& = +\frac{\mathrm{e}^{+\Omega t}}{2\tau\Omega} \alpha \\
\label{res2}
\mathrm{res}\left(\mathrm{e}^{st} \hat{F}(s);-\Omega\right) & = -\frac{\mathrm{e}^{-\Omega t}}{2\tau\Omega} \alpha \,.
\end{align}
Der in Gl. (\ref{alpha}) definierte Vorfaktor $\alpha$ kann aus Gl. (\ref{g}) mit Hilfe der Ausdrücke für $\Omega$ und $\Phi$ aus Gl. (\ref{Omega}) ermittelt werden und ergibt sich zu
\begin{equation}
\label{alphaalt}
\alpha = \left[\Omega^2\tau^2[1-\eta]+1+\eta+2\tau\sqrt{\Omega^2+\eta^2\delta\omega^2} \right] \frac{ 1+\eta+\eta\tau\sqrt{\Omega^2+\delta\omega^2}-\tau\sqrt{\Omega^2+\eta^2\delta\omega^2}}{4\eta\sqrt{1+\tau^2 \delta\omega^2 [1-\eta]^2}} \,.
\end{equation}
Der erste Faktor in eckigen Klammern der Gl. (\ref{alphaalt}) verschwindet, wenn die Singularitäten das Riemann-Blatt verlassen, d.h. wenn die Bedingung (\ref{4neu}) erfüllt ist. Ab diesem Punkt wechselt die Quadratwurzel, die zur Verzweigungslinie führt, ihr Vorzeichen. In diesem Fall nimmt der Zähler des zweiten Faktors den Wert $2[1+\eta]$ an. Insgesamt verschwindet jedoch das Produkt. Für
\begin{equation}
\tau\delta\omega<\frac{1}{\eta}\sqrt{\frac{1+\eta}{1-\eta}}
\end{equation}
können mit Hilfe des Ausdrucks für $\Omega$ in Gl. (\ref{Omega}) die Faktoren der Gl. (\ref{alphaalt}) folgendermaßen vereinfacht werden:
\begin{align}
\Omega^2\tau^2[1-\eta]+1+\eta+2\tau\sqrt{\Omega^2+\eta^2\delta\omega^2} & = 4 \frac{1-\eta\sqrt{1+\tau^2 \delta\omega^2 [1-\eta]^2}}{1-\eta}\\[2ex]
\text{und} \quad 1+\eta+\eta\tau\sqrt{\Omega^2+\delta\omega^2}-\tau\sqrt{\Omega^2+\eta^2\delta\omega^2} & = 2\eta \frac{\eta\sqrt{1+\tau^2 \delta\omega^2 [1-\eta]^2}-\eta}{1-\eta} \,.
\end{align}
Für
\begin{equation}
\tau\delta\omega \geq \frac{1}{\eta}\sqrt{\frac{1+\eta}{1-\eta}}
\end{equation}
sind die Faktoren in Gl. (\ref{alphaalt}) durch Nutzung des entgegengesetzten Vorzeichens der Quadratwurzel im entsprechenden Ausdruck vereinfachbar:
\begin{align}
\Omega^2\tau^2[1-\eta]+1+\eta+2\tau\sqrt{\Omega^2+\eta^2\delta\omega^2} & = 0 \\[2ex]
\text{und} \quad 1+\eta+\eta\tau\sqrt{\Omega^2+\delta\omega^2}-\tau\sqrt{\Omega^2+\eta^2\delta\omega^2} & = 2[1+\eta] \,.
\end{align}
Setzt man diese Resultate in Gl. (\ref{alphaalt}) ein, erkennt man, dass der Vorfaktor $\alpha$ ein rein reeller und positiver Parameter ist, der in der Form
\begin{equation}
\label{alphaneubereich}
\alpha = \left\{ \begin{array}{ll}
\displaystyle{\frac{2}{[1-\eta]^2}\left[1+\eta^2-\eta\frac{2+\tau^2\delta\omega^2[1-\eta]^2}{\sqrt{1+\tau^2\delta\omega^2[1-\eta]^2}} \right]} & \quad\text{für}\quad \displaystyle{ \tau\delta\omega < \frac{1}{\eta}\sqrt{\frac{1+\eta}{1-\eta}}}\\[4ex]
0 & \quad\text{sonst}
\end{array} \right.
\end{equation}
geschrieben werden kann.

Im Critical-Regime fallen beide Singularitäten zusammen, was zu einem einzigen Pol zweiter Ordnung im Punkt $\Omega=0$ im Koordinatenursprung führt. Deshalb wird das Residuum im Sinne der ersten Ableitung in diesem Punkt auf folgende Weise bestimmt:
\begin{align}
\nonumber
\mathrm{res}\left(\mathrm{e}^{st} \hat{F}(s);0\right) & = \left. \frac{\mathrm{d}}{\mathrm{d}s}\left( s^2 \mathrm{e}^{st} \hat{F}(s) \right) \right|_{s=0}\\
\nonumber
& = \left. \frac{\mathrm{d}}{\mathrm{d}s}\left(\mathrm{e}^{st} \frac{1}{[1-\eta]^2 \tau^3} \frac{g(s)}{s^2-\Phi^2} \right) \right|_{s=0}\\
\nonumber
& = \frac{1}{[1-\eta]^2 \tau^3} \left. \frac{\mathrm{d}}{\mathrm{d}s}\left( \frac{\mathrm{e}^{st} g(s)}{s^2-\Phi^2}\right) \right|_{s=0} \\
\label{5}
& = \frac{2[1+\eta]^2}{\tau^3[1-\eta]^2 \Phi^2}t\\
\label{resIR}
& = \frac{[1+\eta]^2}{1+\eta^2} \frac{t}{\tau}\,,
\end{align}
wobei $\Phi$ aus Gl. (\ref{Omega}) in Gl. (\ref{5}) eingesetzt wurde.

Die Integration um die Enden der Verzweigungslinien bei $\pm \text{i}\eta\delta\omega$ lassen sich auswerten, indem in einer $\varepsilon$-Umgebung um die Enden integriert wird. Zur Berechnung des Integrals um $+\text{i}\eta\delta\omega$ von $D$ bis $E$ wählt man die Parametrisierung des Weges als $s=+\text{i}\eta\delta\omega + \varepsilon \text{e}^{\text{i}\phi}$. Damit ergibt sich:
\begin{equation}
\int_D^E \mathrm{d}s\ \mathrm{e}^{st}\hat{F}(s) = -\int_E^D \mathrm{d}s\ \mathrm{e}^{st}\hat{F}(s) = -\text{i} \lim_{\varepsilon \to 0} \varepsilon \int_{\frac{\pi}{2}}^{\frac{5\pi}{2}}\mathrm{d}\phi\ \mathrm{e}^{\left(\text{i}\eta\delta\omega+\varepsilon \text{e}^{\text{i}\phi}\right)}\hat{F}\left(\text{i}\eta\delta\omega+\varepsilon \text{e}^{\text{i}\phi}\right) \text{e}^{\text{i}\phi} =0\,.
\end{equation}
Analog berechnet sich das Integral um das Ende der anderen Verzweigungslinie und letztlich erhält man:
\begin{equation}
\int_D^E \mathrm{d}s\ \mathrm{e}^{st}\hat{F}(s) = 0 = \int_H^I \mathrm{d}s\ \mathrm{e}^{st}\hat{F}(s) \,.
\end{equation}
Um die Integrale längs des Integrationswegs parallel zur imaginären Achse auszuwerten ist es notwendig, die Symmetrieeigenschaften der Funktion $\hat{F}$ aus Gl. (\ref{Fdachallgemein}) bzw. Gl. (\ref{Fdachfraq}) in der $\varepsilon$-Umgebung der Verzweigungslinien zu analysieren. Die Realteile nehmen immer denselben Wert an, während die Imaginärteile das Vorzeichen wechseln:
\begin{align}
\nonumber
&\text{Re}\left[\hat{F}\left(+\varepsilon+\mathrm{i}y\right)\right] \quad\quad\quad\quad \,\,\,\,\, \text{Im}\left[\hat{F}\left(+\varepsilon+\mathrm{i}y\right)\right]\\
\nonumber
=&\text{Re}\left(\hat{F}\left(-\varepsilon+\mathrm{i}y\right)\right) \quad\quad\quad =\, -\text{Im}\left(\hat{F}\left(-\varepsilon+\mathrm{i}y\right)\right)\\
\nonumber
=&\text{Re}\left(\hat{F}\left(+\varepsilon-\mathrm{i}y\right)\right) \quad\quad\quad =\, -\text{Im}\left(\hat{F}\left(+\varepsilon-\mathrm{i}y\right)\right)\\
\nonumber
=&\text{Re}\left(\hat{F}\left(-\varepsilon-\mathrm{i}y\right)\right) \quad\quad\quad= \,\,\,\,\,\text{Im}\left(\hat{F}\left(-\varepsilon-\mathrm{i}y\right)\right) \,.
\end{align}
Werden diese Symmetrierelationen in das Integral um die Verzweigungslinien eingesetzt, verbleiben nur die Imaginärteile und es ergibt sich:
\begin{align}
\nonumber
& \int_C^D \mathrm{d}s\ \mathrm{e}^{st}\hat{F}(s) + \int_E^F \mathrm{d}s\ \mathrm{e}^{st}\hat{F}(s) + \int_G^H \mathrm{d}s\ \mathrm{e}^{st}\hat{F}(s) + \int_I^J \mathrm{d}s\ \mathrm{e}^{st}\hat{F}(s) \\ 
\nonumber
& = 4\mathrm{i} \int_{\eta\,\delta\omega}^\infty \ \sin(yt)\,\, \lim_{\varepsilon\to 0}\ \text{Im}\left(\hat{F}\left(+\varepsilon+\mathrm{i}y\right)\right) \mathrm{d}y \\
\label{Integralbranch}
& =-2\pi\text{i}h(t).
\end{align}
Mit dem allgemeinen Ausdruck für die Funktion $\hat{F}(s)$ aus Gl. (\ref{Fdachallgemein}) folgt für den obigen Grenzwert des Imaginärteils der Ausdruck
\begin{equation}
\nonumber
\lim_{\varepsilon\to 0}\ \text{Im}\left(\hat{F}\left(+\varepsilon+\mathrm{i}y\right)\right) = \left\{ \begin{array}{ll}
\displaystyle{\frac{-[1+\eta]\sqrt{y^2-\eta^2\,\delta\omega^2}}{\left[ \eta\sqrt{\delta\omega^2-y^2} - \frac{1+\eta}{\tau} \right]^2 + y^2 - \eta^2\delta\omega^2}} & \quad\text{für}\quad \displaystyle{\eta \delta\omega < y < \delta\omega}\\[4ex]
\displaystyle{\frac{-[1+\eta]\left[\sqrt{y^2-\eta^2\,\delta\omega^2} + \eta\sqrt{y^2-\delta\omega^2}\right]}{\left[ \sqrt{y^2-\eta^2\,\delta\omega^2} + \eta\sqrt{y^2-\delta\omega^2} \right]^2 + \left[ \frac{1+\eta}{\tau} \right]^2}} & \quad\text{für}\quad \displaystyle{y > \delta\omega} \,.
\end{array} \right.
\end{equation}
Letztlich ergibt sich (mit der Substitution $y=x\,\delta\omega$) für die Funktion $h(t)$ in Gl. (\ref{Integralbranch}):
\begin{align}
\label{h}
h(t) & = \frac{2}{\pi} \int\limits_\eta^1 \mathrm{d}x \,\, \frac{\sin(x\,\delta\omega\,t) \sqrt{x^2 - \eta^2}}{x^2[1-\eta]+ \displaystyle{\frac{1+\eta}{\tau^2\,\delta\omega^2}} - \displaystyle{\frac{2\eta}{\tau\,\delta\omega}\sqrt{1-x^2}}}  \\ \nonumber
& + \frac{2}{\pi} \int\limits_1^\infty 	\mathrm{d}x \,\, \frac{\sin(x\,\delta\omega\, t)\left[ \sqrt{x^2-\eta^2} + \eta\sqrt{x^2-1} \right]}{\displaystyle{\frac{x^2[1+\eta^2] + 2\eta \left[\sqrt{[x^2-\eta^2][x^2-1]} - \eta \right]}{1+\eta}} + \displaystyle{\frac{1+\eta}{\tau^2\,\delta\omega^2}}} \,.
\end{align}

\section*{\normalsize{2.4 Separation der Bloch-Torrey-Gleichung}}
\addcontentsline{toc}{section}{2.4 Separation der Bloch-Torrey-Gleichung}

Da die zu Grunde liegende Bloch-Torrey-Gleichung (\ref{BT}) die Struktur einer Schrödingergleichung hat, ist es möglich, in Analogie zu den Verhältnissen beim Wasserstoffatom einen Separationsansatz in der Form
\begin{equation} 
\label{sep2}
m(\textbf{r},t) = T(t) \cdot R(r) \cdot \Phi(\phi)
\end{equation}
zu wählen. Die Separation des winkelabhängigen Anteils führt auf die Mathieusche Differenzialgleichung
\begin{equation} 
\label{mat}
\partial_{\phi\phi}\Phi_m(\phi)+\left[ k_m^2 - \text{i} \, \frac{\delta \omega R_{\text{C}}^2}{D}  \cos (2\phi) \right]\Phi_m(\phi)=0
\end{equation}
mit der Separationskonstanten $k_m^2$ \cite{Meixner54,McLachlan64}. Weitere interessante Aspekte der Mathieuschen Differenzialgleichung finden sich in den Arbeiten von Strutt \cite{Strutt}, Campbell \cite{Campbell} und Arscott \cite{Arscott}. Die Eigenfunktionen müssen die Anfangsbedingung (\ref{Anfang}) erfüllen, welches genau durch die geraden Mathieu-Funktionen $\text{ce}_m(\phi)$ gewährleistet wird, da die Anfangsbedingung (\ref{Anfang}) eine gerade Funktion ist und die Funktionen $\text{ce}_m(\phi)$ ebenfalls gerade Funktionen sind. Eine visuelle Darstellung der Mathieu-Funktionen findet sich in der Arbeit \cite{Vega}. Weitere Anwendungen der Mathieu-Funktionen sind in der Arbeit \cite{Ruby} zu finden.

Auf Grund der Periodizität $\omega(r,\phi) = \omega(r,\phi + \pi)$ des zweidimensionalen Dipolfeldes in Gl. (\ref{Dipol}) werden nur die $\pi$-periodischen Mathieu-Funktionen $\text{ce}_{2m}(\phi)$ betrachtet \cite{Seeger97}. Deshalb kann die Lösung von Gl. (\ref{mat}) in der Form
\begin{equation} 
\label{Eigen_Phi}
\Phi_m(\phi) = \text{ce}_{2m}(\phi)
\end{equation}
geschrieben werden, wobei die Orthogonalitätsrelation
\begin{equation} 
\label{Ort_Phi}
\int_{0}^{2\pi} \text{d} \phi \, \text{ce}_{2m}(\phi) \text{ce}_{2m^{'}} (\phi) = \pi \delta_{mm^{'}}
\end{equation}
erfüllt ist. 

Die Eigenfunktionen $\text{ce}_{2m}(\phi)$, die der Mathieuschen Differenzialgleichung $\partial_{\phi\phi} \text{ce}_{2m}(\phi) + [a_{2m}-2q\cos(2\phi)]\text{ce}_{2m}(\phi)=0$ genügen, zeigen nur für bestimmte charakteristische Werte $a_{2m}(q)$ die geforderte Periodizität. Um die $\pi$-periodischen Eigenfunktionen $\text{ce}_{2m}(\phi)$ zu erhalten, müssen die charakteristischen Werte $a_{2m}(q)$ mit $q=\text{i}\delta\omega R_{\text{C}}^2 / [2D]$ betrachtet werden. Deshalb ergeben sich die Eigenwerte in Gl. (\ref{mat}) zu
\begin{equation} 
\label{Parameter}
k_m^2 = a_{2m}\left(\frac{\text{i}}{2}\frac{\delta \omega R_{\text{C}}^2}{D} \right)
\end{equation}
und es ergibt sich letztlich mit $\Phi_m(\phi)=\text{ce}_{2m}(\phi)$ aus Gl. (\ref{mat}) für den Winkelanteil die Differenzialgleichung
\begin{equation} 
\label{matdgl}
\partial_{\phi\phi}\text{ce}_{2m}(\phi)+\left[ k_m^2 - \text{i} \, \frac{\delta \omega R_{\text{C}}^2}{D}  \cos (2\phi) \right]\text{ce}_{2m}(\phi)=0 \,.
\end{equation}
Die geraden $\pi$-periodischen Mathieu-Funktionen werden sehr anschaulich in \cite{Hochstadt} beschrieben.

Die Mathieu-Funktionen können in die Fourier-Reihe
\begin{equation} 
\label{FourierReihe}
\text{ce}_{2m}(\phi)=\sum_{r=0}^{\infty} A_{2r}^{(2m)} \cos(2r\phi)
\end{equation}
entwickelt werden, wodurch die Symmetrie $\text{ce}_{2m}(\phi)=\text{ce}_{2m}(-\phi)$ widergespiegelt wird. Beispielsweise können für kleine Parameter $\delta\omega R_C^2/D$ die ersten geraden Mathieu-Funktionen durch
\begin{align} 
\label{MathieuReihe0}
m=0:\; \text{ce}_{0}(\phi) & \approx \frac{1+\left[\delta\omega R_{\text{C}}^2/[8D]\right]^2}{\sqrt{2}} - \frac{\text{i}\delta\omega R_{\text{C}}^2/D}{4\sqrt{2}} \cos(2\phi)-\frac{\left[\delta\omega R_{\text{C}}^2/D\right]^2}{128\sqrt{2}}\cos(4\phi) \,, \\
\label{MathieuReihe2}
m=1:\; \text{ce}_{2}(\phi) & \approx \frac{\text{i} \delta\omega R_{\text{C}}^2/D}{8} + \cos(2\phi) - \frac{\text{i} \delta\omega R_{\text{C}}^2/D}{24} \cos(4\phi)\quad\text{und}\\
\label{MathieuReihe4}
m=2:\; \text{ce}_{4}(\phi) & \approx -\frac{\left[\delta\omega R_{\text{C}}^2/D\right]^2}{768} + \frac{\text{i} \delta\omega R_{\text{C}}^2/D}{24} \cos(2\phi) + \cos(4\phi)
\end{align}
approximiert werden (siehe 20.2.27 und 20.2.28 in \cite{Abramowitz72}). 

Mit der Orthogonalitätsrelation (\ref{Ort_Phi}) folgt für die Fourier-Koeffizienten die Normalisierung:
\begin{equation}
\label{Normalisierung}
2\left[ A_0^{(2m)} \right]^2 + \sum_{r=1}^{\infty} \left[A_{2r}^{(2m)}\right]^2 = 1 \,.
\end{equation}
Setzt man die Entwicklung der Mathieu-Funktionen in der Fourier-Reihe entsprechend Gl. (\ref{FourierReihe}) in die Mathieu-Differenzialgleichung (\ref{matdgl}) ein, erhält man folgende Rekursionsformeln zur Berechnung der Fourier-Koeffizienten und der charakteristischen Werte:
\begin{align}
\label{rekursion1}
2k_m^2 A_{0}^{(2m)} & = \text{i} \frac{\delta\omega R_C^2 }{D} A_{2}^{(2m)}\,,\\
\label{rekursion2}
2\left[ k_m^2 - 4 \right] A_{2}^{(2m)} & = \text{i} \frac{\delta\omega R_C^2}{D} \left[2 A_{0}^{(2m)}+ A_{4}^{(2m)} \right]\quad\text{und}\\
\label{rekursion3}
2\left[ k_m^2 - 4 r^2 \right] A_{2r}^{(2m)} & = \text{i} \frac{\delta\omega R_C^2}{D} \left[ A_{2r-2}^{(2m)}+ A_{2r+2}^{(2m)} \right] \quad \text{für}\quad r \geq 2\,.
\end{align}
Diese Rekursionsformeln lassen sich elegant als Eigenwertproblem in Matrix-Schreibweise darstellen \cite{Chaos}:
\begin{equation}
\nonumber
\left( \begin{array}{llllllll}
0 & \sqrt{2}\text{i}\frac{\delta\omega R_C^2}{2D} &  & \\[1.5ex]
\sqrt{2}\text{i}\frac{\delta\omega R_C^2}{2D} & 4 & \text{i}\frac{\delta\omega R_C^2}{2D} & \\[1.5ex]
 & \text{i}\frac{\delta\omega R_C^2}{2D} & 16 & \text{i}\frac{\delta\omega R_C^2}{2D} \\[1.5ex]
 & & \text{i}\frac{\delta\omega R_C^2}{2D} & 36 & \text{i}\frac{\delta\omega R_C^2}{2D}\\[1.5ex]
& & &\ddots & \ddots & \ddots\\[1.5ex]
& & & & \text{i}\frac{\delta\omega R_C^2}{2D} & 4r^2 &
\text{i}\frac{\delta\omega R_C^2}{2D} \\[1.5ex]
& & & & & &
\ddots
\end{array} \right) 
\left(\begin{array}{l} \sqrt{2}A_0^{(2m)} \\[1.5ex]
A_2^{(2m)}\\[1.5ex] A_4^{(2m)}\\[1.5ex] A_6^{(2m)}\\[1.5ex] \vdots\\[1.5ex] A_{2r}^{(2m)}\\[1.5ex] \vdots
\end{array}\right) = k_m^2 \left(\begin{array}{l} \sqrt{2}A_0^{(2m)} \\[1.5ex]
A_2^{(2m)}\\[1.5ex] A_4^{(2m)}\\[1.5ex] A_6^{(2m)}\\[1.5ex] \vdots\\[1.5ex] A_{2r}^{(2m)}\\[1.5ex] \vdots
\end{array}\right) \,,
\end{equation}
wobei die Eigenwerte der obigen Matrix die charakteristischen Werte $k_m^2$ sind und die Fourier-Koeffizienten in den Eigenvektoren enthalten sind. Für numerische Berechnungen muss also obiges Eigenwertproblem gelöst werden \cite{Ikebe}. 

Die Fourier-Koeffizienten der Fourier-Reihe in Gl. (\ref{FourierReihe}) werden für die weiteren Betrachtungen wichtig (siehe (5.3.4) in \cite{Mechel}):
\begin{equation}
\label{a02m}
A_0^{(2m)} = \frac{1}{2\pi} \int_0^{2\pi} \text{d} \phi \, \text{ce}_{2m}(\phi) \quad\text{und}\quad A_{2r}^{(2m)} = \frac{1}{\pi} \int_0^{2\pi} \text{d} \phi \, \text{cos}(2r\phi)\text{ce}_{2m}(\phi) \,.
\end{equation}
Diese ersten Fourier-Koeffizienten erfüllen die Parseval-Relation
\begin{equation}
\label{ParsevalFourier}
\sum_{m=0}^{\infty} \left[ A_0^{(2m)} \right]^2 = \frac{1}{2}
\end{equation}
(siehe Gl. (20) in \cite{Seeger97} und Gl. (18.4.1) in \cite{Mechel}). Diese Relation kann genutzt werden, um die Anzahl der Koeffizienten abzuschätzen, die für eine geforderte numerische Genauigkeit benötigt werden.

Näherungsweise kann die Iteration berechnet werden, indem man nur die ersten beiden Fourier-Koeffizienten $A_{0}^{(2m)}$ und $A_{2}^{(2m)}$ berücksichtigt, also $A_{2r}^{(2m)}=0$ für $r \geq 2$ annimmt \cite{Seeger97,Hunter81}. Damit ergeben sich aus Gl. (\ref{rekursion1}) und Gl. (\ref{rekursion2}) die ersten Eigenwerte als Lösung der Gleichung $k_m^2 [k_m^2-4]+[R_C^2 \delta\omega]^2/[2D^2]=0$ zu
\begin{align}
\label{Approximationk0k1}
k_0 \approx \sqrt{2-\sqrt{4-\frac{1}{2} \left[\frac{R_C^2 \delta\omega}{D}\right]^2}} \quad \text{und} \quad
k_1 \approx \sqrt{2+\sqrt{4-\frac{1}{2} \left[\frac{R_C^2 \delta\omega}{D}\right]^2}}\,.
\end{align}
\begin{figure}
\begin{center}
\includegraphics[width=13.2cm]{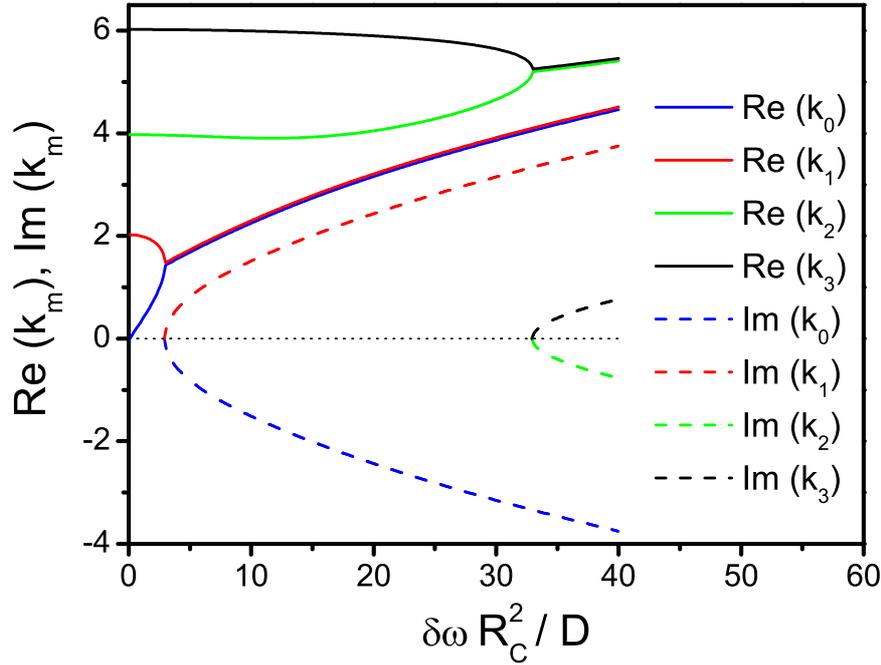}\vspace{-0.5cm}
\vspace{-0.5cm}
\caption[Realteil und Imaginärteil des charakteristischen Wertes]{\label{Fig:k}
\footnotesize Realteil und Imaginärteil des charakteristischen Wertes $k_m^2 = a_{2m}(\text{i}\delta \omega R_{\text{C}}^2/[2D])$ entsprechend Gl. (\ref{Parameter}). Für $\delta\omega R_{\text{C}}^2/D > p_0 \approx 2,\!93754$ können die charakteristische Wert konjugiert komplex sein. Für $\delta\omega R_{\text{C}}^2/D < p_0 \approx 2,\!93754$ verschwindet der Imaginärteil für alle $m$.}
\end{center}
\end{figure}
Ähnlich wird in \cite{Hochstadt} vorgegangen, wobei allerdings die ersten drei Gleichungen der Rekursion (\ref{rekursion1}) bis (\ref{rekursion1}) berücksichtigt werden. In Abb. \ref{Fig:k} sind Realteil und Imaginärteil des charakteristischen Wertes $k_m$ in Abhängigkeit des Parameters $\delta\omega R_{\text{C}}^2 / D$ dargestellt, wobei für sämtliche Darstellungen die Rekursionsformeln (\ref{rekursion1}) bis (\ref{rekursion3}) numerisch ausgewertet wurden \cite{Chaos,Ikebe}. Des Weiteren gilt für die charakteristischen Werte der Zusammenhang (siehe 28.7.1 in \cite{Olver})
\begin{equation}
\sum_{m=0}^{\infty} \left[ k_m^2 - 4 m^2 \right] = 0 \,,
\end{equation}
der auch zur Abschätzung der numerischen Genauigkeit genutzt werden kann. Die numerische Genauigkeit kann erhöht werden, indem mehr Gleichungen in der Rekursion (\ref{rekursion1}) bis (\ref{rekursion3}) berücksichtigt werden. Die Berechnung der charakteristischen Werte für den Fall eines rein imaginären Parameters $q$ in der Mathieu-Differenzialgleichung wurde in der Arbeit \cite{Blanch69} vorgenommen. Für $q/\text{i} < p_0/2 \approx 1,\!46876861 $ verschwinden alle Imaginärteile und der charakteristische Wert $k_m$ nimmt rein reelle Werte an. Mit dem Zusammenhang $q=\text{i}\delta \omega R_{\text{C}}^2/[2D]$ ergibt sich, dass die charakteristischen Werte $k_m$ rein reell sind, wenn $\delta\omega R_{\text{C}}^2/D < p_0 \approx 2,\!93754$ gilt. Aus Abb. \ref{Fig:k} ist ersichtlich, dass die Eigenwerte ab gewissen Verzweigungspunkten konjugiert komplex sind. Aus diesem Grund sind die Eigenwerte auch nicht reell, sondern im Allgemeinen komplex. Dies folgt aus der Tatsache, dass der Differenzialoperator in der Bloch-Torrey-Gleichung (\ref{BT}) nicht selbstadjungiert ist. Die Lage der Verzweigungspunkte wurde in der Arbeit \cite{Hunter81} untersucht und ist in Tabelle \ref{Tab:1} zusammengestellt. Näherungsweise lassen sich die Verzweigungspunkte an den Stellen $p_l\approx16,\!336\,l^2 + 13,\!64\,l + 2,\!94$ finden.
\begin{table*}
\begin{center}
\begin{tabular}{rrrrr}
$l$ \vline & $0$ \vline & $1$  \vline & $2$ \vline & $3$ \\ \hline
$p_l$ \vline & 2,94  \vline  & 32,94 \vline  & 95,62 \vline  & 190,96 \\ \hline
$k_{2l}=k_{2l+1}$ \vline & 1,45  \vline  & 5,23 \vline  & 8,98 \vline  & 12,73
\end{tabular}
\caption{\label{Tab:1}\footnotesize Lage der Verzweigungspunkte.}
\end{center}
\end{table*}
Die ersten Eigenwerte können für kleine Parameter $\delta \omega R_{\text{C}}^2/D$ näherungsweise angegeben werden (siehe 20.2.25 in \cite{Abramowitz72}):
\begin{align} 
\label{k_naeherung}
m=0:\; k_0 & \approx \frac{\delta \omega R_{\text{C}}^2/D}{2 \sqrt{2}} + \frac{7[\delta \omega R_{\text{C}}^2/D]^3}{1024 \sqrt{2}} + \frac{3271[\delta \omega R_{\text{C}}^2/D]^5}{9437184 \sqrt{2}} \quad \text{für} \; \frac{\delta\omega R_C^2}{D} < p_0 \approx 2,94 \,,\\
m=1:\; k_1 & \approx 2-\frac{5 [\delta \omega R_{\text{C}}^2/D]^2}{192}-\frac{913 [\delta \omega R_{\text{C}}^2/D]^4}{884736} \quad \quad\quad\quad\;\;\,\text{für} \; \frac{\delta\omega R_C^2}{D} < p_0 \approx 2,94 \,,\\
m=2:\; k_2 & \approx 4 -\frac{[\delta \omega R_{\text{C}}^2/D]^2}{960}+\frac{209 [\delta \omega R_{\text{C}}^2/D]^4}{55296000} \quad \quad\quad\quad\quad\;\text{für} \; \frac{\delta\omega R_C^2}{D} < p_1 \approx 32,94 \,,\\
m=3:\; k_3 & \approx 6 -\frac{[\delta \omega R_{\text{C}}^2/D]^2}{3360}+\frac{1123 [\delta \omega R_{\text{C}}^2/D]^4}{75866112000} \quad \quad\quad\quad\;\;\;\text{für} \; \frac{\delta\omega R_C^2}{D} < p_1 \approx 32,94 \,.
\end{align}
Diese Näherungen gelten allerdings nur bis zum jeweiligen Verzweigungspunkt, während die Approximation für $k_0$ und $k_1$ in Gl. (\ref{Approximationk0k1}) für sämtliche Werte des Parameters $\delta\omega R_C^2/D$ gilt. Die Entwicklung für die höheren charakteristischen Werte findet sich z. B. in 17.1.5 in \cite{Mechel}:
\begin{equation}
k_m \approx 2m - \frac{[\delta \omega R_{\text{C}}^2/D]^2}{32m[4m^2-1]} \quad \text{für} \quad \delta\omega R_C^2/D < p_{\lfloor\frac{m}{2}\rfloor}\,.
\end{equation}
Die höheren Eigenwerte sind $k_m \approx 2m$. Im Motional-Narrowing-Grenzfall entsprechen die Eigenwerte den geraden natürlichen Zahlen: $\lim_{ \delta \omega R_{\text{C}}^2/D \to 0}k_m = 2m$. 

Für große Werte des Parameters $\delta \omega R_{\text{C}}^2/D$ können die Eigenwerte durch
\begin{align}
\label{k0k1}
k_{1} & = k_{0}^{\displaystyle{*}} \quad \text{für} \quad \frac{\delta\omega R_C^2}{D} > p_0 \approx 2,94 \\
\label{k2lk2lp1}
k_{2l+1} & = k_{2l}^{\displaystyle{*}} \quad \text{für} \quad \frac{\delta\omega R_C^2}{D} > p_l \quad \text{mit}\\
k_{2l}^2 & \approx [4l + 1][1+\text{i}]\sqrt{\frac{\delta\omega R_{\text{C}}^2}{D}}-\text{i} \frac{\delta \omega R_{\text{C}}^2}{D}-2l^2-l-\frac{1}{4} \quad \text{für} \quad \frac{\delta\omega R_C^2}{D} > p_l
\end{align}
berechnet werden (siehe 28.8.1	in \cite{Olver}), wobei die Werte für $p_l$ in Tabelle \ref{Tab:1} gegeben sind. Die Tatsache, dass die charakteristischen Werte nach dem jeweiligen Verzweigungspunkt konjugiert komplex sind hat noch weitere Konsequenzen: Betrachtet man die ursprüngliche Mathieusche Differenzialgleichung (\ref{matdgl}) für $m=2l+1$ und bildet auf beiden Seiten der Gleichung das konjugiert Komplexe, so ergibt sich mit Gl. (\ref{k2lk2lp1}):
\begin{equation}
\partial_{\phi\phi}\text{ce}_{4l+2}^{\displaystyle{*}}(\phi)+\left[ k_{2l}^2 + \text{i} \, \frac{\delta \omega R_{\text{C}}^2}{D}  \cos (2\phi) \right]\text{ce}_{4l+2}^{\displaystyle{*}}(\phi)=0 \quad \text{für} \quad \frac{\delta\omega R_C^2}{D} > p_l \,.
\end{equation}
Des Weiteren lässt sich durch eine einfache Variablentransformation in der ursprünglichen Mathieuschen Differenzialgleichung (\ref{matdgl}) zeigen, dass 
\begin{equation}
\partial_{\phi\phi}\text{ce}_{2m}(\pi/2-\phi)+\left[ k_m^2 + \text{i} \, \frac{\delta \omega R_{\text{C}}^2}{D}  \cos (2\phi) \right]\text{ce}_{2m}(\pi/2-\phi)=0
\end{equation}
gilt. Setzt man in dieser Beziehung $m=2l$, so lässt sich erkennen, dass
\begin{equation}
\label{Symmetriece}
\text{ce}_{4l+2}(\phi)=\text{ce}_{4l}^{\displaystyle{*}}(\pi/2-\phi) \quad \text{für} \quad \delta\omega R_C^2/D > p_l \,.
\end{equation}

\begin{figure}
\begin{center}
\includegraphics[width=13.2cm]{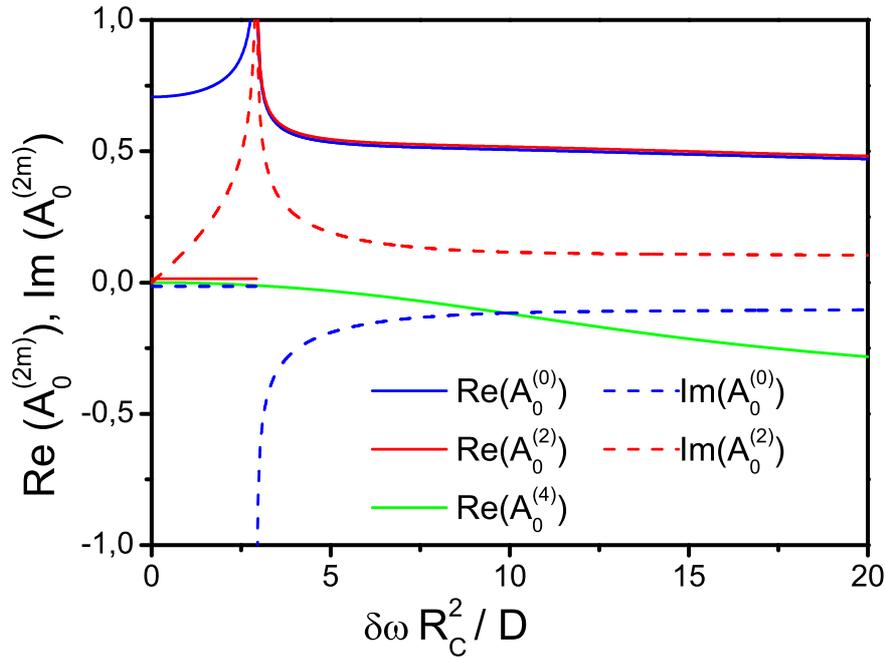}\vspace{-0.5cm}
\vspace{-0.5cm}
\caption[Realteil und Imaginärteil des Fourier-Koeffizienten]{\label{Fig:a}
\footnotesize Realteil und Imaginärteil des Fourier-Koeffizienten $A_0^{(2m)}$ entsprechend Gl. (\ref{a02m}).}
\end{center}
\end{figure}

Für die Fourier-Koeffizienten gilt nach Gl. (\ref{rekursion1}) ganz allgemein:
\begin{equation}
\label{A0A2allgemein}
\frac{A_0^{(2m)}}{A_2^{(2m)}} = \frac{\text{i}}{2} \frac{R_C^2 \delta\omega}{k_m^2 D} \,.
\end{equation}
Betrachtet man nun wieder die Rekursionsformeln und bricht analog zu oben die Rekursion ab, indem man $A_{2r}^{(2m)}=0$ für $r \geq 2$ annimmt, so ergibt sich aus Gl. (\ref{rekursion1}) unter Beachtung der Normalisierung in Gl. (\ref{Normalisierung}):
\begin{equation}
\label{A0A2}
\left[A_0^{(2m)}\right]^2 \approx \frac{1}{2}\left[ 1 -\left[ A_2^{(2m)} \right]^2 \right] \approx \frac{1}{2} \frac{1}{1-2\left[ \frac{k_m^2 D}{R_C^2 \delta\omega} \right]^2} \quad \text{für} \quad m=0,1 \,.
\end{equation}
Daraus ergeben sich analog zu den Ausdrücken in Gl. (\ref{Approximationk0k1}) Näherungen für die ersten Fourier-Koeffizienten, die für den gesamten Parameterbereich gültig sind:
\begin{align}
\label{A02mSeeger}
A_{0}^{(0)} \approx \frac{1}{\sqrt{2}} \sqrt{\frac{1}{1-2\left[\frac{k_0^2 D}{R_C^2\delta\omega}\right]^2}} \quad \text{und} \quad
A_{0}^{(2)} \approx \frac{1}{\sqrt{2}} \sqrt{\frac{1}{1-2\left[\frac{k_1^2 D}{R_C^2\delta\omega}\right]^2}}  \,.
\end{align}
Die zweiten Fourier-Koeffizienten folgen aus Gl. (\ref{A0A2allgemein}):
\begin{align}
\label{A22mSeeger}
A_{2}^{(0)} = -2\text{i} \frac{k_0^2 D}{R_C^2 \delta\omega} A_{0}^{(0)}
\quad \text{und} \quad
A_{2}^{(2)} = -2\text{i} \frac{k_1^2 D}{R_C^2 \delta\omega} A_{0}^{(2)} \,.
\end{align}
Für beliebige Werte des Parameters $\delta\omega R_C^2 / D$ sind die ersten Fourier-Koeffizienten $A_0^{(2m)}$ in Abb. \ref{Fig:a} dargestellt. 
Es lässt sich erkennen, dass die Fourier-Koeffizienten für kleine Werte des Parameters entweder rein reell oder rein imaginär sind. Ab den Verzweigungspunkten sind die Entwicklungskoeffizienten konjugiert komplex. Insgesamt ergibt sich:
\begin{equation}
\label{A02A00}
A_{0}^{(2)}=A_{0}^{(0) \displaystyle{*}} \quad \text{und} \quad A_{2}^{(2)}=-A_{2}^{(0) \displaystyle{*}} \quad \text{für} \quad \delta\omega R_C^2/D > p_0 \approx 2,94 \,,
\end{equation}
wobei für den zweiten Fourier-Koeffizienten die Beziehung (\ref{A0A2allgemein}) und Gl. (\ref{k0k1}) genutzt wurde. Dieses Ergebnis lässt sich verallgemeinern, indem man die allgemeine Fourier-Entwicklung der Mathieu-Funktionen aus Gl. (\ref{FourierReihe}) in die Beziehung (\ref{Symmetriece}) einsetzt. Damit ergibt sich:
\begin{equation}
\label{A4lA4lp2}
A_{2r}^{(4l+2)}=(-1)^r A_{2r}^{(4l) \displaystyle{*}} \quad \text{für} \quad \delta\omega R_C^2/D > p_l \,.
\end{equation}
Mit der Reihendarstellung der Funktionen $\text{ce}_{2m}$ (siehe 20.2.27 und 20.2.28 in \cite{Abramowitz72}), die auch in den Gln. (\ref{MathieuReihe0}) - (\ref{MathieuReihe4})) dargestellt ist, ergeben sich für die ersten Fourier-Koeffizienten:
\begin{align}
m=0:\; A_0^{(0)} & \approx [1 +[\delta \omega R_{\text{C}}^2/[8D]]^2]/\sqrt{2} \quad \text{für} \quad \delta\omega R_C^2/D < p_0 \approx 2,94\,, \\
m=1:\; A_0^{(2)} & \approx \text{i}\delta \omega R_{\text{C}}^2/[8D] \quad\quad\quad\quad\quad\;\;\; \text{für} \quad \delta\omega R_C^2/D < p_0 \approx 2,94 \quad\text{und}\\
m=2:\; A_0^{(4)} & \approx -[\delta \omega R_{\text{C}}^2/D]^2/768 \quad\quad\quad\;\;\, \text{für} \quad \delta\omega R_C^2/D < p_1 \approx 32,94\,,
\end{align} 
wobei die Näherungen wieder nur bis zum jeweiligen Verzweigungspunkt gelten. Diese Näherungen für die Fourier-Koeffizienten lassen sich auch allgemein angeben:
\begin{alignat}{4}
&A_0^{(0)} & & \approx \frac{1}{\sqrt{2}}\left[1 + \left[\frac{\delta \omega R_{\text{C}}^2}{8D}\right]^2\right] \quad & &\text{für} \quad m=0=r\quad & &\text{und} \quad \delta\omega R_C^2/D < p_0 \approx 2,94 \,, \\[1.0ex]
&A_{2r}^{(0)} & & \approx \frac{2[-\text{i}]^r}{[r!]^2\sqrt{2}} \left[ \frac{\delta\omega R_C^2}{8D} \right]^r \quad & &\text{für} \quad m=0 \quad & &\text{und} \quad \delta\omega R_C^2/D < p_0 \approx 2,94 \,,\\[1.0ex]
&A_{2r}^{(2m)} & & \approx \frac{\text{i}^{m-r}[2m]!}{[r-m]![r+m]!} \left[ \frac{\delta\omega R_C^2}{8D} \right]^{r-m} \quad & &\text{für} \quad r>m\geq 1 \quad & &\text{und} \quad \delta\omega R_C^2/D < p_{\lfloor\frac{m}{2}\rfloor} \,,\\[1.0ex]
&A_{2}^{(2)} & & \approx 1+\frac{19}{1152} \left[ \frac{\delta\omega R_C^2}{8D} \right]^{2} \quad & &\text{für} \quad r=m= 1 \quad & &\text{und} \quad \delta\omega R_C^2/D < p_0 \approx 2,94 \,,\\[1.0ex]
&A_{2m}^{(2m)} & & \approx 1 + \frac{4m^2+1}{\left[4m^2-1\right]^2} \left[\frac{\delta\omega R_C^2}{8D}\right]^2 \quad & &\text{für} \quad m> 1 \quad & &\text{und} \quad \delta\omega R_C^2/D < p_{\lfloor\frac{m}{2}\rfloor} \quad \text{und}\\[1.0ex]
&A_{2r}^{(2m)} & & \approx \frac{\text{i}^{m-r}[m+r-1]!}{[m-r]![2m-1]!} \left[ \frac{\delta\omega R_C^2}{8D} \right]^{m-r} \quad & &\text{für} \quad m>r \quad & &\text{und} \quad \delta\omega R_C^2/D < p_{\lfloor\frac{m}{2}\rfloor} \,.
\end{alignat}
Insbesondere folgt aus der letzten Näherungsformel:
\begin{equation}
A_0^{(2m)} \approx \text{i}^m\frac{[\delta \omega R_{\text{C}}^2/[8D]]^m}{[2m-1]!m} \quad \text{für} \quad \delta\omega R_C^2/D < p_{\lfloor\frac{m}{2}\rfloor}\,.
\end{equation}
Weitere brauchbare Eigenschaften der Fourier-Koeffizienten finden sich in 28.4.21 bis 28.4.24 in \cite{Olver} bzw. in 18.1.5, 18.1.6, 18.1.7, 18.1.9, 18.1.10, 18.2.1 und 18.2.2 in \cite{Mechel}. So folgt beispielsweise aus 18.2.1 in \cite{Mechel}
\begin{equation}
A_{0}^{(0)} \approx \left[ 2\pi^2 \frac{\delta\omega R_C^2}{D} \right]^{-\frac{1}{8}} \text{e}^{-\text{i}\frac{\pi}{16}} \quad \text{für} \quad \frac{\delta\omega R_C^2}{D} > p_0 \approx 2,94
\end{equation}
und $A_{0}^{(2)}=A_{0}^{(0) \displaystyle{*}}$ für $\delta\omega R_C^2/D > p_0 \approx 2,94$ wie schon in Gl. (\ref{A02A00}) gezeigt wurde. Für die höheren Koeffizienten gilt:
\begin{equation}
A_{0}^{(4l)} \approx [-1]^l\frac{\sqrt{[4l]!}}{[2l]!4^l} \left[ \frac{\pi^2}{2} \frac{\delta\omega R_C^2}{D} \right]^{-\frac{1}{8}} \text{e}^{-\text{i}\frac{\pi}{16}} \quad \text{für} \quad \frac{\delta\omega R_C^2}{D} > p_l
\end{equation}
und $A_{0}^{(4l+2)}=A_{0}^{(4l) \displaystyle{*}}$ für $\delta\omega R_C^2/D > p_l$ wie schon in Gl. (\ref{A4lA4lp2}) gezeigt wurde. Allgemein gilt:
\begin{equation}
A_{2r}^{(4l)} \approx [-1]^{r+l} \frac{\sqrt{[4l]!}}{[2l]!4^l} \left[ \frac{\pi^2}{128} \frac{\delta\omega R_C^2}{D} \right]^{-\frac{1}{8}} \text{e}^{-\text{i}\frac{\pi}{16}} \quad \text{für} \quad \frac{\delta\omega R_C^2}{D} > p_l \,.
\end{equation}

\begin{figure}
\begin{center}
\includegraphics[width=11.5cm]{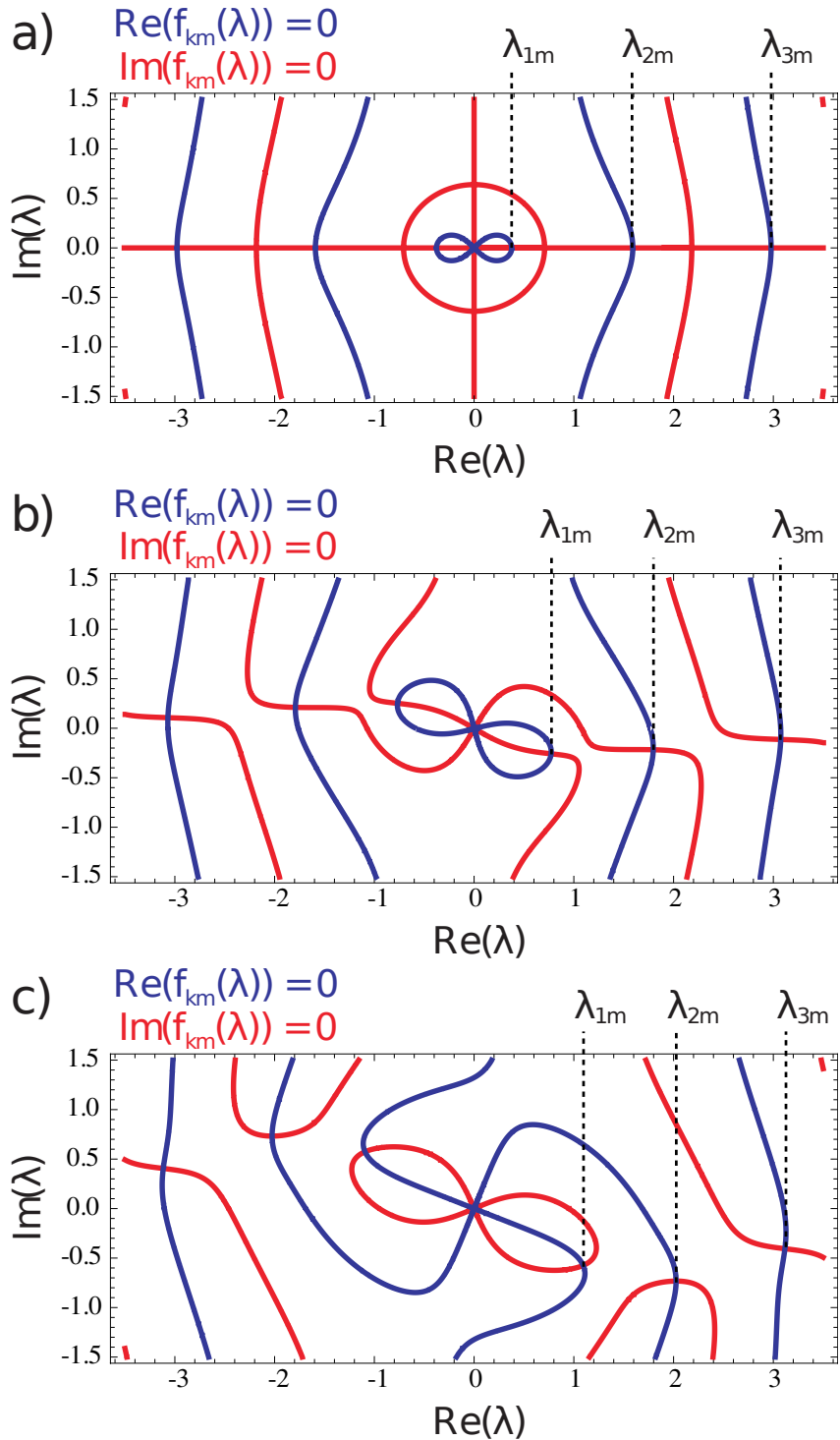}
\vspace{-0.5cm}
\caption[Bestimmung der Eigenwerte]{\label{Fig:lnm}{\footnotesize Bestimmung der Eigenwerte $\lambda_{nm}$ für $m=0$ und $\eta=0,\!1$. Im Fall a) beträgt $\delta\omega R_{\text{C}}^2/D =2$ und der Index $k_m$ ergibt sich aus Gl. (\ref{Parameter}) zu $k_m = 0,\!76$. An den Schnittpunkten der blauen Kurven (Realteil) mit den roten Kurven (Imaginärteil) sind Realteil und Imaginärteil der Funktion $f_{k_{m}}(\lambda)$ nach Gl. (\ref{lambdan}) gleichzeitig Null. Diese Schnittpunkte liegen im Falle eines rein reellen Index $k_m$ auf der reellen Achse und entsprechen den gesuchten Eigenwerten. Im Fall b) beträgt $\delta\omega R_{\text{C}}^2/D =4$ und der Index $k_m$ ergibt sich aus Gl. (\ref{Parameter}) zu $k_m = 1,\!58 - 0,\!59 \text{i}$. Im Fall c) beträgt $\delta\omega R_{\text{C}}^2/D =10$ und der Index $k_m$ ergibt sich aus Gl. (\ref{Parameter}) zu $k_m = 2,\!27 - 1,\!51 \text{i}$.}}
\end{center}
\end{figure}

Für den Radialanteil ergibt sich aus dem Separationsansatz (\ref{sep2}) die Besselsche Differenzialgleichung
\begin{equation} \nonumber
\frac{\partial^2}{\partial r^2} R_{nm}(r) + \frac{1}{r}\frac{\partial}{ \partial r} R_{nm}(r)+ \left[ \frac{\lambda_{nm}^2}{R_{\text{C}}^2} - \frac{k_{m}^2}{r^2} \right] \! R_{nm}(r)=0
\end{equation}
mit der Separationskonstanten $\lambda_{nm}^2$. Die Lösung dieser Besselschen Differenzialgleichung wird als Linearkombination der Besselfunktionen $J_{k_{m}}$ mit der Neumannfunktionen $Y_{k_{m}}$ geschrieben: $R_{nm}(r)=a_{nm} J_{k_{m}}\left( \lambda_{nm} r/R_{\text{C}} \right) + b_{nm} Y_{k_{m}}\left( \lambda_{nm} r/R_{\text{C}} \right) $, wobei die Koeffizienten $a_{nm}$ und $b_{nm}$ durch die reflektierenden Randbedingungen an der Oberfläche der Kapillare $\partial_r R_{nm}(r)|_{r=R_{\text{C}}} = 0$ und am Rande des Dephasierungsvolumens $\partial_r R_{nm}(r)|_{r=R} = 0$ (siehe Gl. (\ref{rand})) festgelegt werden. Die reflektierende Randbedingung an der Kapillaroberfläche wird durch Wahl der Koeffizienten entsprechend $a_{nm} = + Y_{k_{m}}^{'} \left( \lambda_{nm} \right)$ und $b_{nm} = - J_{k_{m}}^{'} \left( \lambda_{nm} \right)$ erfüllt. Damit ergeben sich die radialen Eigenfunktionen zu
\begin{equation}
\label{Eigenr}
R_{nm}(r) = Y_{k_{m}}^{'} \left( \lambda_{nm} \right) J_{k_{m}}\left( \frac{\lambda_{nm}}{R_{\text{C}}} r\right) - J_{k_{m}}^{'} \left( \lambda_{nm} \right) Y_{k_{m}}\left( \frac{\lambda_{nm}}{R_{\text{C}}} r\right) \,,
\end{equation}
welche die folgende Orthogonalitätsrelation erfüllen (siehe 11.4.2 in \cite{Abramowitz72} bzw. Gl. (11) in Abschnitt 5.11 in \cite{Watson95}):
\begin{align} \label{Ort_Rad}
&\frac{1}{R_{\text{C}}^2} \int_{R_{\text{C}}}^{R} \text{d} r \, r R_{nm}(r) R_{n^{'}m}(r) = N_{nm}\delta_{nn^{'}}\quad\text{mit}\\
&N_{nm} = \frac{1}{R_{\text{C}}^2} \int_{R_{\text{C}}}^{R} \text{d} r \, r R_{nm}^2(r) =\frac{1}{2} \left[\frac{1}{\eta}-\frac{k_m^2}{\lambda_{nm}^2} \right] R_{nm}^2\left(\frac{R_{\text{C}}}{\sqrt{\eta}} \right) - \frac{2}{\pi^2 \lambda_{nm}^2}\left[1-\frac{k_m^2}{\lambda_{nm}^2} \right] \\
\label{Nnm}
&\quad\;\;= \frac{q_{nm}^2}{2} \left[\frac{1}{\eta}-\frac{k_m^2}{\lambda_{nm}^2} \right] - \frac{2}{\pi^2 \lambda_{nm}^2} \left[1-\frac{k_m^2}{\lambda_{nm}^2} \right] \quad \text{und} \\
\label{pkm} 
& q_{nm} = Y_{k_{m}}^{'} \left( \lambda_{nm} \right) J_{k_{m}}\left( \frac{\lambda_{nm}}{\sqrt{\eta}} \right) - J_{k_{m}}^{'} \left( \lambda_{nm} \right) Y_{k_{m}}\left( \frac{\lambda_{nm}}{\sqrt{\eta}} \right) \,.
\end{align}
Dabei ist es wichtig, die Abelsche Identität für die Wronski-Determinante in der Form 
\begin{equation}
\label{Wronski}
Y_{k_{m}}^{'}(\lambda_{nm}) J_{k_{m}}(\lambda_{nm}) - J_{k_{m}}^{'}(\lambda_{nm}) Y_{k_{m}}(\lambda_{nm}) = \frac{2}{\pi\lambda_{nm}}
\end{equation}
zu nutzen (siehe 6. 58. in \cite{Spiegel76a}), womit aus Gl. (\ref{Eigenr}) direkt $R_{nm}(R_{\text{C}}) = 2/[\pi \lambda_{nm}]$ folgt. Die Normierungskonstanten $N_{nm}$ sind einheitenlos und hängen nur vom regionalen Blutvolumenverhältnis $\eta$ und dem Parameter $\delta\omega R_{\text{C}}^2 / D$ ab. An dieser Stelle ist es wichtig darauf hinzuweisen, dass für $\delta\omega R_{\text{C}}^2 / D > p_0 \approx 2,\!93754$ der Index $k_m$ der Bessel-Funktionen komplexe Werte annehmen kann. Dies beeinflusst jedoch nicht die Theorie der Bessel-Funktionen, obwohl diese dann selbst komplexe Werte annehmen können \cite{Watson95}. Die reflektierende Randbedingung an der Oberfläche des Versorgungszylinders in Gl. (\ref{rand}) überträgt sich auf die radialen Eigenfunktionen in der Form $\partial_r R_{nm}(r)|_{r=R}=0$. Daraus ergibt sich mit den radialen Eigenfunktionen aus Gl. (\ref{Eigenr}) die Bestimmungsgleichung für die Eigenwerte $\lambda_{nm}$:
\begin{equation} 
\label{lambdan}
0 = f_{k_{m}}(\lambda_{nm}) \quad\quad \text{mit} \quad\quad f_{k_{m}}(\lambda) = Y_{k_{m}}^{'} \left( \lambda \right) J_{k_{m}}^{'} \left(\frac{\lambda}{\sqrt{\eta}} \right) - J_{k_{m}}^{'} \left(\lambda \right)Y_{k_{m}}^{'} \left(\frac{\lambda}{\sqrt{\eta}} \right) \,.
\end{equation}
Diese Eigenwerte müssen numerisch ermittelt werden. Hilfreich ist es auch, die Eigenschaften der Funktion $f_{km}(\lambda)$ zu kennen:
\begin{align}
f_{k_{m}}(\lambda) & = f_{k_{m}}(-\lambda) \,, \\
f_{k_{m}}(\lambda) & = f_{-k_{m}}(\lambda) \quad\text{und}\\
\label{Eigenschaftk}
f_{k_{m}^{\displaystyle{*}}}(\lambda) & = f_{k_{m}}^{\displaystyle{*}}(\lambda^{\displaystyle{*}})\,.
\end{align}
Oben wurde gezeigt, dass die charakteristischen Werte konjugiert komplex sein können. So ist z. B. $k_1=k_0^{\displaystyle{*}}$ für $\delta\omega R_C^2/D > p_0 \approx 2,\!94$ (siehe Gl. (\ref{k0k1}) und Abb. \ref{Fig:k}). Auf Grund der letzten Eigenschaft in Gl. (\ref{Eigenschaftk}) folgt demnach, dass in diesem Fall auch die Eigenwerte konjugiert komplex sind: $\lambda_{n1}=\lambda_{n0}^{\displaystyle{*}}$. Allgemein folgt aus Gl. (\ref{k2lk2lp1}) die Beziehung
\begin{equation}
\label{Eigenschaftl}
\lambda_{n \, 2l+1} = \lambda_{n \, 2l}^{\displaystyle{*}} \quad \text{für} \quad \delta\omega R_C^2/D > p_l \,,
\end{equation}
die sich auf die Eigenfunktionen überträgt:
\begin{equation}
\label{EigenschaftR}
R_{n \, 2l+1} (r)= R_{n \, 2l}^{\displaystyle{*}} (r)\quad \text{für} \quad \delta\omega R_C^2/D > p_l \,.
\end{equation}
Für $\delta\omega R_{\text{C}}^2 / D \leq p_0$ sind die Eigenwerte $k_m$ rein reell und in Abb. \ref{Fig:lnm} a) sind Realteil und Imaginärteil der Funktion $f_{k_{m}}(\lambda) = 0$ in der komplexen Ebene dargestellt. Man erkennt, dass für $\delta\omega R_{\text{C}}^2 / D \leq p_0$ also auch die Eigenwerte $\lambda_{nm}$ rein reell sind.

Ab dem ersten Verzweigungspunkt, d.h. für $\delta\omega R_{\text{C}}^2/D > p_0$ können die Eigenwerte $k_m$ komplex werden. Für ein komplexes $k_m$ sind in Abb. \ref{Fig:lnm} b) und c) Realteil und Imaginärteil der Funktion $f_{k_{m}}(\lambda) = 0$ in der komplexen Ebene dargestellt. Es zeigt sich, dass die Eigenwerte $\lambda_{nm}$ nun auch komplex sind.

\begin{figure}
\begin{center}
\includegraphics[width=12cm]{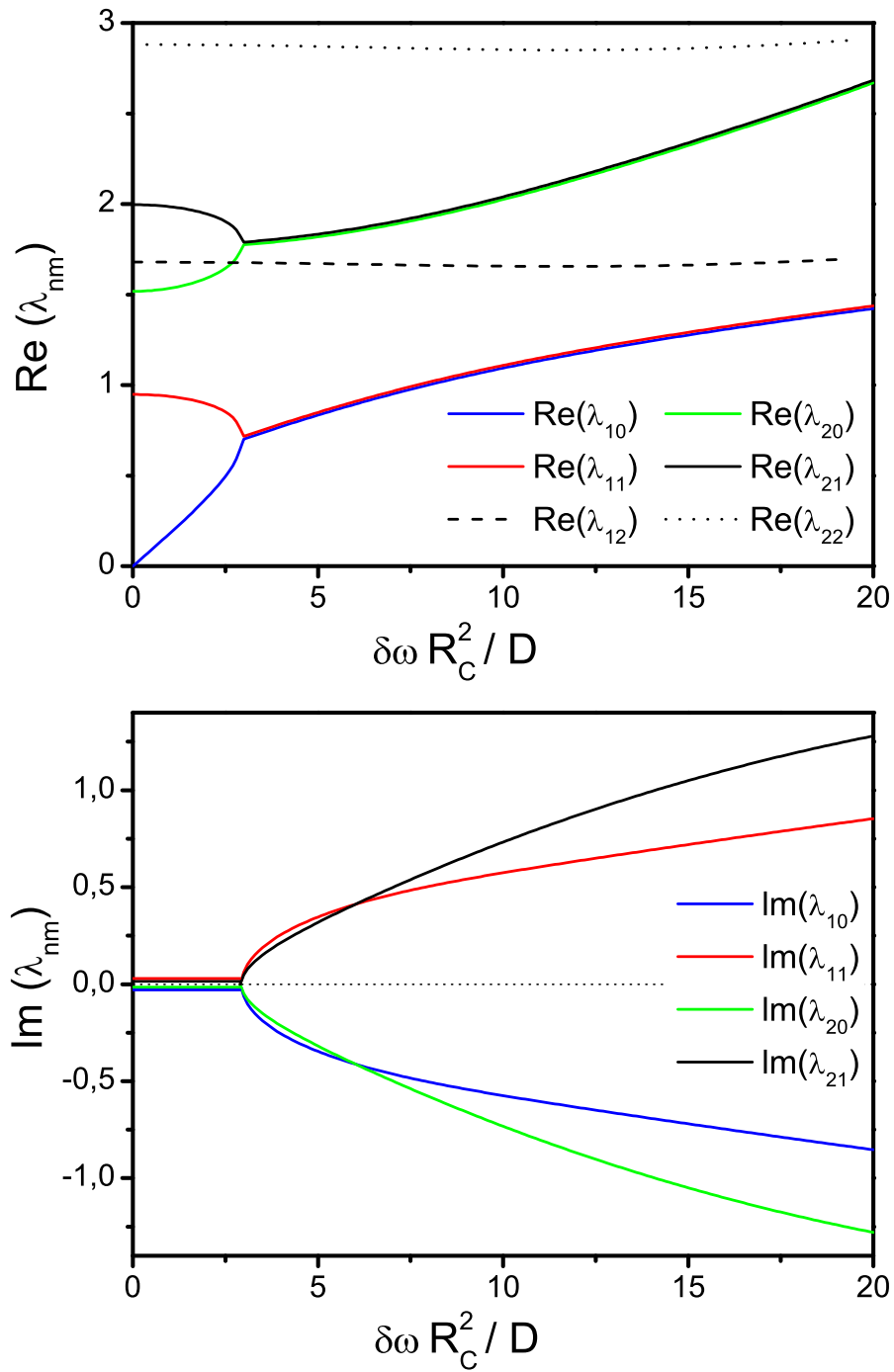}
\caption[Eigenwertspektrum]{\label{Fig:Spektrum-l1m}{\footnotesize Eigenwertspektrum als Lösung der Bestimmungsgleichung (\ref{lambdan}) für das regionale Blutvolumenverhältnis $\eta=0,\!1$. Auf Grund der Beziehungen (\ref{approxlambda1m}) sehen die Kurvenverläufe für $\lambda_{1m}$ ähnlich aus wie die Kurvenverläufe für $k_{m}$in Abb. \ref{Fig:k}.}}
\end{center}
\end{figure}

Der erste Eigenwert, der durch die Eigenwertgleichung (\ref{lambdan}) festgelegt wird kann durch eine Taylor-Entwicklung der Besselfunktionen für kleine Argumente erhalten werden:
\begin{equation}
\label{l1m}
\lambda_{1m}^2 \approx \frac{4 \eta k_m [1-k_m^2][\eta ^{k_m}-1]}{[k_m^2+k_m-2][1-\eta][1+\eta^{k_m}]-2k_m[\eta^{k_m}-\eta]}.
\end{equation}
Für kleine $k_m$ lässt sich dieser Ausdruck noch vereinfachen. Letztlich ergibt sich als Näherungsausdruck für die gesuchten Eigenwerte:
\begin{align}
\label{approxlambda1m}
\lambda_{1m} & \approx k_m \frac{\eta \text{ln}(\eta)}{\eta-1} \quad \text{für} \quad n=1 \quad \text{und}\\
\label{approxlambdanm}
\lambda_{nm} & \approx \pi \sqrt{\eta} \frac{n-1}{1-\sqrt{\eta}} + \frac{1-\sqrt{\eta}}{n-1} \frac{4k_m^2+3}{8\pi} \quad \text{für} \quad n\geq2\,,
\end{align}
wobei die Näherung (\ref{approxlambda1m}) aus der Taylor-Entwicklung der Näherung (\ref{l1m}) stammt und die Beziehung (\ref{approxlambdanm}) in Abschnitt 10.21.49 in \cite{Olver} beschrieben ist. Approximationen für den ersten und die weiteren Eigenwerte im Falle großer Blutvolumenverhältnisse $\eta \approx 1$ sind in den Gleichungen (A.5) und (A.4) in der Arbeit \cite{Gottlieb} aufgeführt. Im Allgemeinen ergeben sich die Eigenwerte $\lambda_{nm}$ als numerische Lösung der Bestimmungsgleichung (\ref{lambdan}), wobei die Näherungswerte aus (\ref{approxlambda1m}) und (\ref{approxlambdanm}) als Startpunkt für die numerische Nullstellensuche dienen können. Das Eigenwertspektrum für die ersten Eigenwerte ist in Abb. \ref{Fig:Spektrum-l1m} dargestellt.

Der Separationsansatz (\ref{sep2}) führt schließlich zu der Differenzialgleichung 
\begin{equation}
\frac{R_{\text{C}}^2}{D} \, \left[\frac{\partial}{\partial t} + \frac{1}{T_2} \right] \,T_{nm}(t) = -\lambda_{nm}^2 \, T_{nm}(t) 
\end{equation}
für den zeitabhängigen Anteil und hat die Lösung
\begin{equation} 
\label{Zeitanteil}
T_{nm}(t) = \text{e}^{-t \left[ \lambda_{nm}^2 \frac{D}{R_{\text{C}}^2} +\frac{1}{T_2} \right]} \,.
\end{equation}

\begin{figure}
\begin{center}
\includegraphics[width=12cm]{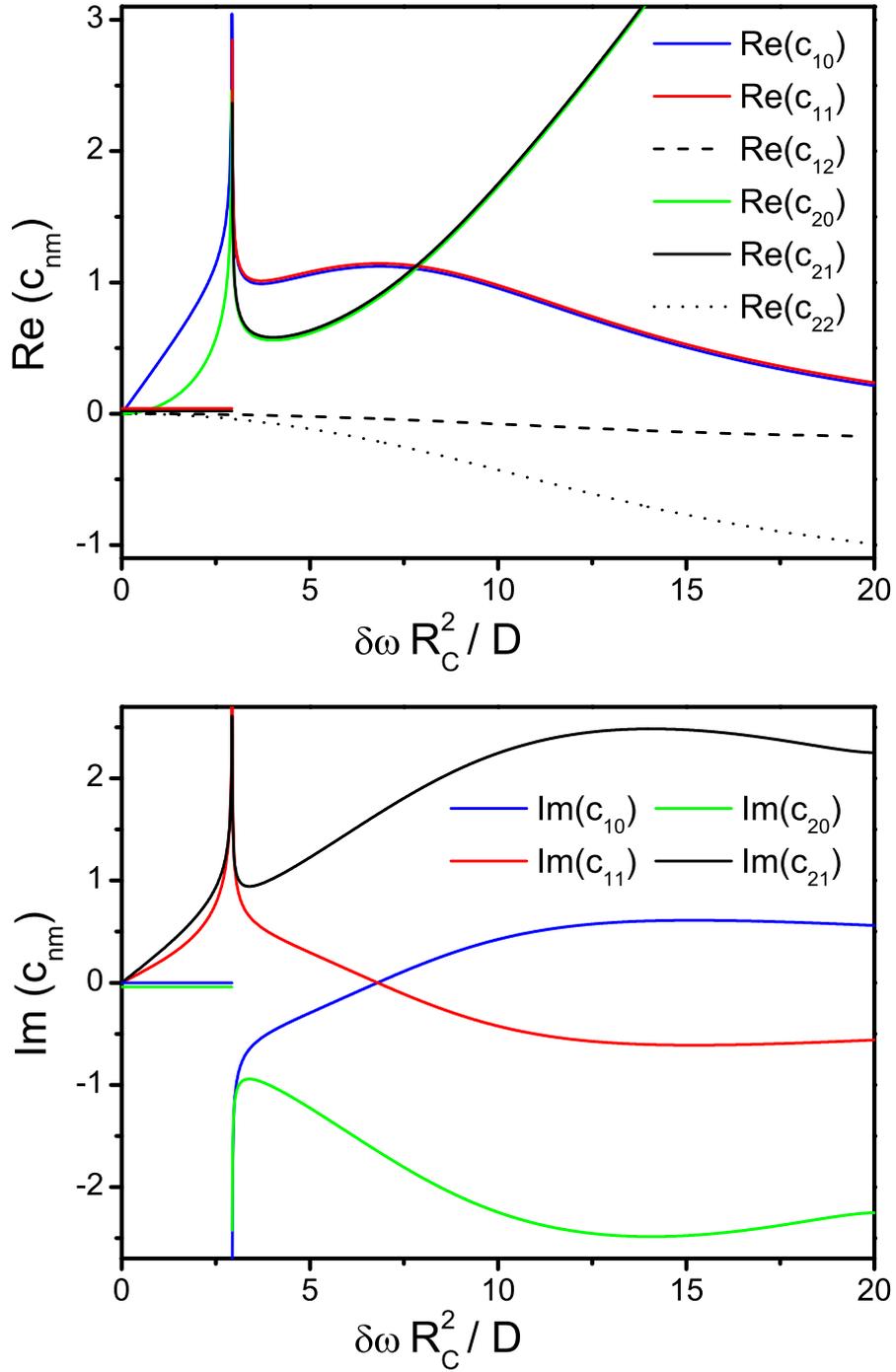}
\caption[Entwicklungskoeffizienten der Magnetisierung]{\label{Fig:Spektrum-cnm}{\footnotesize Entwicklungskoeffizienten als Lösung der Bestimmungsgleichung (\ref{cnm}) für das regionale Blutvolumenverhältnis $\eta=0,\!1$. Man erkennt, dass entsprechend Gl. (\ref{Eigenschaftc}) für $l=0$ gilt: $c_{n1}=c_{n0}^{\displaystyle{*}}$ für $\delta\omega R_C^2/D > p_0 \approx 2,94$. So haben beispielsweise  für $\delta\omega R_C^2/D > p_0 \approx 2,94$ die Entwicklungskoeffizienten $c_{10}$ und $c_{11}$ gleiche Realteile und die Imaginärteile unterscheiden sich nur im Vorzeichen.}}
\end{center}
\end{figure}

Zur Konstruktion der gesamten Lösung werden die drei Anteile (\ref{Eigen_Phi}), (\ref{Eigenr}) und (\ref{Zeitanteil}) in den Separationsansatz (\ref{sep2}) eingesetzt. Damit kann die Lösung der Bloch-Torrey-Gleichung als Summe über alle Eigenwerte geschrieben werden:
\begin{equation}
\label{Endergebnis}
\frac{m(\mathbf{r},t)}{m_0} = \sum_{m=0}^{\infty} \sum_{n=1}^{\infty} c_{nm} \text{ce}_{2m}(\phi) \left[Y_{k_{m}}^{'} \left( \lambda_{nm} \right) J_{k_{m}}\left( \frac{\lambda_{nm}}{R_{\text{C}}} r\right) - J_{k_{m}}^{'} \!\left( \lambda_{nm} \right) Y_{k_{m}}\left( \frac{\lambda_{nm}}{R_{\text{C}}} r\right)\right]\text{e}^{- t \left[\lambda_{nm}^2 \frac{D}{R_{\text{C}}^2} + \frac{1}{T_2}\right]}.
\end{equation}
Die Entwicklungskoeffizienten $c_{nm}$ können mit Hilfe der Anfangsbedingung $m(\mathbf{r},t=0) = m_0$ und mit den Orthogonalitätsrelationen (\ref{Ort_Phi}) und (\ref{Ort_Rad}) bestimmt werden:
\begin{equation}
\label{int_R}
c_{nm} = \frac{2 A_0^{(2m)}}{N_{nm}R_{\text{C}}^2} \int_{R_{\text{C}}}^R \text{d}r \,r R_{nm}(r) \,.
\end{equation}
Zur Berechnung der Entwicklungskoeffizienten muss das Integral in Gl. (\ref{int_R}) gelöst werden. Nutzt man das regionale Blutvolumenverhältnis $\eta=R_{\text{C}}^2/R^2$ ergibt sich mit Hilfe der Gl. (5) aus Abschnitt 10.74 in \cite{Watson95} der Ausdruck
\begin{align}
M_{nm}=\frac{1}{R_{\text{C}}^2} \int_{R_{\text{C}}}^{R} \!\!\text{d} r \, r R_{nm}(r) = \frac{q_{nm}}{\lambda_{nm} \sqrt{\eta}} s_{1,k_{m}}^{'} \left( \frac{\lambda_{nm}}{\sqrt{\eta}} \right) - \frac{2}{\pi \lambda_{nm}^2} s_{1,k_{m}}^{'} \left( \lambda_{nm} \right) \,,
\end{align}
wobei die Größe $q_{nm}$ in Gl. (\ref{pkm}) gegeben ist und 
\begin{equation}
\label{Lommel}
s_{\mu,k_{m}}(x) = \frac{x^{\mu+1}}{[\mu+1]^2 - k_m^2} \; _1F_2\left(1;\frac{\mu+3-k_m}{2},\frac{\mu+3+k_m}{2};-\frac{x^2}{4}\right)
\end{equation}
die Lommelfunktion darstellt \cite{Lommel75,Magnus}, wobei $_1F_2$ die in Gl. (\ref{BEHF}) definierte verallgemeinerte hypergeometrische Funktion ist. Eine äquivalente Definition der Lommelfunktion findet sich beispielsweise in 11.9.3 in \cite{Olver}. Anstatt der Lommelfunktion $s$ kann man auch die zweite Lommelfunktion $S$ verwenden (siehe 11.9.5 in \cite{Olver}). Nutzt man den Zusammenhang $s_{1,k_m}=1+k_m^2 s_{-1,k_m}$ (siehe \cite{Magnus}, Seite 112), folgt:
\begin{equation}
s_{1,k_{m}}^{'}(x) = \frac{2x}{4 - k_m^2} \; _1F_2\left(2;2-\frac{k_m}{2},2+\frac{k_m}{2};-\frac{x^2}{4}\right) \,.
\end{equation}
Letztlich folgt als analytischer Ausdruck für die Entwicklungskoeffizienten:
\begin{equation}
\label{cnm}
c_{nm} =2A_0^{(2m)} \frac{M_{nm}}{N_{nm}}= 4 \pi \eta A_0^{(2m)} \lambda_{nm}^2 \frac{2s_{1,k_{m}}^{'} \left( \lambda_{nm} \right) - \pi q_{nm} \frac{\lambda_{nm}}{\sqrt{\eta}} s_{1,k_{m}}^{'} \left( \frac{\lambda_{nm}}{\sqrt{\eta}} \right)}{4\eta [\lambda_{nm}^2-k_m^2] - \pi^2 \lambda_{nm}^2 q_{nm}^2 [\lambda_{nm}^2 - \eta k_m^2]} \,.
\end{equation}
Diese Entwicklungskoeffizienten sind nur vom regionalen Blutvolumenverhältnis $\eta$ und dem Parameter $\delta\omega R_C^2/D$ abhängig. Die Eigenschaften der Eigenwerte aus Gl. (\ref{Eigenschaftl}) übertragen sich auf die Entwicklungskoeffizienten:
\begin{equation}
\label{Eigenschaftc}
c_{n\, 2l+1}=c_{n\,2l}^{\displaystyle{*}} \quad \text{für} \quad \delta\omega R_C^2/D > p_l \,.
\end{equation}
In Abb. \ref{Fig:Spektrum-cnm} sind die Entwicklungskoeffizienten $c_{nm}$ in Abhängigkeit vom Parameter $\delta\omega R_C^2/D$ dargestellt.

Betrachtet man die Magnetisierung an der Oberfläche der Kapillare, so ergibt sich mit Gl. (\ref{Wronski}) aus Gl. (\ref{Endergebnis})
\begin{equation}
m(r=R_C,\phi,t) = \frac{2m_0}{\pi}\sum_{m=0}^{\infty} \sum_{n=1}^{\infty} \frac{c_{nm}}{\lambda_{nm}} \text{ce}_{2m}(\phi) \text{e}^{- t \left[\lambda_{nm}^2 \frac{D}{R_{\text{C}}^2} + \frac{1}{T_2}\right]}.
\end{equation}
Da zum Zeitpunkt $t=0$ die Magnetisierung den konstanten Anfangswert $m(r=R_C,\phi,t=0) = m_0$ annimmt, erhält man mit der Orthogonalitätsrelation (\ref{Ort_Phi}) und den ersten Fourier-Koeffizienten aus Gl. (\ref{a02m}) die Relation
\begin{equation}
\label{testn1}
\sum_{n=1}^{\infty} \frac{M_{nm}}{\lambda_{nm}N_{nm}}= \frac{\pi}{2} \quad \text{bzw.} \quad \sum_{n=1}^{\infty} \frac{2s_{1,k_{m}}^{'} \left( \lambda_{nm} \right) - \pi q_{nm} \frac{\lambda_{nm}}{\sqrt{\eta}} s_{1,k_{m}}^{'} \left( \frac{\lambda_{nm}}{\sqrt{\eta}} \right)}{4\eta [\lambda_{nm}^2-k_m^2] - \pi^2 \lambda_{nm}^2 q_{nm}^2 [\lambda_{nm}^2 - \eta k_m^2]} \lambda_{nm} = \frac{1}{4\eta} \,.
\end{equation}

Ähnlich kann man vorgehen, wenn man die Magnetisierung an der Oberfläche des Dephasierungszylinders betrachtet. Mit der Definition (\ref{pkm}) folgt aus Gl. (\ref{Endergebnis})
\begin{equation}
m(r=R,\phi,t) = m_0 \sum_{m=0}^{\infty} \sum_{n=1}^{\infty} c_{nm} q_{nm} \text{ce}_{2m}(\phi) \text{e}^{- t \left[\lambda_{nm}^2 \frac{D}{R_{\text{C}}^2} + \frac{1}{T_2}\right]}.
\end{equation}
Da auch hier zum Zeitpunkt $t=0$ die Magnetisierung den Wert $m(r=R,\phi,t=0) = m_0$ annimmt, erhält man analog zu oben mit der Orthogonalitätsrelation (\ref{Ort_Phi}) und den ersten Fourier-Koeffizienten aus Gl. (\ref{a02m}) die Relation
\begin{equation}
\label{testn2}
\sum_{n=1}^{\infty} \frac{q_{nm}M_{nm}}{N_{nm}}= 1 \quad \text{bzw.} \quad
\sum_{n=1}^{\infty} \frac{2s_{1,k_{m}}^{'} \left( \lambda_{nm} \right) - \pi q_{nm} \frac{\lambda_{nm}}{\sqrt{\eta}} s_{1,k_{m}}^{'} \left( \frac{\lambda_{nm}}{\sqrt{\eta}} \right)}{4\eta [\lambda_{nm}^2-k_m^2] - \pi^2 \lambda_{nm}^2 q_{nm}^2 [\lambda_{nm}^2 - \eta k_m^2]} q_{nm} \lambda_{nm}^2 = \frac{1}{2\pi\eta} \,.
\end{equation}

Setzt man in Gl. (\ref{Endergebnis}) $t=0$ und integriert über das Dephasierungsvolumen, erhält man mit Gl. (\ref{a02m}) und Gl. (\ref{int_R}) die Parseval-Relation
\begin{equation} 
\label{Parsevalcnm}
\sum_{m=0}^{\infty} \sum_{n=1}^{\infty} c_{nm}^2 N_{nm} = \frac{1}{\eta}-1 \,.
\end{equation}
Diese Relation kann genutzt werden, um die Anzahl der Koeffizienten $n$ abzuschätzen, die für eine ausreichende numerische Genauigkeit notwendig ist. Die Anzahl der Koeffizienten $m$ kann aus der analogen Relation (\ref{ParsevalFourier}) erhalten werden.

\begin{figure}
\begin{center}
\includegraphics[width=12cm]{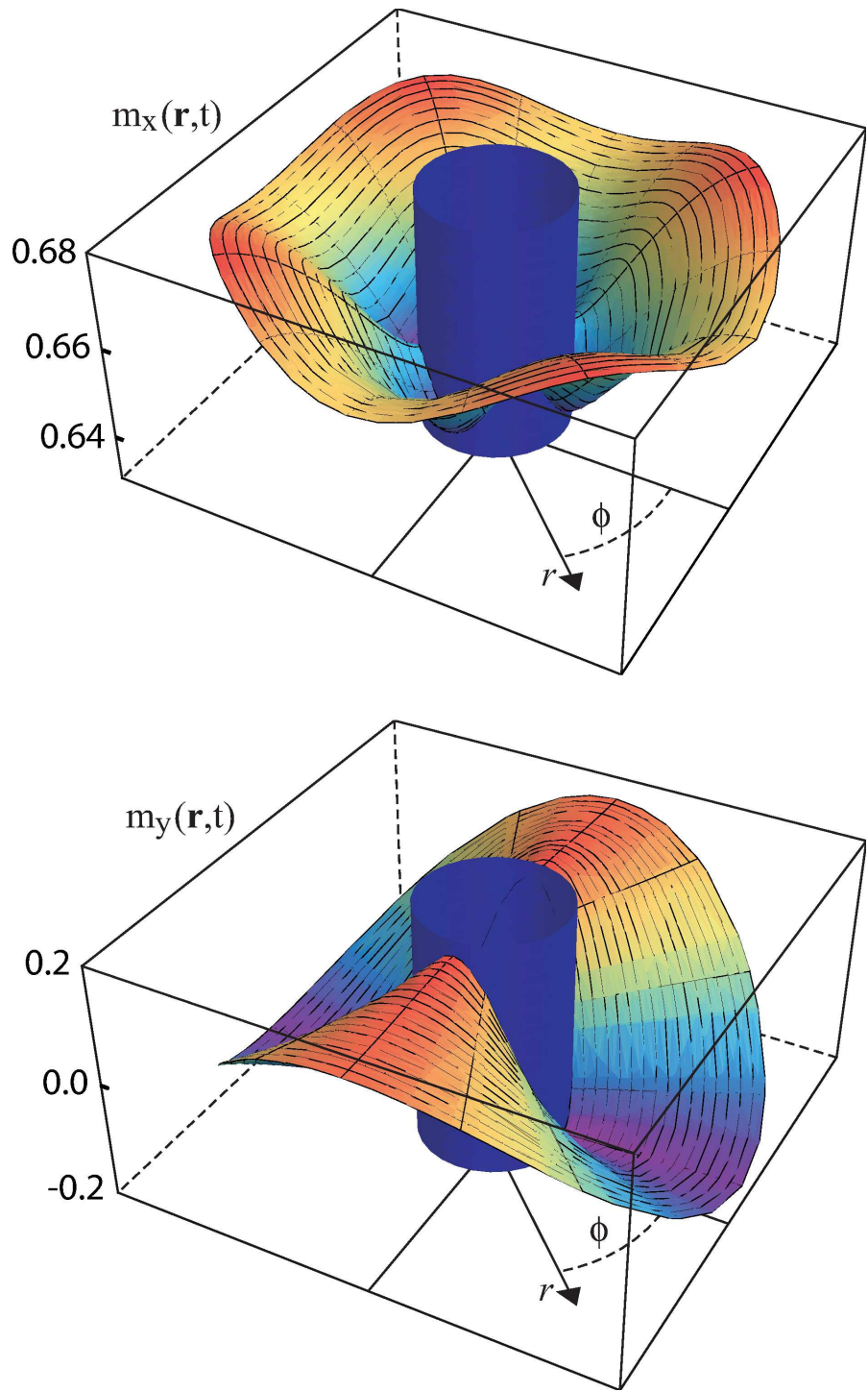}
\caption[Magnetisierung um eine Kapillare]{\label{mvonrundt}{\footnotesize Transversale Komponenten $m_x(\mathbf{r},t)$ und $m_y(\mathbf{r},t)$ nach Gl. (\ref{Endergebnis}) um eine Kapillare des Myokards (mit $R_{\text{C}}=2,75\,\mu\text{m}$, $T_2=57\,\text{ms}$, $\delta\omega=151\,\text{s}^{-1}$, $D=1\, \mu\text{m}^2 / \text{ms}$ und $\eta=0,084$) für $t=20\,\text{ms}$ nach dem Anregungspuls. Durch den Anregungspuls wurde die Anfangsmagnetisierung $m_0=1$, also nur eine Komponente in $x$-Richtung erzeugt. Die Magnetisierung endet am Rand des Versorgungszylinders, der zwecks Verbesserung der Erkennbarkeit nicht eingezeichnet wurde. Zur Veranschaulichung der Polarkoordinaten $r$ und $\phi$ wurde ein quaderförmiges Voxel eingezeichnet.}}
\end{center}
\end{figure}

In Abb. \ref{mvonrundt} sind die $x$- und $y$-Komponenten der Magnetisierung nach $t=20 \, \text{ms}$ für eine Kapillare im Myokard dargestellt. In der Nähe der Kapillare besteht ein starker Suszeptibilitätseinfluss, und die Spins dephasieren sehr schnell. Dementsprechend ist die Magnetisierung hier nach kurzer Zeit schon stark abgefallen. Auf Grund der $1/r^2$-Abhängigkeit der lokalen Resonanzfrequenz (siehe Gl. (\ref{Dipol})) werden die Spins am Rande des Voxels nur minimal von der Dephasierung beeinflusst und die Magnetisierung fällt weniger schnell ab.

Das Endergebnis für die Magnetisierung $m(\mathbf{r},t)$ in Gl. (\ref{Endergebnis}) ist für jeden Wert des Parameters $\delta\omega R_C^2/D$ anwendbar. Allerdings zeigen die einzelnen Größen, die in Gl. (\ref{Endergebnis}) für $\delta\omega R_C^2/D > p_0$ gewisse Symmetrieeigenschaften. So gilt beispielsweise für die charakteristischen Werte $k_1=k_0^{\displaystyle{*}}$ (siehe Gl. (\ref{k0k1}) und Abb. \ref{Fig:k}) und für die Eigenwerte $\lambda_{n1}=\lambda_{n0}^{\displaystyle{*}}$ (siehe Gl. (\ref{Eigenschaftl})). Werden nun die allgemeinen Eigenschaften der Mathieufunktionen (siehe Gl. (\ref{Symmetriece})), der Eigenwerte (siehe Gl. (\ref{Eigenschaftl})), der Eigenfunktionen (siehe Gl. (\ref{EigenschaftR})) und der Entwicklungskoeffizienten (siehe Gl. (\ref{Eigenschaftc})) berücksichtigt, ergibt sich aus Gl. (\ref{Endergebnis})
\begin{align}
\nonumber
\frac{m(\mathbf{r},t)}{m_0} 
=& \sum_{m=0}^{l} \sum_{n=1}^{\infty} \left[ c_{n\,2m} \text{ce}_{4m}(\phi) R_{n\,2m}(r) \,\text{e}^{- t \left[\lambda_{n\,2m}^2 \frac{D}{R_{\text{C}}^2} + \frac{1}{T_2}\right]}+ c_{n\,2m}^{\displaystyle{*}} \text{ce}_{4m}^{\displaystyle{*}} \left(\frac{\pi}{2}-\phi\right) R_{n\,2m}^{\displaystyle{*}}(r) \, \text{e}^{- t \left[\lambda_{n\,2m}^{2\displaystyle{*}} \frac{D}{R_{\text{C}}^2} + \frac{1}{T_2}\right]} \right]\\
+& \sum_{m=2l+2}^{\infty} \sum_{n=1}^{\infty} c_{nm} \text{ce}_{2m}(\phi) R_{nm} (r)\text{e}^{- t \left[\lambda_{nm}^2 \frac{D}{R_{\text{C}}^2} + \frac{1}{T_2}\right]} \quad \text{für} \quad p_{l}<\frac{\delta\omega R_C^2}{D}<p_{l+1} \,.
\end{align}
Analog zur Relation (\ref{Parsevalcnm}) erhält man
\begin{equation} 
2\sum_{m=0}^{l} \sum_{n=1}^{\infty}\text{Re}(c_{n\,2m}^2 N_{n\,2m}) + \sum_{m=2l+2}^{\infty} \sum_{n=1}^{\infty} c_{nm}^2 N_{nm} = \frac{1}{\eta}-1 \quad \text{für} \quad p_{l}<\frac{\delta\omega R_C^2}{D}<p_{l+1} \,.
\end{equation}

\section*{\normalsize{2.5 Experimentelle Untersuchung}}
\addcontentsline{toc}{section}{2.5 Experimentelle Untersuchung}
Das Kroghsche Kapillarmodell wurde ursprünglich zur Beschreibung von Skelettmuskelgewebe entwickelt. Der Skelettmuskel eignet sich zur Untersuchung der Dephasierung im Muskelgewebe als Modellsystem. Deshalb wurden Messungen sowohl am Skelettmuskel als auch am Herzmuskel durchgeführt. Die Messung am Skelettmuskel bietet allerdings den Vorteil, dass keine Bewegungsartefakte wie bei schlagenden Herzen auftreten. 

\paragraph{Skelettmuskelgewebe:} Der Freie Induktionszerfall im Skelettmuskel wurde in der Muskulatur des Oberschenkels einer Ratte gemessen. Die geeignete Methode zur exakten Vermessung des Freien Induktionszerfalls ist die voxelselektive PRESS-Sequenz (englisch: Point RESolved Spectroscopy) \cite{Bottomley84,Bottomley87}. Die Experimente am Skelettmuskel wurden an einem MR-Tomographen vom Typ Bruker-Biospec bei einer Feldstärke von $7\,\mathrm{T}$ durchgeführt. Die Bildgebungsgradienten erzeugen eine maximale Gradientenstärke von $397\, \mathrm{mT}/\mathrm{m}$. Zum Senden und Empfangen der Hochfrequenzsignale wurde ein $72 \,\mathrm{mm}$ Quadratur-Birdcage-Resonator genutzt. 

Für die Messungen wurde eine $300\,\mathrm{g}$ schwere und gesunde weibliche Wistar-Ratte (Charles River Laboratories, Research Model and Services, Germany GmbH, Charles River, Sulzfeld) mit $4\%$ Isofluran in Sauerstoff anästhesiert. Das Tier wurde im MR-Scanner so positioniert, dass der Skelettmuskel des Oberschenkels im Zentrum des Birdcage-Resonators liegt. Atmung und Herzaktion wurden mit einem pneumatischen Sensor (Respiration Sensor, Graseby Medical Limited, Watford, UK) überwacht, der auf dem Bauch der Ratte angebracht war. 

\begin{figure}
\begin{center}
\includegraphics[width=13cm]{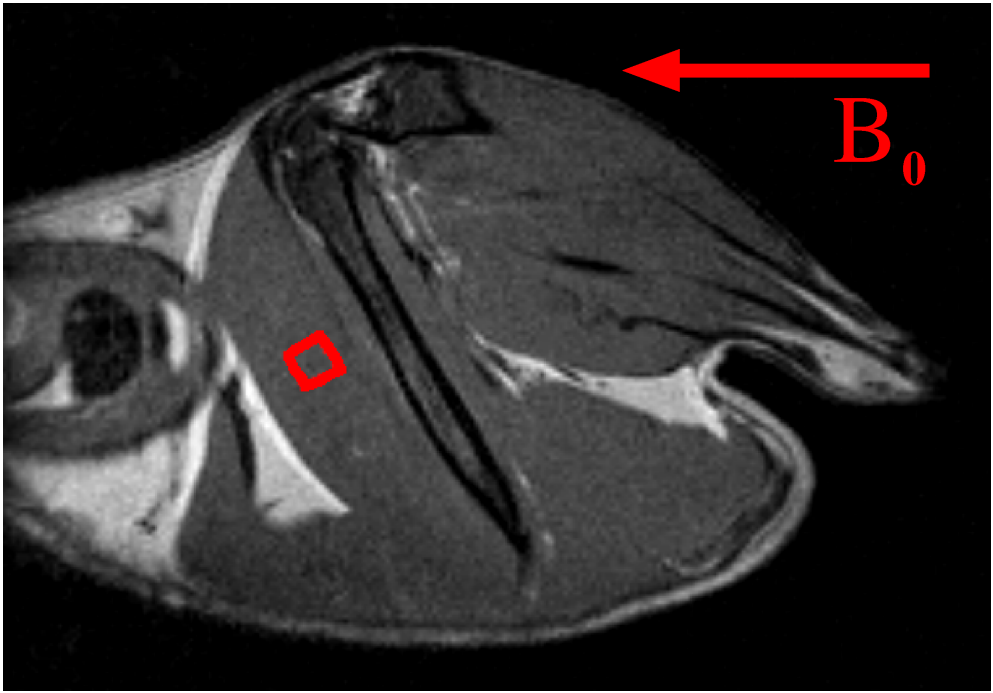}
\vspace{-0.5cm}
\caption[Messung im Skelettmuskel]{\label{fig:Ratte}{\footnotesize Repräsentative Schicht des Oberschenkels einer Ratte mit der Lokalisation des ausgewählten PRESS-Voxels.}}
\end{center}
\end{figure}

Nach der Lokalisationssequenz (RARE, Repetitionszeit $T_R = 2,\!5 \,\mathrm{s}$, effektive Echozeit $T_{Eeff} = 9,\!4\,\mathrm{ms}$, RARE-Faktor $=2$), wurde mit der FASTMAP-Sequenz \cite{Gruetter93} lokal auf ein würfelförmiges, $5\, \mathrm{mm} \times 5\, \mathrm{mm} \times 5\, \mathrm{mm}$ großes Voxel im Zentrum der linken Hüftmuskulatur geshimmt. Die Längsachse des Voxels war um $60^{\circ}$ gegen das äußere Magnetfeld geneigt. Innerhalb des würfelförmigen Shimmvoxels würde ein kleineres würfelförmiges $2,\!5\, \mathrm{mm} \times 2,\!5\, \mathrm{mm} \times 2,\!5\, \mathrm{mm}$ großes Voxel ausgewählt, um den Freien Induktionszerfall mit einer PRESS-Sequenz zu messen (siehe Abb. \ref{fig:Ratte}). Die geeigneten Parameter der Sequenz waren: Repetitionszeit $T_R = 6\,\mathrm{s}$, Echozeit $T_E = 9,\!9 \,\mathrm{ms}$, wobei acht Mittelungen durchgeführt wurden ($NA = 8$). Größe und Lokalisation der Voxel wurden so gewählt, um Einflüsse und daraus resultierende Artefakte der umgebenden Knochen bzw. Gewebegrenzen zu minimieren. Des Weiteren wurde mit der PRESS-Sequenz die Relaxationszeit $T_2$ im Muskel gemessen, wobei eine Messung ($NA = 1$) bei einer Repetitionszeit von $T_R = 6\,\mathrm{s}$ mit acht verschiedenen Echozeiten $T_E = (12, 15, 20, 25, 35, 50, 80, 120)\,\mathrm{ms}$ aufgenommen wurde.

\begin{figure}
\begin{center}
\includegraphics[width=10cm]{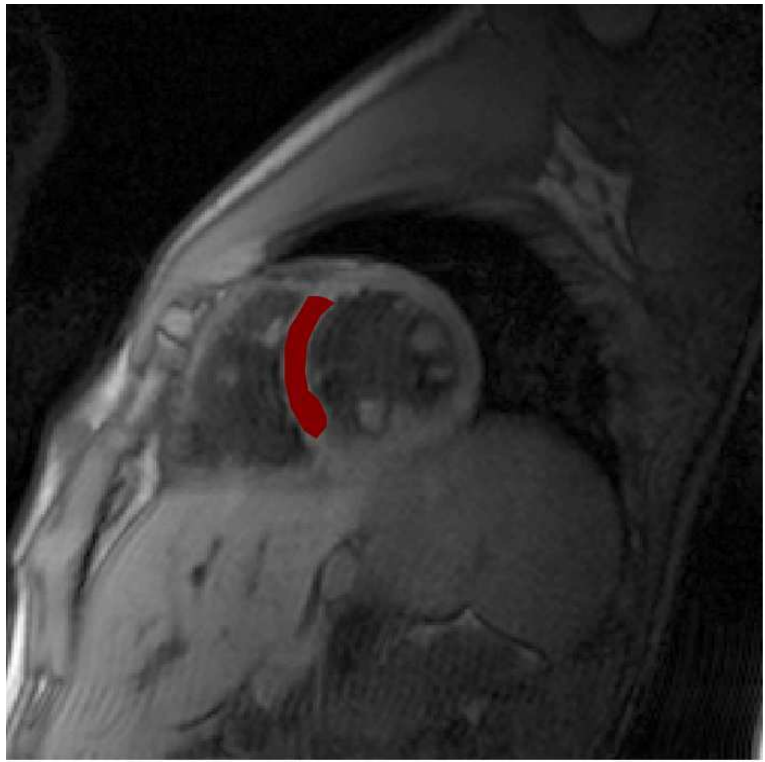}
\vspace{-0.5cm}
\caption[Areal im Septum interventriculare]{\label{Fig:MRBild}{\footnotesize Areal im Septum interventriculare (rot markierter Bereich), aus dem der Signal-Zeit-Verlauf analysiert wurde.}}
\end{center}
\end{figure}

\paragraph{Herzmuskelgewebe:} Zur experimentellen Untersuchung des Signal-Zeit-Verlaufs im Myokard wurde ein gesunder Proband in einem MR-Tomographen vom Typ Siemens Avanto mit einer Feldstärke von $1,\!5 \,\text{T}$ untersucht. Der freie Induktionszerfall wird durch eine Gradientenechosequenz zu verschiedenen Echozeiten ausgelesen. Um Bewegungsartefakte zu vermeiden, wird auf die R-Zacke des EKG getriggert. Wie in Abb. \ref{Fig:MRBild} dargestellt, wird das Signal aus einem umschriebenen Bereich im Septum interventriculare aufgenommen. In diesem Bereich des Herzens sind die Kapillaren nahezu senkrecht zum äußeren Magnetfeld orientiert. Des Weiteren können Suszeptibilitätseinflüsse von umliegenden Geweben, insbesondere der Lunge, minimiert werden, da das gewählte Areal im Herzen nur von Blut umgeben ist. Allerdings entstehen durch die Blutströmungen in den angrenzenden Ventrikeln Flussartefakte, die jedoch durch eine Black-Blood-Methode unterdrückt wurden.
Für die transversale Spin-Spin-Relaxationszeit wird der für das Myokard typische Wert $T_2=57 \,\text{ms}$ angenommen \cite{Schachner}. Der Neigungswinkel $\theta$ der Kapillaren beeinflusst die Stärke der Spindephasierung (siehe Gl. (\ref{Dipol})). Da die Kapillaren parallel zu den Kardiomyozyten verlaufen, ist deren Neigungswinkel mit dem Neigungswinkel der Muskelfasern identisch. Durch Diffusionstensorbildgebung kann die Faserarchitektur des Myokards untersucht werden \cite{Reese95} und es zeigt sich, dass im Septum interventriculare des Herzens die Fasern nahezu senkrecht zum äußeren Magnetfeld verlaufen.

\chapter*{\label{Kap.Ergebnisse}\vspace{-3cm} 3 Ergebnisse}
\addcontentsline{toc}{chapter}{3 Ergebnisse}

\section*{\normalsize{3.1 Signalevolution in der Strong-Collision-Näherung}}
\addcontentsline{toc}{section}{3.1 Signalevolution in der Strong-Collision-Näherung}

Mit den in Abschnitt 2.3 dargestellten Methoden ist es nun möglich, die Ergebnisse zu einem Gesamtbild zusammenzufügen. Das zu Grunde liegende Diffusionsregime wird durch das Produkt $\tau \delta\omega$ festgelegt, welches durch die funktionellen Parameter des Myokards bestimmt wird. Die Lage der Punkte $+ \Omega$ und $- \Omega$ in Abb. \ref{fig:4_neu} ist eng mit dem Wert $\tau \delta\omega$, also dem zu Grunde liegenden Diffusionsregime, verbunden. Deshalb ist es sinnvoll, die Reise dieser Punkte in der komplexen Ebene zu betrachten, wenn das Diffusionsregime verändert wird.

\paragraph{Motional-Narrowing-Regime:} Im Motional-Narrowing-Grenzfall können die Suszeptibilitätseffekte vernachlässigt werden ($\tau \, \delta\omega = 0 $). In diesem Fall nimmt die Frequenz aus Gl. (\ref{Omega}) den Wert $\Omega=1/\tau$ an. Dementsprechend befinden sich die Singularitäten auf der reellen Achse an den Positionen $+ \Omega = + 1/\tau$ und $- \Omega = - 1/\tau$. Um jedoch den Magnetisierungs-Zeit-Verlauf zu erhalten, ist es zweckmäßig, die Funktion $\hat{M}(s)$ in diesem Grenzfall $\tau \delta\omega = 0$ (siehe Gl. (\ref{MdachMN})) zu betrachten. Dies führt direkt zu
\begin{equation}
M(t)=1 \,.
\end{equation}
Das Motional-Narrowing-Regime ist auch das zu Grunde liegende Diffusionsregime wenn das regionale Blutvolumenverhältnis $\eta$ gegen Null strebt. In diesem Fall wird der Radius des Dephasierungszylinders unendlich groß und die Diffusion ist uneingeschränkt. Die Mehrzahl der  Spins ist weit von der Kapillare entfernt und präzedieren fast immer mit der gleichen Resonanzfrequenz. Wie erwartet gibt es keinen Magnetisierungszerfall auf Grund der Dephasierung im Motional-Narrowing-Regime. Der experimentell beobachtbare Magnetisierungszerfall wird nur durch die intrinsische transversale Spin-Spin-Relaxation verursacht: $S(t) = S(0) \exp{(-t/T_2)}$.

\paragraph{Fast-Diffusion-Regime:} Mit abnehmenden Diffusionskoeffizienten nimmt das Produkt $\tau \, \delta\omega$ zu und entsprechend Gl. (\ref{1}) nimmt die Frequenz $\Omega$ ab; sie bleibt dabei aber reell. Demzufolge bewegen sich die Singularitäten auf der reellen Achse aufeinander zu und konvergieren im Koordinatenursprung, d.h. dieses Regime wird durch die Bedingung
\begin{equation}
\label{Bedingungfastdiff}
0 < \tau \, \delta \omega < \frac{1+\eta}{2\eta}
\end{equation}
charakterisiert. 

\begin{figure}
\begin{center}
\includegraphics[width=\textwidth]{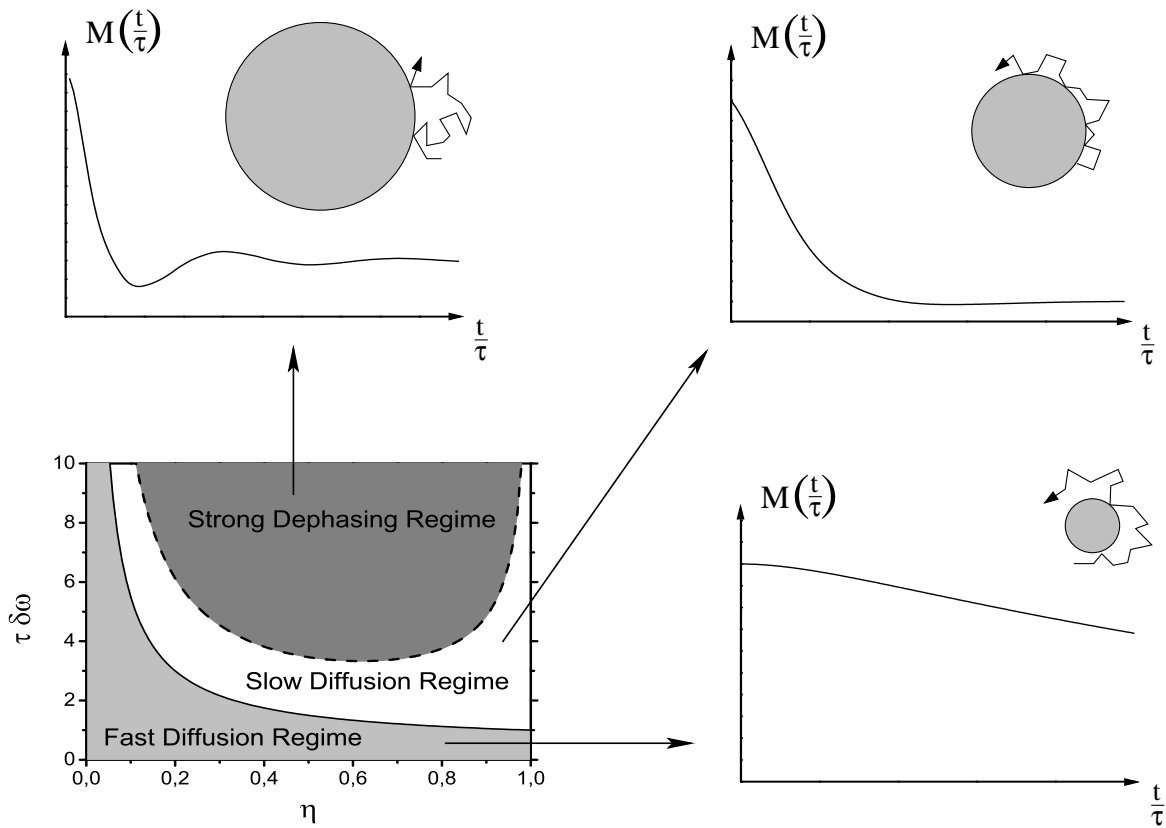}
\vspace{-1.0cm}
\caption[Veranschaulichung der Diffusionsregime]{\footnotesize Veranschaulichung der Diffusionsregime. Die linke untere Abbildung zeigt die Einteilung der Diffusionsregime nach Abb. \ref{fig:Bereiche}. Im Fast-Diffusion-Regime (rechte untere Abbildung) diffundiert ein Spin fast um die ganze dünne Kapillare und sieht somit viele verschiedene Resonanzfrequenzen, was in einen langsamen Abfall der Magnetisierung resultiert. Im Strong-Diffusion-Regime (linke obere Abbildung) befindet sich der Spin fast immer an der gleichen Stelle bezüglich der dicken Kapillare und die Magnetisierung zeigt oszillierende Anteile. Das Slow-Diffusion-Regime liegt zwischen den beiden gezeigten Regimen und ist in der rechten oberen Abbildung dargestellt.} 
\label{fig:Bereicheneu}
\end{center}
\end{figure}

Im Fast-Diffusion-Regime ist die dynamische Frequenz $1/\tau$ viel größer als die statische Frequenz $\delta\omega$. Dies ist bei einer dünnen Kapillare, einem großen Diffusionskoeffizienten und bei einer kleinen statischen Frequenz gegeben. Unter diesen Bedingungen hat die Trajektorie eines diffundierenden Spins während des Dephasierungsprozesses genug Zeit, um eine große Anzahl unterschiedlicher Resonanzfrequenzen zu sehen. Dies führt zu einem Mittelungsprozess in der akkumulierten Phase und demzufolge zu einem langsamen Abfall der gesamten Magnetisierung (siehe das rechte untere Schema in Abb. \ref{fig:Bereicheneu}).

\begin{figure}
\begin{center}
\includegraphics[width=\textwidth]{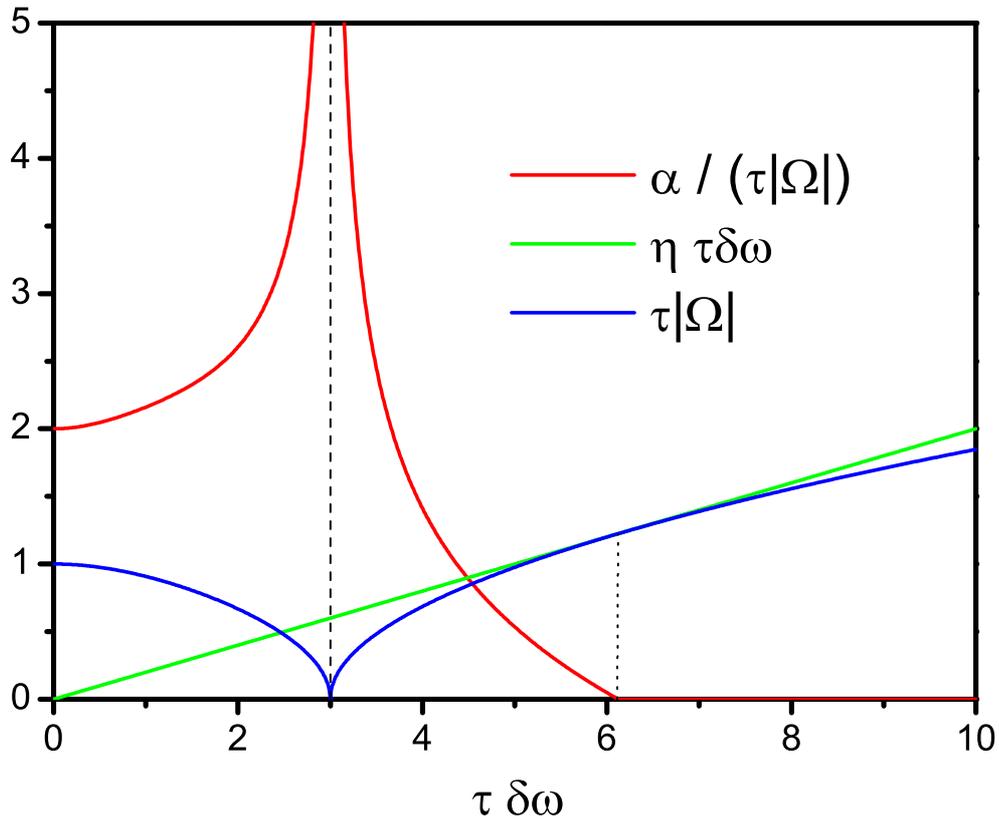}
\vspace{-1.0cm}
\caption[Abhängigkeit des Koeffizienten $\alpha/(\tau|\Omega|)$ vom Diffusionsregime]{\footnotesize Abhängigkeit des Koeffizienten $\alpha/(\tau|\Omega|)$ vom Diffusionsregime. Der Koeffizient $\alpha/(\tau|\Omega|)$ ist nach Gl. (\ref{alphaneu}) berechnet (rote Kurve). Im Critical-Regime (siehe Gl. (\ref{2})) strebt der Koeffizient gegen unendlich, analog dem Fall der kritischen Dämpfung beim gedämpften Oszillator. Der Absolutbetrag der Frequenz $\Omega$ ergibt sich aus Gl. (\ref{Omega}) (blaue Kurve). Im Fast-Diffusion-Regime (\ref{1}) ist diese Frequenz rein reell und im Slow-Diffusion-Regime (\ref{3}) ist die Frequenz rein imaginär. Im Critical-Regime (\ref{2}) verschwindet die Frequenz. Der Abstand der Verzweigungslinien von der reellen Achse ist $\eta\delta\omega$ (grüne Gerade). Die Kurven sind für das regionale Blutvolumenverhältnis $\eta=0,\!2$ ermittelt, d.h. das Critical-Regime liegt bei $\tau\delta\omega=3$ (gestrichelte Linie) und die Singularitäten erreichen die Verzweigungslinien bei $\tau\delta\omega=6,\!12$ (gepunktete Linie).} 
\label{fig:3}
\end{center}
\end{figure}

\begin{figure}
\begin{center}
\includegraphics[width=15cm]{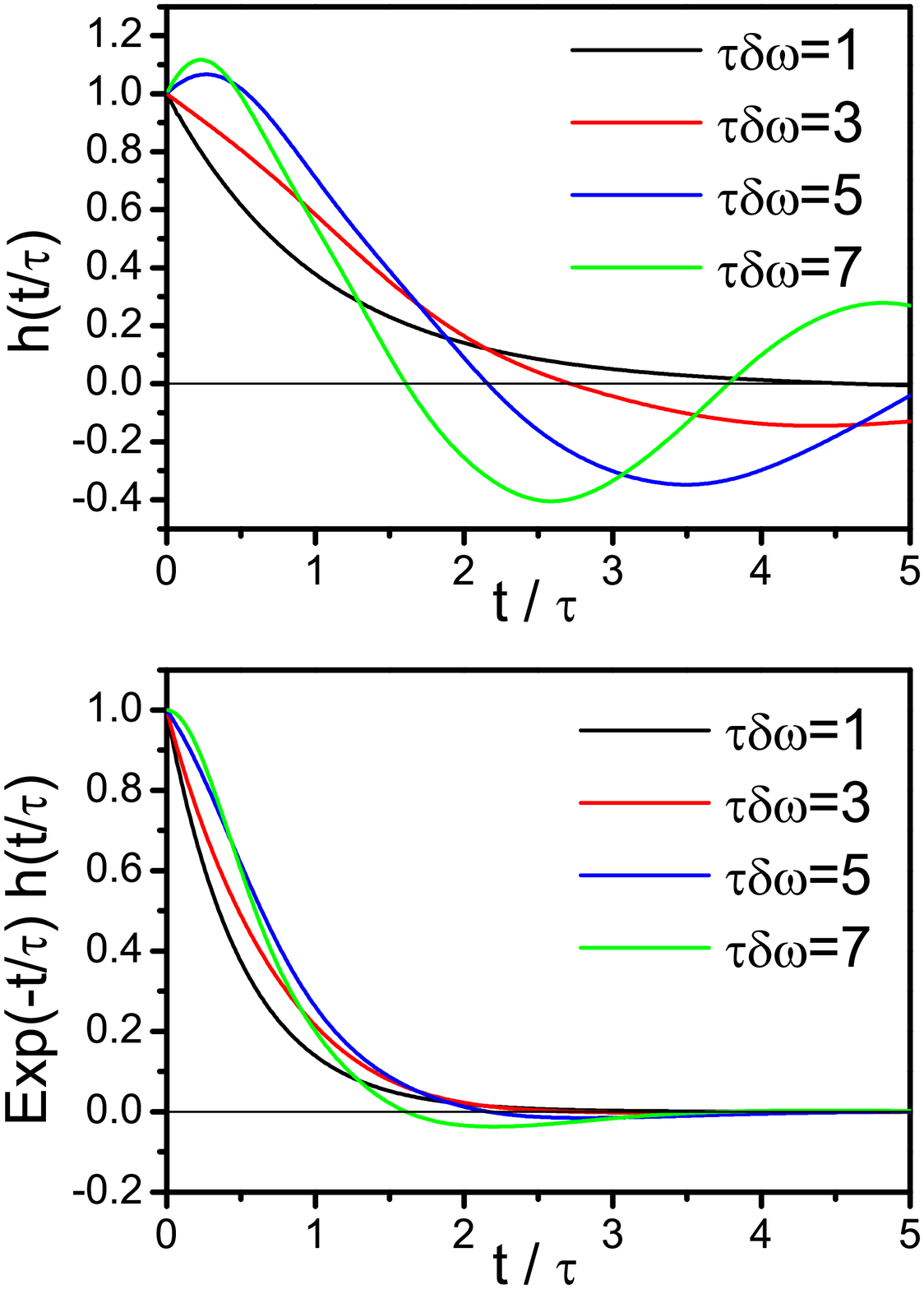}
\vspace{-1.0cm}
\caption[Schaubild der Funktionen $h(t/\tau)$ bzw. $\text{exp}(-t/\tau)h(t/\tau)$.]{\footnotesize Abhängigkeit der Funktionen $h(t/\tau)$ (obere Abbildung) bzw. $\text{exp}(-t/\tau)h(t/\tau)$ (untere Abbildung) von der normalisierten Zeit nach Gl. (\ref{hneu}) für ein regionales Blutvolumenverhältnis von $\eta=0,\!2$. Wenn die Bedingung $0<\tau\delta\omega<3$ erfüllt ist (siehe Gl. (\ref{Bedingungfastdiff})), liegt das Fast-Diffusion-Regime zu Grunde (schwarze Kurve) und die Funktion fällt monoton. Im Critical-Regime ($\tau\delta\omega=3$, Gl. (\ref{BedingungIR})) sind noch keine Oszillationen sichtbar (rote Kurve). Im Slow-Diffusion-Regime ($3<\tau\delta\omega<6,\!12$, siehe Gl. (\ref{Bedingungslowdiff})) und im Strong-Dephasing-Regime ($6,\!12 \leq \tau\delta\omega< \infty$, siehe Gl. (\ref{BedingungStrongDephas})) treten die typischen Oszillationen auf (blaue und grüne Kurve).} 
\label{fig:4}
\end{center}
\end{figure}

\begin{figure}
\begin{center}
\includegraphics[width=\textwidth]{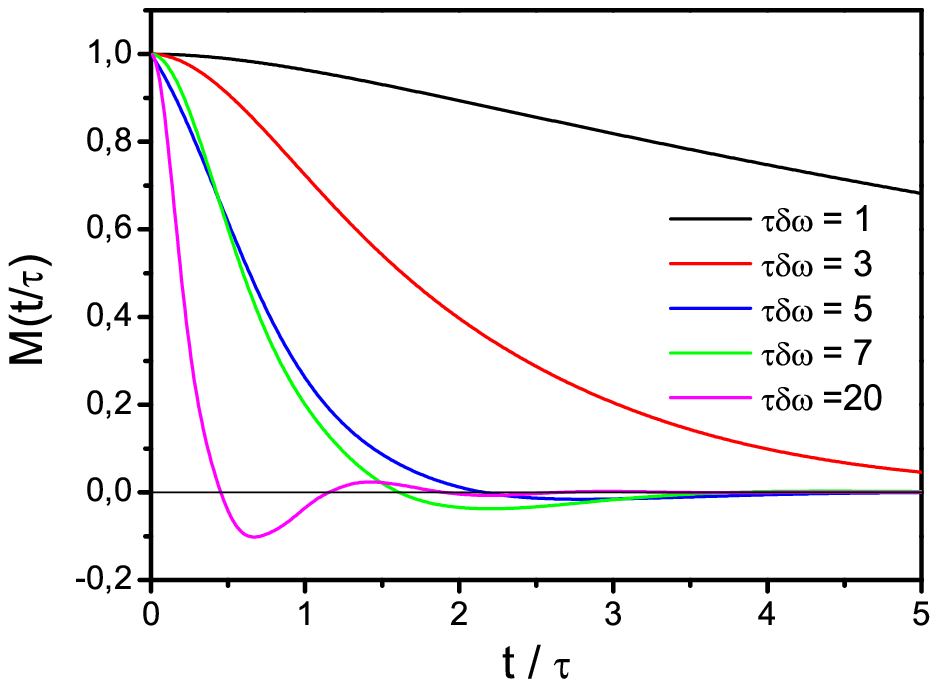}
\vspace{-1.0cm}
\caption[Magnetisierungs-Zeit-Verlauf]{\footnotesize Magnetisierungs-Zeit-Verlauf für ein regionales Blutvolumenverhältnis von $\eta=0,\!2$. Im Fast-Diffusion-Regime ($0<\tau\delta\omega<3$) fällt die Magnetisierung entsprechend Gl. (\ref{ErgebnisFDR}) monoton. Die Magnetisierung im Critical-Regime ($\tau\delta\omega=3$) folgt aus Gl. (\ref{ErgebnisIR}). Im Slow-Diffusion-Regime ($3<\tau\delta\omega<6,\!12$) ist das oszillierende Verhalten der Sinusfunktion in Gl. (\ref{ErgebnisSDR}) erkennbar. Im Strong-Dephasing-Regime ($6,\!12 \leq \tau\delta\omega< \infty$) stimmt die grüne Kurve in dieser Abbildung mit der grünen Kurve in der unteren Abb. \ref{fig:4} überein, da in diesem Fall $M(t)=\text{e}^{-t/\tau}h(t)$ gilt. Für größere Werte ($\tau\delta\omega=20$) sind die charakteristischen Oszillationen klar erkennbar.}
\label{fig:5}
\end{center}
\end{figure}

Die Residuen in Gl. (\ref{Mres}) wurden in Gl. (\ref{res1}) und Gl. (\ref{res2}) berechnet. Die Integration um die Verzweigungslinien erfolgte in Gl. (\ref{Integralbranch}), und das Ergebnis kann in Gl. (\ref{hint}) eingesetzt werden, was zu einem zusätzlichen Term $h(t)$ führt. Letztlich kann die Magnetisierung in der Form
\begin{align}
\label{ErgebnisFDR}
M(t) = \mathrm{e}^{-\,\frac{t}{\tau}} \left[\frac{\alpha}{\tau\Omega} \mathrm{sinh}(\Omega t) +h(t)\right]
\end{align}
geschrieben werden, wobei der Vorfaktor $\alpha$ ein rein reeller und positiver Parameter ist, der in Gl. (\ref{alphaneubereich}) angegeben wurde:
\begin{equation}
\label{alphaneu}
\alpha = \frac{2}{[1-\eta]^2}\left[1+\eta^2-\eta\frac{2+\tau^2\delta\omega^2[1-\eta]^2}{\sqrt{1+\tau^2\delta\omega^2[1-\eta]^2}} \right] \,.
\end{equation}
Die Frequenz $\Omega$ wurde in Gl. (\ref{Omega}) im Abschnitt 2.3 berechnet:
\begin{equation}
\label{Omeganeu}
\Omega \!=\! \frac{\sqrt{\!1 \!+\! \eta^2 \!-\!2\eta\sqrt{1 + \tau^2 \delta\omega^2[1-\eta]^2}}}{\tau[1-\eta]}
\end{equation}
und die Funktion $h(t)$, wurde schon in Gl. (\ref{h}) im Abschnitt 2.3 angegeben:
\begin{align}
\label{hneu}
h(t) & = \frac{2}{\pi} \int\limits_\eta^1 \mathrm{d}x \,\, \frac{\sin(x\,\delta\omega\,t) \sqrt{x^2 - \eta^2}}{x^2[1-\eta]+ \displaystyle{\frac{1+\eta}{\tau^2\,\delta\omega^2}} - \displaystyle{\frac{2\eta}{\tau\,\delta\omega}\sqrt{1-x^2}}}  \\ \nonumber
& + \frac{2}{\pi} \int\limits_1^\infty 	\mathrm{d}x \,\, \frac{\sin(x\,\delta\omega\, t)\left[ \sqrt{x^2-\eta^2} + \eta\sqrt{x^2-1} \right]}{\displaystyle{\frac{x^2[1+\eta^2] + 2\eta \left[\sqrt{[x^2-\eta^2][x^2-1]} - \eta \right]}{1+\eta}} + \displaystyle{\frac{1+\eta}{\tau^2\,\delta\omega^2}}} \,.
\end{align}
Der Koeffizient $\alpha$ hängt nur vom Produkt $\tau\delta\omega$ ab; diese Abhängigkeit ist in Abb. \ref{fig:3} dargestellt. 
Wenn die Singularitäten die Verzweigungslinien erreichen, wird $\alpha=0$ (siehe die gepunktete Linie in Abb. \ref{fig:3}). Für größere Werte von $\tau\delta\omega$ verlassen die Singularitäten das Riemann-Blatt, in dem der Integrationsweg liegt, und bewegen sich auf dem zweiten Riemann-Blatt entsprechend dem zweiten Vorzeichen der Quadratwurzel. Die Funktionen $h(t/\tau)$ bzw. $\text{exp}(-t/\tau)h(t/\tau)$ hängen ebenfalls nur vom Produkt $\tau\delta\omega$ ab und zeigen einen monoton fallenden Verlauf im Fast-Diffusion-Regime (schwarze Kurven in Abb. \ref{fig:4}). Auf Grund der zwei Residuen zeigt die Magnetisierung ein biexponentielles Verhalten, das zu einer Sinushyperbolicus-Funktion mit exponentiellem Zerfall zusammengefasst werden kann. In Analogie zu dem gedämpften Oszillator entspricht dieser Fall dem Kriechfall. Die resultierende Magnetisierung entsprechend Gl. (\ref{ErgebnisFDR}) ist in Abb. \ref{fig:5} dargestellt (schwarze Kurve). 

\paragraph{Critical-Regime:} Wenn die Frequenz $\Omega$ den Wert $\Omega=0$ annimmt, fallen die beiden Singularitäten im Koordinatenursprung zusammen (siehe Abb. \ref{fig:4_neu}). Aus Gl. (\ref{2}) lässt sich erkennen, dass dies eintritt, wenn
\begin{equation} 
\label{BedingungIR}
\tau \, \delta \omega = \frac{1+\eta}{2\eta}\,.
\end{equation}
Das entsprechende Residuum des verbleibenden Pols zweiter Ordnung ist in Gl. (\ref{resIR}) berechnet. Die Integration um die Verzweigungslinien gleicht den oben genannten Fällen. Demzufolge kann die Magnetisierung in der Form
\begin{align}
\label{ErgebnisIR}
M(t) = \mathrm{e}^{-\,\frac{t}{\tau}} \left[\frac{[1+\eta]^2}{1+\eta^2} \frac{t}{\tau} +h(t)\right]
\end{align}
geschrieben werden. Dieses Ergebnis kann auch erhalten werden, wenn die Bedingung (\ref{BedingungIR}) in den Ausdruck (\ref{alphaneu}) eingesetzt und der Sinushyperbolikus für kleine Argumente entwickelt wird. In Analogie zum gedämpften Oszillator entspricht dieser Fall der kritischen Dämpfung. In Abb. \ref{fig:5} ist der Magnetisierungs-Zeit-Verlauf im Critical-Regime dargestellt (rote Kurve). Für größere Werte $\tau\delta\omega$ zeigt die Magnetisierung ein oszillierendes Verhalten, wie im nächsten Abschnitt gezeigt wird.

\paragraph{Slow-Diffusion-Regime:} Dieses Regime ist durch die Bedingung
\begin{equation}
\label{Bedingungslowdiff}
\frac{1+\eta}{2\eta} < \tau \, \delta \omega < \frac{1}{\eta}\sqrt{\frac{1+\eta}{1-\eta}}
\end{equation}
charakterisiert. Die Dephasierung geschieht schnell genug, um ein effektives Mitteln der Resonanzfrequenzen auf Grund der Diffusion zu verhindern. Während der Dephasierung besucht die Trajektorie des diffundierenden Spins nicht so viele Resonanzfrequenzen wie im Fast-Diffusion-Regime. Dies führt zu einem schnelleren Abfall der Magnetisierung
(siehe rechtes oberes Schema in Abb. \ref{fig:Bereicheneu}).

Entsprechend Gl. (\ref{3}) nimmt im Slow-Diffusion-Regime die Frequenz $\Omega$ rein imaginäre Werte $\Omega=\mathrm{i}|\Omega|$ an. Die Singularitäten bewegen sich auf der imaginären Achse (siehe Abb. \ref{fig:4_neu}). In diesem Fall können die Residuen zu einer Sinusfunktion kombiniert werden und für die Magnetisierung ergibt sich der Ausdruck
\begin{align}
\label{ErgebnisSDR}
M(t) = \mathrm{e}^{-\,\frac{t}{\tau}} \left[\frac{\alpha}{\tau|\Omega|} \text{sin}(|\Omega| t) +h(t) \right] \,.
\end{align}
Dieses Ergebnis kann auch aus Gl. (\ref{ErgebnisFDR}) mit $\Omega=\mathrm{i}|\Omega|$ erhalten werden. In Analogie zum gedämpften Oszillator entspricht dieser Fall der Unterdämpfung. Der entsprechende Zeitverlauf der Magnetisierung ist in Abb. \ref{fig:5} dargestellt (blaue Kurve). Ein minimaler oszillierender Anteil ist erkennbar, da die Magnetisierung ab einer gewissen Zeit auch negative Werte annimmt.

\paragraph{Strong-Dephasing-Regime:} Wenn die Frequenz $\Omega$ die Verzweigungslinien erreicht ($\Omega=\mathrm{i}\eta\delta\omega$), tragen die Residuen nicht zum Integral (\ref{BromwichIntegral}) bei. Dies geschieht, wenn die Bedingung
\begin{equation}
\label{BedingungStrongDephas}
\frac{1}{\eta}\sqrt{\frac{1+\eta}{1-\eta}} \leq \tau \, \delta \omega < \infty
\end{equation}
erfüllt ist (siehe auch Bedingung (\ref{4neu})). Wie in Abb. \ref{fig:3} zu sehen ist, wird der Vorfaktor $\alpha=0$, wenn die Bedingung (\ref{BedingungStrongDephas}) erfüllt ist. In diesem Fall trägt nur das Integral um die Verzweigungslinien zur Magnetisierung bei:
\begin{align}
\label{MagnetisierungStrongDephas}
M(t) = \mathrm{e}^{-\,\frac{t}{\tau}} h(t) \,.
\end{align}
Der entsprechende Magnetisierungs-Zeit-Verlauf ist ebenfalls in Abb. \ref{fig:5} dargestellt (grüne Kurve). Das typische oszillierende Verhalten ist erkennbar. Im Strong-Dephasing-Regime ist die Dephasierung hauptsächlich durch die statische Frequenz $\delta\omega$ bestimmt. Während des Dephasierungsprozesses besucht die Trajektorie eines Spins nur einen kleinen Teil aller möglichen Resonanzfrequenzen und eine Diffusionsmittelung findet kaum statt. Das oszillierende Verhalten der Magnetisierung ergibt sich aus der spezifischen Form der lokalen Resonanzfrequenz. 

\paragraph{Static-Dephasing-Regime:} Im Static-Dephasing-Grenzfall gibt es keine Diffusionseffekte ($D \to 0$), und dementsprechend folgt
\begin{equation}
\tau \, \delta\omega \to \infty \,.
\end{equation}
Die signalgebenden Spins befinden sich immer an der gleichen Stelle und bewegen sich nicht. Demzufolge gibt es keinen Mittelungseffekt auf Grund der Diffusion. Die Korrelationszeit $\tau$ nach Gl. (\ref{eEq19}) strebt gegen unendlich ($\tau \to \infty$). Die Frequenz in Gl. (\ref{Omega}) strebt gegen den Grenzwert $\Omega \to \mathrm{i}\infty$ und die Singularitäten streben auf der imaginären Achse gegen $+\Omega \to + \mathrm{i}\infty$ und $-\Omega \to - \mathrm{i}\infty$. Ausgangspunkt für den Magnetisierungs-Zeit-Verlauf ist Gl. (\ref{MagnetisierungStrongDephas}) des Strong-Dephasing-Regimes. Im Grenzfall $\tau \to \infty$ wird der Vorfaktor zu $\mathrm{e}^{-\,\frac{t}{\tau}} \to 1$. Führt man in der Funktion $h(t)$ in Gl. (\ref{hneu}) den Grenzübergang $\tau \, \delta\omega \to \infty$ durch, ergibt sich letztlich für den Magnetisierungs-Zeit-Verlauf im Static-Dephasing-Regime:
\begin{align}
\label{MagnetisierungStaticDephas}
M(t) = \frac{2}{\pi} \left[\int\limits_\eta^1 \!\!\! \mathrm{d}x \frac{\sin(x\,\delta\omega\,t) \sqrt{x^2 - \eta^2}}{x^2[1-\eta]} + \int\limits_1^\infty \!\!\!	\mathrm{d}x \frac{\sin(x\,\delta\omega\, t)\left[ \sqrt{x^2-\eta^2} + \eta\sqrt{x^2-1} \right][1+\eta]}{\displaystyle{x^2[1+\eta^2] + 2\eta \left[\sqrt{[x^2-\eta^2][x^2-1]} - \eta \right]}}\right] \,,
\end{align}
die mit dem in Gl. (\ref{FIDcyl}) angegebenen Magnetisierungs-Zeit-Verlauf übereinstimmt. Des Weiteren stimmt der Magnetisierungs-Zeit-Verlauf im Static-Dephasing-Grenzfall mit den Ergebnissen überein, die sich aus der Frequenzverteilung um eine Kapillare ergeben \cite{Ziener05MAGMA}. Aus der Darstellung der Magnetisierung nach Gl. (\ref{MagnetisierungStaticDephas}) folgen auch leicht die Grenzfälle $\eta \to 0$: $M(t)=1$ und $\eta \to 1$: $M(t)=J_0(\delta\omega t)$, die mit den Ergebnissen des Static-Dephasing-Regimes in \cite{Ziener05MAGMA} übereinstimmen.

\section*{\normalsize{3.2 Analytische Lösung der Bloch-Torrey-Gleichung}}
\begin{figure}
\begin{center}
\includegraphics[width=12cm]{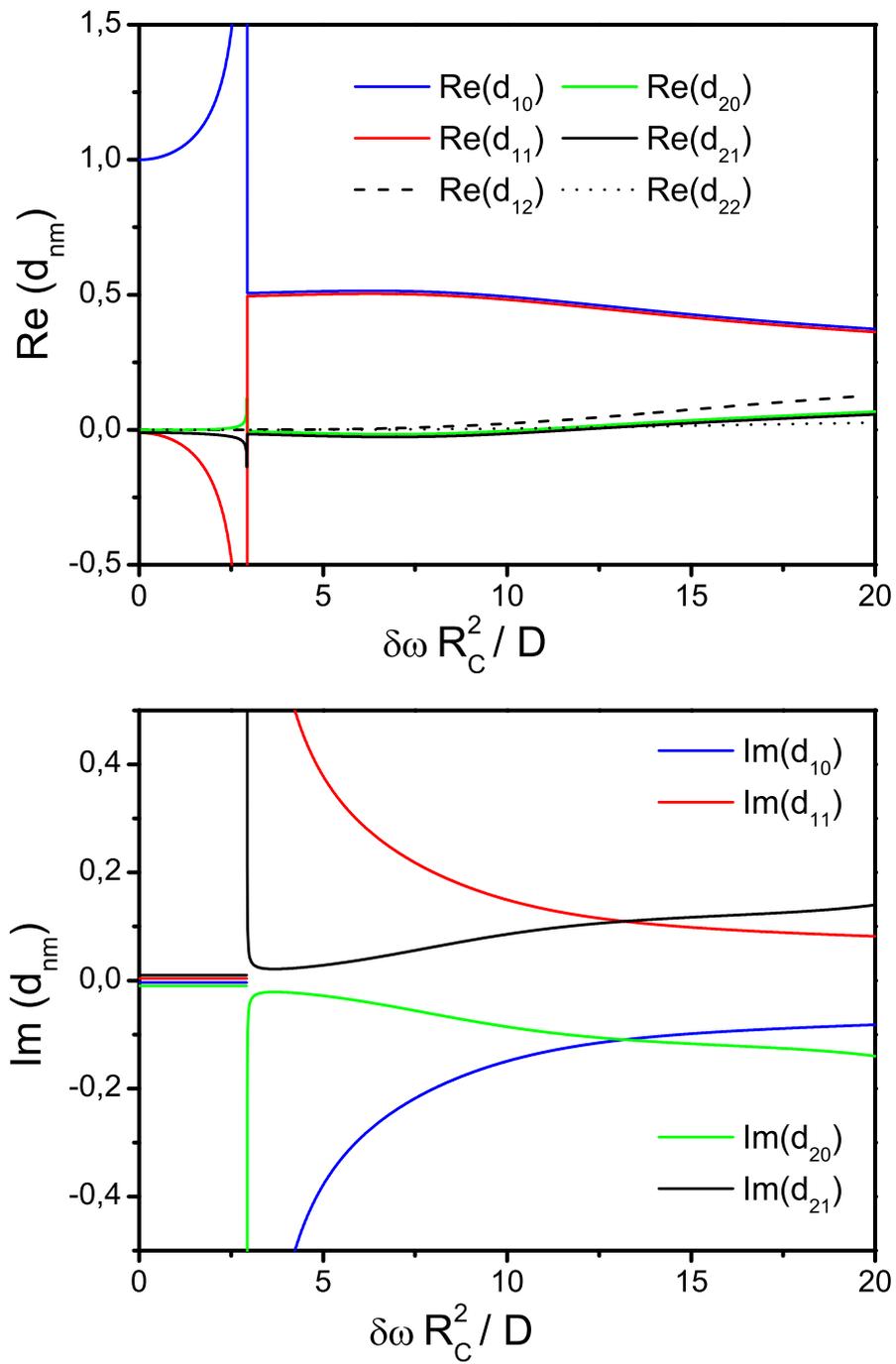}
\caption[Entwicklungskoeffizienten des Signals]{\label{Fig:Spektrumdnm}{\footnotesize Entwicklungskoeffizienten als Lösung der Bestimmungsgleichung (\ref{dnm}) für das regionale Blutvolumenverhältnis $\eta=0,\!1$.}}
\end{center}
\end{figure}

\addcontentsline{toc}{section}{3.2 Analytische Lösung der Bloch-Torrey-Gleichung}
Die in Abschnitt 2.4 untersuchte Magnetisierung $m(\mathbf{r},t)$ induziert in einer Spule, die sich auf dem Brustkorb des Patienten befindet, eine Spannung, die letztlich gemessen wird. Allerdings kann nur die gesamte Spannung, d.h. das Signal aus einem Voxel, experimentell ermittelt werden. Messtechnisch zugänglich ist also der Beitrag aller Spins entsprechend
\begin{equation}
S(t) = \int \text{d}^2 \textbf{r} \, m(\textbf{r},t) \,,
\end{equation}
wobei über das gesamte Dephasierungsgebiet zu integrieren ist. Das Signal zum Zeitpunkt $t=0$ ist demnach $S(0)=\pi m_0 [R^2-R_{\text{C}}^2]$. Mit der Darstellung (\ref{Endergebnis}) ergibt sich letztlich 
\begin{align} 
\label{Signalmess}
\frac{S(t)}{S(0)} = \sum_{m=0}^{\infty} \sum_{n=1}^{\infty} d_{nm} \text{e}^{-t \left[ \lambda_{nm}^2 \frac{D}{R_{\text{C}}^2}+ \frac{1}{T_2}\right]}
\end{align}
mit den Koeffizienten $d_{nm} = \eta N_{nm} c_{nm}^{2}/[1-\eta]$, die sich mit Hilfe der Entwicklungskoeffizienten $c_{nm}$ aus Gl. (\ref{cnm}) und der Normierungskonstanten $N_{nm}$ aus Gl. (\ref{Nnm}) in der Form 
\begin{equation}
\label{dnm}
d_{nm} = \frac{2}{\eta-1} \left[2 \eta A_0^{(2m)}\right]^2 \frac{\left[ 2s_{1,k_{m}}^{'} \left( \lambda_{nm} \right) - \pi q_{nm} \frac{\lambda_{nm}}{\sqrt{\eta}} s_{1,k_{m}}^{'} \left( \frac{\lambda_{nm}}{\sqrt{\eta}} \right) \right]^2}{4\eta [\lambda_{nm}^2-k_m^2] - \pi^2 \lambda_{nm}^2 q_{nm}^2 [\lambda_{nm}^2 - \eta k_m^2]}
\end{equation}
schreiben lassen, wobei für $\delta\omega R_{\text{C}}^2/D \leq p_0 \approx 2,\!93754$ die $d_{nm}$ rein reell sind. Analog zu den Eigenschaften der Entwicklungskoeffizienten zur Berechnung der Magnetisierung in Gl. (\ref{Eigenschaftc}) gilt für die Entwicklungskoeffizienten zur Berechnung des Signals:
\begin{equation}
\label{Eigenschaftd}
d_{n\, 2l+1}=d_{n\,2l}^{\displaystyle{*}} \quad \text{für} \quad \delta\omega R_C^2/D > p_l \,.
\end{equation}
In Abb. \ref{Fig:Spektrumdnm} sind die Entwicklungskoeffizienten $d_{nm}$ dargestellt.
Werden nun die Eigenschaften der Entwicklungskoeffizienten (siehe Gl. (\ref{Eigenschaftd})) bzw. der Eigenwerte (siehe Gl. (\ref{Eigenschaftl})) berücksichtigt, so lässt sich schreiben:
\begin{align}
\nonumber
\frac{S(t)}{S(0)}= & 2\sum_{m=0}^{l} \sum_{n=1}^{\infty} \left[ \text{Re}(d_{n\,2m}) \text{cos}\left( \text{Im}(\lambda_{n\,2m}^2)\frac{t}{\tau} \right) + \text{Im}(d_{n\,2m}) \text{sin}\left( \text{Im}(\lambda_{n\,2m}^2)\frac{t}{\tau} \right) \right] \text{e}^{-t \left[ \text{Re}(\lambda_{n\,2m}^2) \frac{D}{R_{\text{C}}^2}+ \frac{1}{T_2}\right]} \\
\label{SignalnachVerzweigung}
+  &\sum_{m=2l+2}^{\infty} \sum_{n=1}^{\infty} d_{nm} \text{e}^{-t \left[ \lambda_{nm}^2 \frac{D}{R_{\text{C}}^2}+ \frac{1}{T_2}\right]} \quad \text{für} \quad p_{l}<\frac{\delta\omega R_C^2}{D}<p_{l+1} \,.
\end{align}

Das Signal zum Zeitpunkt $t=0$ führt zu der Parseval-Relation 
\begin{equation}
\label{Parsevaldnm}
\sum_{m=0}^{\infty} \sum_{n=1}^{\infty} d_{nm} = 1 \,,
\end{equation}
bzw.
\begin{align}
2\sum_{m=0}^{l} \sum_{n=1}^{\infty}\text{Re}(d_{n\,2m}) + \sum_{m=2l+2}^{\infty} \sum_{n=1}^{\infty} d_{nm} = 1 \quad \text{für} \quad p_{l}<\frac{\delta\omega R_C^2}{D}<p_{l+1} \,,
\end{align}
die zur Abschätzung der numerischen Genauigkeit genutzt werden kann. 

\section*{\normalsize{3.3 Vergleich der theoretischen mit den experimentellen Ergebnissen}}
\begin{figure}
\begin{center}
\includegraphics[width=12cm]{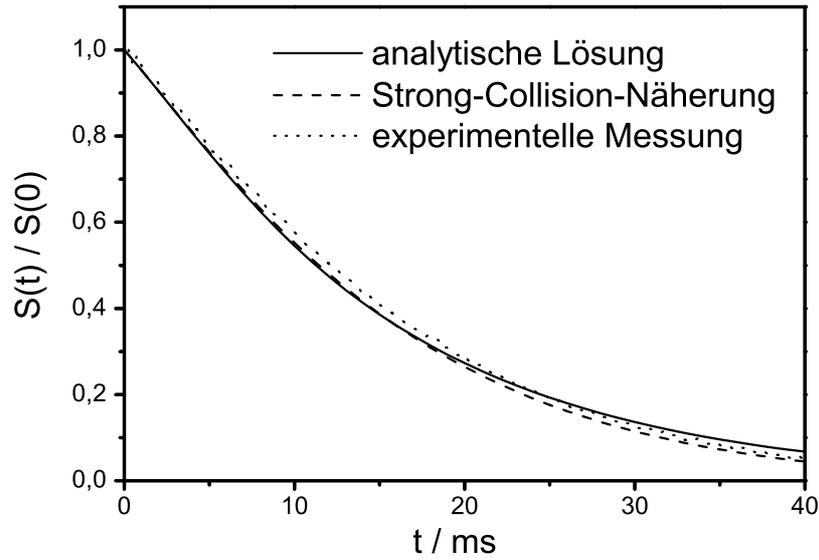}
\vspace{-1.0cm}
\caption[Messung im Skelettmuskel]{\label{fig:exp}{\footnotesize Messung im Skelettmuskel und Vergleich des gemessenen Freien Induktionszerfalls (gepunktete Linie) mit der Strong-Collision-Näherung (gestrichelte Linie nach Gl. (\ref{SignalFDR})) und dem analytischen Ergebnis (durchgezogene Linie nach Gl. (\ref{SignalnachVerzweigung})) für die Parameter $\delta\omega=330\,\mathrm{s}^{-1}$, $R_C=4\,\mu\mathrm{m}$, $\eta=0,\!08$, und $D=1\, \mu\mathrm{m}^2/\mathrm{ms}$. Die intrinsische Spin-Spin-Relaxationszeit ist $T_2=21,\!8\,\mathrm{ms}$.}}
\end{center}
\end{figure}

\addcontentsline{toc}{section}{3.3 Vergleich der theoretischen mit den experimentellen Ergebnissen}
\paragraph{Skelettmuskelgewebe:} Der mit der Abschnitt 2.5 beschriebenen Sequenz gemessene Freie Induktionszerfall ist in der gepunkteten Linie in Abb. \ref{fig:exp} dargestellt. Ein monoexponentieller Fit des gemessenen Signals in der Form $S(t)=S(0)\text{exp}(-t/T_{2}^{*})$ ergibt die Relaxationszeit $T_{2}^{*}=14,\!7\,\mathrm{ms} \pm 0,\!2\,\mathrm{ms}$. Die intrinsische Spin-Spin-Relaxationszeit im Voxel ergab sich durch das exponentielle Anfitten der Amplituden der acht aufgenommenen Echos zu $T_2=21,\!8\, \text{ms} \pm 0,\!8\,\text{ms}$. Dem entsprechend ist der Beitrag der Dephasierung $T_{2}^{'}=45,\!1\,\mathrm{ms} \pm 5,\!0\,\mathrm{ms}$. Die Kapillaren des Skelettmuskels habe einen Durchmesser von $2R_C=8\,\mu\mathrm{m}$ und das regionale Blutvolumenverhältnis beträgt $\eta=0,\!08$ \cite{Krogh70}. Der Diffusionskoeffizient beträgt $D=1\, \mu\mathrm{m}^2/\mathrm{ms}$ und demzufolge beträgt die Korrelationszeit nach Gl. (\ref{eEq19}) $\tau=11\,\mathrm{ms}$. Mit diesen Werten erhalten wir aus Gl. (\ref{T2strich}) die charakteristische Frequenz $\delta\omega=330\,\mathrm{s}^{-1}$. Da $\tau\delta\omega=3,\!63$ und $[1+\eta]/[2\eta]=6,\!75$ ist das Fast-Diffusion-Regime das zu Grunde liegende Diffusionsregime (siehe Gl. (\ref{Bedingungfastdiff})). Die Magnetisierung, die auf Grund der Spindephasierung durch den Suszeptibilitätsunterschied zwischen der blutgefüllten Kapillare und dem umgebenden Gewebe entsteht, wurde in Gl. (\ref{ErgebnisFDR}) angegeben. Zusätzlich ist noch die intrinsische Spin-Spin-Relaxation mit der Relaxationszeit $T_2$ durch den Faktor $\text{e}^{-t/T_2}$ zu berücksichtigen. Das gesamte messbare Signal ergibt sich somit in der Strong-Collision-Näherung zu
\begin{align}
\label{SignalFDR}
\frac{S(t)}{S(0)} = \text{e}^{-t \left[ \frac{1}{\tau} + \frac{1}{T_2} \right] } \left[\frac{\alpha}{\tau\Omega} \mathrm{sinh}(\Omega t) +h(t)\right]\,.
\end{align}
In der gestrichelten Linie in Abb. \ref{fig:exp} ist der Freie Induktionszerfall entsprechend der Strong-Collision-Näherung dargestellt.

Mit den Parametern des Skelettmuskels ($\delta\omega=330\,\mathrm{s}^{-1}$, $R_C=4\,\mu\mathrm{m}$, $D=1\, \mu\mathrm{m}^2/\mathrm{ms}$, $\eta=0,\!08$ und $T_2=21,\!8\,\mathrm{ms}$) ergibt sich für den Parameter $\delta\omega R_C^2/D = 5,\!28$. Wie aus Tabelle \ref{Tab:1} hervorgeht, befindet man sich also zwischen dem erstem Verzweigungspunkt $p_0$ und dem zweiten Verzweigungspunkt $p_1$, also $p_0 < \delta\omega R_C^2/D < p_1$. Der analytisch ermittelte Signal-Zeit-Verlauf entsprechend Gl. (\ref{SignalnachVerzweigung}) ist in der durchgezogenen Linie in Abb. \ref{fig:exp} dargestellt.

\paragraph{Herzmuskelgewebe:} Wie in Abschnitt 2.5 beschrieben, wird das Signal zu verschiedenen Echozeiten aus einem Areal im Septum interventriculare (siehe Abb. \ref{Fig:MRBild}) gemessen. In Tabelle \ref{Tab:2} sind die Echozeiten und die gemessenen Signale dargestellt.
\begin{table*}
\begin{center}
\begin{tabular}{rrrrrr}
$t$/ms \vline &3,3  &8,0   &12,0    &20,0    &30,0  \\ \hline
$S(t)/S(0)$ \vline &0,9226   &0,8109   &0,7093  &0,5570   &0,3962 
\end{tabular}
\caption{\label{Tab:2} \footnotesize Gradientenechozeiten und gemessene Signale. Für einen monoexponentiellen Abfall der Form $S(t)/S(0)=\text{exp}(-t/T_2^{*})$ ergibt sich damit die transversale Relaxationszeit zu $T_2^{*}=32,\!0\,\mathrm{ms}$.}
\end{center}
\end{table*}
Im Myokard haben die Kapillaren einen Radius von $R_C=2,\!75\,\mu\text{m}$ und der Diffusionskoeffizient beträgt $D=1\, \mu\mathrm{m}^2/\mathrm{ms}$. Mit dem regionalen Blutvolumenverhältnis $\eta=0,\!084$ ergibt sich nach Gl. (\ref{eEq19}) die Korrelationszeit im Myokard zu $\tau=5,\!1\,\text{ms}$. Die Dephasierung der Spins wird, wie in Abschnitt 2.2 beschrieben, durch die charakteristische Frequenz $\delta\omega=151\,\text{s}^{-1}$ verursacht. Damit ergibt sich für das Produkt $\tau\delta\omega=0,\!77$. Mit dem regionalen Blutvolumenverhältnis $\eta=0,\!084$, ergibt sich der obere Grenzwert des Fast-Diffusion-Regimes zu $[1+\eta]/[2\eta]=6,\!45$. Entsprechend der Bedingung (\ref{Bedingungfastdiff}) liegt im Myokard also das Fast-Diffusion-Regime zu Grunde und das Signal lässt sich wie im Skelettmuskel in der Strong-Collision-Näherung durch Gl. (\ref{SignalFDR}) beschreiben, wobei die intrinsische Spin-Spin-Relaxationszeit $T_2=57\,\text{ms}$ beträgt (siehe Abschnitt 2.5).

Um die gemessenen Werte mit der analytischen Lösung der Bloch-Torrey-Gleichung zu vergleichen, muss zuerst der Parameter $\delta\omega R_C^2/D$ berechnet werden. Dieser ergibt sich mit obigen Werten zu $\delta\omega R_C^2/D=1,\!14$ und liegt damit vor dem ersten Verzweigungspunkt (siehe Tabelle \ref{Tab:1}).  Demzufolge lässt sich das Signal entsprechend Gl. (\ref{Signalmess}) darstellen.

\begin{figure}
\begin{center}
\includegraphics[width=12.2cm]{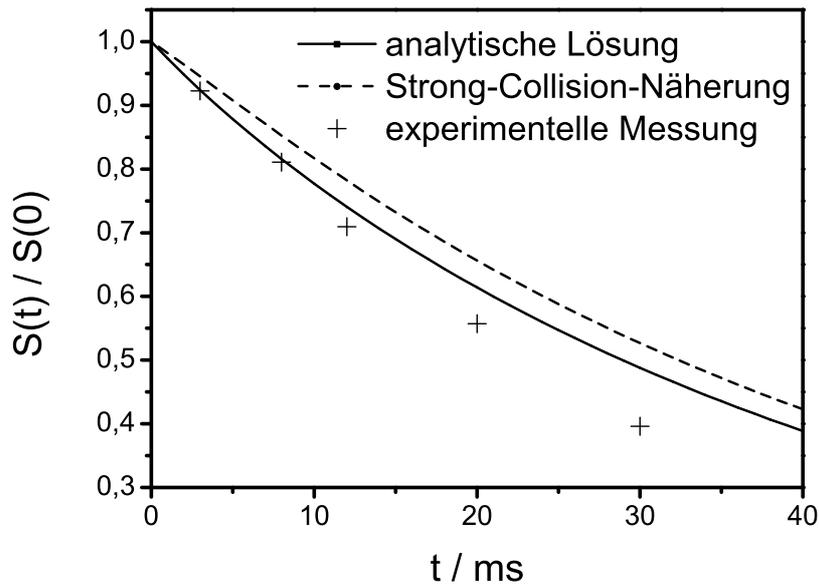}
\vspace{-1.0cm}
\caption[Messung im Herzmuskel]{\label{Fig:Signalexpanalyt}{\footnotesize Messung im Herzmuskel und Vergleich mit analytischer Lösung und Näherung. Die Messwerte wurden, wie in Abschnitt 2.5 beschrieben, aus dem rot markierten Areal in Abb. \ref{Fig:MRBild} gewonnen und sind in Tabelle \ref{Tab:2} zu finden. Die analytische Lösung basierend auf der Lösung der Bloch-Torrey-Gleichung wurde aus Gl. (\ref{Signalmess}) erhalten. Der Signal-Zeit-Verlauf in der Strong-Collision-Näherung folgt aus Gl. (\ref{SignalFDR}).}}
\end{center}
\end{figure}

In Abb. \ref{Fig:Signalexpanalyt} werden die Messwerte mit der analytischen Lösung der Bloch-Torrey-Gleichung und mit der Strong-Collision-Näherung verglichen.
Es ist zu erkennen, dass der experimentell ermittelte Signal-Zeit-Verlauf insbesondere für kleine Echozeiten gut mit der analytischen Näherung übereinstimmt. Für größere Echozeiten liegt die analytische Lösung auch näher an den experimentellen Werten als die Strong-Collision-Näherung. 

Für die klinische Praxis ist die Relaxationszeit $T_2^{*}$ eine wichtige Größe, die sich aber nur für einen exponentiellen Abfall des Signals in der Form $S(t)/S(0)=\text{exp}(-t/T_2^{*})$ angeben lässt. Jedoch kann dem Signal-Zeit-Verlauf nach Gl. (\ref{Signalmess}) entsprechend der Mean-Relaxation-Time-Näherung \cite{Bauer99} die Relaxationszeit $T_2^* = \int_{0}^{\infty} \text{d}t S(t)/S(0)$ zugeordnet werden. Da im Myokard die Entwicklungskoeffizienten $d_{nm}$ rein reell sind, folgt aus Gl. (\ref{Signalmess}) für die Relaxationszeit:
\begin{equation} \label{T2stern}
T_{2}^{*} = \sum_{m=0}^{\infty} \sum_{n=1}^{\infty} \frac{d_{nm}}{\lambda_{nm}^2\frac{D}{R_{\text{C}}^2} +\frac{1}{T_2}} \,.
\end{equation}
Damit ergibt sich mit den typischen Parametern des Myokards eine Relaxationszeit von $T_2^*=42,\!8 \,\text{ms}$. Dies stimmt im Rahmen der Messgenauigkeit mit der durch die Gradientenechosequenz gemessenen Relaxationszeit $T_2^*=32,\!0 \,\text{ms}$ überein (siehe Tabelle \ref{Tab:2}). Den größten Beitrag liefert der Eigenwert $\lambda_{10}$. Berücksichtigt man in Gl. (\ref{T2stern}) nur diesen Eigenwert im Nenner, erhält man mit Hilfe der Parseval-Relation (\ref{Parsevaldnm}) den Ausdruck $1/T_{2}^{*} \approx 1/T_2+\lambda_{10}^2 D/R_{\text{C}}^2$, wobei $\lambda_{10}$ aus Gl. (\ref{l1m}) und $k_0$ aus Gl. (\ref{k_naeherung}) folgt. Damit ergibt sich $T_2^* \approx 44,\!6 \, \text{ms}$. Da im Myokard der Entwicklungskoeffizient $d_{10}$ den größten Beitrag liefert, kann er durch $d_{10} \approx 1$ approximiert werden. Im Myokard beträgt der Radius einer Kapillare etwa $R_{\text{C}}=2,\!75 \, \mu \text{m}$ und damit ergibt sich aus dem regionalen Blutvolumenverhältnis $\eta=0,\!084$ der Radius des Dephasierungszylinders zu $R=9,\!49 \, \mu \text{m}$. Der Diffusionskoeffizienten beträgt etwa $D=1\, \mu\text{m}^2 / \text{ms}$ und der Frequenzshift beträgt etwa $\delta\omega = 151 \, \text{s}^{-1}$. Damit ergibt sich für das Produkt $R_{\text{C}}^2 \delta\omega / D = 1,\!14$. Für diesen Wert verschwindet der Imaginärteil aller Eigenwerte $k_m$, und damit bleibt auch der Index der Besselfunktionen reell.

\chapter*{\label{Kap.Diskussion}\vspace{-3cm} 4 Diskussion}
\addcontentsline{toc}{chapter}{4 Diskussion}
Die Dephasierung der signalgebenden Spins im lokalen inhomogenen Magnetfeld einer Kapillare des Myokards wurde mit verschiedenen Methoden untersucht. Als Gewebemodell für das Myokard diente das Kroghsche Kapillarmodell, das sich auf die Betrachtung einer Kapillare, die von einem Gewebezylinder umgeben wird, beschränkt.

Verschiedene Faktoren beeinflussen das bildgebende MR-Signal, das im Experiment gemessen werden kann. Die Diffusion der Wassermoleküle um die Kapillare wird durch den Diffusionskoeffizienten $D$ beschrieben. Kapillarradius $R_{\text{C}}$ und regionales Blutvolumenverhältnis $\eta$ charakterisieren als Mikrozirkulationsparameter die Architektur des Myokards. 

Die Diffusion der Wassermoleküle, die die Kapillare umgeben, kann im Sinne der Brownschen Molekularbewegung als stochastischer Prozess aufgefasst werden. Mikroskopisch betrachtet bewegt sich ein spintragendes Wassermolekül geradlinig, bis es mit einem anderen Wassermolekül zusammenstößt. Die Strong-Collision-Näherung zur Beschreibung der Dephasierung ersetzt das ursprüngliche Modell des Diffusionsprozesses durch ein einfacheres Modell eines stochastischen Prozesses. Mit Hilfe dieser Näherung konnte in der Arbeit \cite{Bauer99} ein Zusammenhang zwischen den Parametern des Myokards und der Laplace-Transformierten der Magnetisierung $\hat{M}(s)$ hergestellt werden (siehe Gl. (\ref{Mdachallgemein})). In Abschnitt 2.3 wurde die Möglichkeit aufgezeigt, die ursprüngliche Magnetisierung $M(t)$ direkt zu untersuchen. Es stellte sich heraus, dass sich in der Strong-Collision-Näherung die Magnetisierung um eine Kapillare analog zum Schwingungsverhalten eines gedämpften harmonischen Oszillators verhält. Die entscheidende Größe, die das zu Grunde liegende Diffusionsregime bestimmt, ist das Produkt aus Korrelationszeit und Frequenzshift $\tau\delta\omega$. Im Fast-Diffusion-Regime hat die Diffusion einen größeren Einfluss als die Suszeptibilitätseffekte; dies entspricht dem Kriechfall des harmonischen Oszillators und die Magnetisierung fällt monoton, aber nicht monoexponentiell. Das Slow-Diffusion-Regime und das Strong-Dephasing-Regime zeigen analog dem Schwingfall des gedämpften Oszillators ein oszillierendes Verhalten. In diesen Regimen mit Oszillationsverhalten überwiegen die Dephasierungseffekte auf Grund der magnetischen Eigenschaften des Blutes, während die Diffusionseffekte einen geringeren Einfluss haben. Analog zum aperiodischen Grenzfall beim gedämpften Oszillator trennt das Critical-Regime zwischen den überwiegend oszillierenden Signalen und den überwiegend monoton fallenden Signalen.

Allgemein lässt sich die Magnetisierung $M(t)$ als Fourier-Transformierte der Frequenzverteilung $p(\omega)$ ausdrücken:
\begin{align} \label{FT} 
M(t) = \int_{-\infty}^{+\infty} \mathrm{d} \omega \,\, p(\omega) \, \mathrm{e}^{{\mathrm{i} \, \omega \, t}} \,.
\end{align} 
In früheren Arbeiten wurde der Einfluss der Diffusion auf die Frequenzverteilung unter Benutzung der Strong-Collision-Näherung untersucht \cite{Ziener07PRE}. Die Frequenzverteilung um eine Kapillare kann auch durch die Parameter des Myokards (regionales Blutvolumenverhältnis $\eta$, Korrelationszeit $\tau$ und Frequenzshift $\delta\omega$) ausgedrückt werden:
\begin{align}
\nonumber
p(\omega) = \frac{\tau}{\pi} \text{Re} \frac{\sqrt{1 + \left[ \frac{\eta\tau\delta \omega}{1+\mathrm{i}\tau\omega}\right]^2 } - \eta \sqrt{1 + \left[ \frac{\tau\delta \omega}{1+\mathrm{i}\tau\omega}\right]^2}}{[1-\eta][1+\mathrm{i}\tau\omega]-\sqrt{1 + \left[ \frac{\eta\tau\delta \omega}{1+\mathrm{i}\tau\omega}\right]^2 } + \eta \sqrt{1 + \left[ \frac{\tau\delta \omega}{1+\mathrm{i}\tau\omega}\right]^2}} \,.
\end{align}
Wie erwartet stimmt der Magnetisierungs-Zeit-Verlauf, der unter Verwendung der Frequenzverteilung gewonnen wurde, exakt mit dem in dieser Arbeit erhaltenen Signal-Zeit-Verlauf in der Strong-Collision-Näherung überein. Abhängig vom Parameter $\tau\delta\omega$ verändert sich die Form der Frequenzverteilung qualitativ. Für kleine Werte $\tau\delta\omega$ (dies entspricht einem großen Diffusionskoeffizienten) besitzt die Frequenzverteilung nur einen einzigen Peak, während im entgegengesetzten Grenzfall großer Werte $\tau\delta\omega$ zwei separate Peaks entstehen (siehe Abb. 4 in \cite{Ziener07PRE}). Diese zwei Peaks in der Frequenzverteilung führen zu einem Schwebungsverhalten der Magnetisierung, während ein einziger Peak einem monotonen Abfall entspricht. Mit den Ergebnissen dieser Arbeit ist es nun möglich vorherzusagen, ob die Magnetisierung monoton abfällt oder ein oszillierendes Verhalten aufweist. Aus diesem Grunde wurden in dieser Arbeit Wertebereiche für den Parameter $\tau\delta\omega$ angegeben, wodurch die Einteilung in verschiedene Diffusionsregime mit charakteristischen Magnetisierungs-Zeit-Verläufen möglich wurde. Besonders bei Untersuchungen in starken Magnetfeldern ist das oszillierende Verhalten interessant, da die Verwendung einer einfachen Relaxationszeit einen monoexponentiellen Abfall voraussetzt. Demzufolge kann das oszillierende Verhalten zu signifikanten Abweichungen bei der Bestimmung der Relaxationszeit führen.

\chapter*{\label{Kap.Zusammenfassung}\vspace{-3cm} 5 Zusammenfassung}
\addcontentsline{toc}{chapter}{5 Zusammenfassung}
In dieser Arbeit wurde der Zusammenhang zwischen den Parametern der Mikrostruktur des Myokards und der Spindephasierung, die zur Entstehung des MR-Signals aus dem Myokard beiträgt, hergestellt. Zur Beschreibung der Mikrostruktur des Myokards wurde das Kroghsche Kapillarmodell genutzt. In diesem Modell wird das Myokard auf eine einzige Kapillare reduziert, die von einem konzentrischen Gewebszylinder umgeben ist. In dem betrachteten Gewebszylinder findet die Dephasierung der signalgebenden Protonen statt. Mathematisch wird die Dephasierung durch die Bloch-Torrey-Gleichung beschrieben -- eine parabolische Differenzialgleichung zweiter Ordnung. Die Dephasierung konnte auf zwei verschieden Wegen analysiert werden: durch die Strong-Collision-Näherung und durch die analytische Lösung der Bloch-Torrey-Gleichung mittels eines Separationsansatzes. 

Die Strong-Collision-Näherung zur Beschreibung der Dephasierung im Myokard wurde zuerst von Bauer et al. untersucht \cite{Bauer99}, wobei ein Zusammenhang zwischen der Mikrostruktur des Myokards und der Relaxationszeit $T_2^*$ gefunden wurde. Die Angabe solch einer Relaxationszeit setzt aber voraus, dass das Signal aus dem Gewebszylinder exponentiell in der Form $\propto \text{e}^{-t/T_2^*}$ abfällt. Da diese exponentielle Abnahme der Signalstärke jedoch nicht a priori vorausgesetzt werden kann, wurde in dieser Arbeit der genaue Signal-Zeit-Verlauf untersucht. Es stellte sich heraus, dass das Signal aus dem Gewebszylinder nicht rein monoexponentiell abfällt, sondern mit einer hyperbolischen Sinusfunktion moduliert ist. In Analogie zu Kriechfall, Schwingfall und aperiodischen Grenzfall des gedämpften harmonischen Oszillators ist es möglich, beim Magnetisierungszerfall um eine Kapillare verschiedene Diffusionsregime zu unterscheiden. Durch Analyse der inversen Laplace-Transformierten des Magnetisierungs-Zeit-Verlaufs konnten Kriterien für die Einteilung in das jeweilige Diffusionsregime gegeben werden. Auf Grund der typischen Mikrostrukturparameter lässt sich die Dephasierung im Myokard dem Fast-Diffusion-Regime zuordnen.

Im Gegensatz zur Strong-Collision-Näherung, die den ursprünglichen Diffusionsprozess um die Kapillare durch einen anderen stochastischen Prozess ersetzt, ist es allerdings auch möglich, die ursprüngliche Bloch-Torrey-Gleichung exakt zu lösen. In Analogie zur Quantentheorie des Wasserstoffatoms ist es möglich, durch einen Separationsansatz Eigenfunktionen zu finden, die dann zu einer Gesamtlösung kombiniert werden können. Mit dieser Lösung kann beschrieben werden, wie groß die Magnetisierung an einer bestimmten Position im Gewebszylinder zu einer bestimmten Zeit nach dem Anregungspuls ist. Durch Integration über den gesamten Gewebszylinder erhält man eine Aussage über das experimentell messbare Signal.

Zur experimentellen Bestätigung wurde der Signal-Zeit-Verlauf aus dem Myokard aufgenommen und mit den analytischen Ergebnissen der beiden Methoden verglichen. Wie zu erwarten, stimmen die experimentellen Werte besser mit der analytischen Lösung der Bloch-Torrey-Gleichung überein als mit den Ergebnissen der Strong-Collision-Näherung. 

Im Gegensatz zu bisherigen Untersuchungen kann nicht nur das gesamte MR-Signal eines Gradienten-Echos analysiert werden, sondern die ortsaufgelöste Magnetisierung die um eine Kapillare innerhalb des Versorgungszylinders herrscht. Dies ermöglicht es, auch andere Sequenzen, wie z.B. die Spin-Echo-Sequenz zu analysieren. 

Das in Abschnitt 2.1 beschriebene regionale Blutvolumenverhältnis ist durch bereits etablierte Messverfahren bestimmbar. Mit den in dieser Arbeit vorgestellten Methoden ist nun auch der Zusammenhang zwischen Kapillarradius und gemessener Relaxationszeit bzw. Freien Induktionszerfall bekannt. Dies ermöglicht es, die Änderung der Kapillardichte im Myokard zu ermitteln und daraus auf den Grad der Mikrozirkulationsversorgung zu schließen. So kann beispielsweise quantifiziert werden, wie sich die Mikrozirkulation in physiologischen und pathologischen Hypertrophiemodellen des Herzens unterscheiden.

\listoffigures

\addcontentsline{toc}{chapter}{Abbildungsverzeichnis}

\thispagestyle{plain}

\chapter*{Publikationsliste}
\addcontentsline{toc}{chapter}{Publikationsliste}
\thispagestyle{empty}
\pagestyle{empty}

\section*{Veröffentlichungen als Erstautor}
\begin{enumerate}
\item {\bf C. H. Ziener}, S. Glutsch, and F. Bechstedt. {\it RKKY interaction in semiconductors:   Effects of magnetic field and screening}, Phys. Rev. B 70, 075205 (2004).
\item {\bf C. H. Ziener}, W. R. Bauer, and P. M. Jakob. {\it Transverse Relaxation of Cells Labeled with Magnetic Nanoparticles}, Magn. Reson. Med. 54, 702-706 (2005).
\item {\bf C. H. Ziener}, W. R. Bauer, and P. M. Jakob. {\it Frequency distribution and signal formation around a vessel}, Magn. Reson. Mater. Phy. 18, 225-230 (2005).
\item {\bf C. H. Ziener}, W. R. Bauer, G. Melkus, T. Weber, V. Herold, P. M. Jakob. {\it Structure-specific magnetic field inhomogeneities and its effect on the correlation time}, Magn. Reson. Imaging 24, 1341-1347 (2006).
\item {\bf C. H. Ziener}, T. Kampf, G. Melkus, P. M. Jakob, W. R. Bauer. {\it Scaling Laws for Transverse Relaxation Times}, J. Magn. Reson. 184, 169-175 (2007).
\item {\bf C. H. Ziener}, T. Kampf, W. R. Bauer, P. M. Jakob, S. Glutsch, F. Bechstedt. {\it Quantum Beats in Semiconductors}, International Journal of Modern Physics B 21, Nos. 8-9, 1621-1625 (2007).
\item{\bf C. H. Ziener}, T. Kampf, G. Melkus, V. Herold, T. Weber, G. Reents, P. M. Jakob, W. R. Bauer. {\it Local frequency density of states around field inhomogeneities in magnetic resonance imaging: Effects of diffusion}, Phys. Rev. E 76, 031915 (2007).
\item {\bf C. H. Ziener}, T. Kampf, V. Herold, P. M. Jakob, W. R. Bauer, W. Nadler. {\it Frequency autocorrelation function of stochastically fluctuating fields caused by specific magnetic field inhomogeneities}, J. Chem. Phys. 129, 014507 (2008).
\item {\bf C. H. Ziener}, S. Glutsch, P. M. Jakob, W. R. Bauer. {\it Spin dephasing in the dipole field around capillaries and cells: Numerical solution}, Phys. Rev. E 80, 046701 (2009).
\item {\bf C. H. Ziener}, T. Kampf, P. M. Jakob, W. R. Bauer. {\it Diffusion effects on the CPMG relaxation rate in a dipolar field}, J. Magn. Reson. 202, 38-42 (2010).
\end{enumerate}

\vspace{0.5cm}

\section*{Veröffentlichungen als Zweitautor}
\begin{enumerate}
\setcounter{enumi}{10}
\item W. R. Bauer, {\bf C. H. Ziener}, and P. M. Jakob. {\it Non-Gaussian spin dephasing}, Phys. Rev. A 71, 053412 (2005).
\item T. Weber, {\bf C. H. Ziener}, T. Kampf, V. Herold, W. R. Bauer, P. M. Jakob. {\it Measurement of Apparent Cell Radii Using a Multiple Wave Vector Diffusion Experiment}, Magn. Reson. Med. 61, 1001-1006 (2009).
\end{enumerate}

\vspace{0.5cm}

\section*{Sonstige Veröffentlichungen}
\begin{enumerate}
\setcounter{enumi}{12}
\item G. Klug, T. Kampf, {\bf C. H. Ziener}, M. Parczyk, E. Bauer, V. Herold, E. Rommel, P. M. Jakob, W. R. Bauer. {\it Murine atherosclerotic plaque imaging with the USPIO Ferumoxtran-10}, Frontiers in Biosci. 14, 2546-2552 (2009).
\item V. Herold, J. Wellen, {\bf C. H. Ziener}, T. Weber, K.-H. Hiller, P. Nordbeck, E. Rommel, A. Haase, W. R. Bauer, P. M. Jakob, S. K. Sarkar. {\it In vivo comparison of atherosclerotic plaque progression with vessel wall strain and blood flow velocity in apoE-/- mice with MR microscopy at 17.6 T}, Magn. Reson. Mater. Phy. 22, 159-166 (2009).
\item V. Herold, M. Parczyk, P. Mörchel, {\bf C. H. Ziener}, G. Klug, W. R. Bauer, E. Rommel, P. M. Jakob. {\it In vivo Measurement of Local Aortic Pulse-Wave Velocity in Mice with MR Microscopy at 17.6 T}, Magn. Reson. Med. 61, 1293-1299 (2009).
\item G. Klug, T. Kampf, S. Bloemer, J. Bremicker, {\bf C. H. Ziener}, A. Heymer, U. Gbureck, E. Rommel, U. Nöth, W. A. Schenk, P. M. Jakob, W. R. Bauer. {\it Intracellular and Extracellular T1 and T2 Relaxivities of Magneto-Optical Nanoparticles at Experimental High-Fields}, Magn. Reson. Med. 64, 1607-1615 (2010).
\end{enumerate}

\newpage

\vspace{0.5cm}

\section*{Diplomarbeit in Physik}
{\bf C. H. Ziener}. {\it Spinquantenschwebungen in semimagnetischen Halbleitern}, Jena (2003).

\vspace{0.5cm}

\section*{Doktorarbeit in Physik}
{\bf C. H. Ziener}. {\it Suszeptibilitätseffekte in der Kernspinresonanzbildgebung}, Würzburg (2009).

\section*{Wissenschaftspreise}
\begin{enumerate}
\item {\bf C. H. Ziener}. {\it Frequency distribution in a vascular network}, Young Investigator Award der ESMRMB, zweiter Preis, Basel (2005). \\
\item {\bf C. H. Ziener}. {\it Suszeptibilitätseffekte in der Kernspinresonanzbildgebung}, Gorter- Award der Deutschen Sektion der ISMRM, erster Preis, Frankfurt (2008). \\
\item {\bf C. H. Ziener}. Wilhelm-Conrad-Röntgen-Wissenschaftspreis der Fakultät für Physik und Astronomie der Universität Würzburg, Würzburg (2009).
\end{enumerate}


\section*{Vorträge}
\begin{enumerate}
\item {\bf C. H. Ziener}, W. R. Bauer, and P. M. Jakob. {\it Frequency distribution in a vascular network}, ESMRMB, Vortrag 148 (Basel 2005).
\item {\bf C. H. Ziener}, W. R. Bauer, P. M. Jakob. {\it Skalierungsgesetze für transversale Relaxationszeiten}, 8. Jahrestreffen der Deutschen Sektion der ISMRM (Münster 2005).
\item {\bf C. H. Ziener}, T.Kampf, G. Melkus, W. R. Bauer, and P. M. Jakob. {\it SSFP signal analysis}, ESMRMB, Vortrag 42 (Warschau 2006).
\item {\bf C. H. Ziener}, T.Kampf, G. Melkus, W. R. Bauer, P. M. Jakob. {\it Diffusionsabhängige Frequenzverteilungen}, 9. Jahrestreffen der Deutschen Sektion der ISMRM (Jena 2006).
\item {\bf C. H. Ziener}, T. Weber, W. R. Bauer, and P. M. Jakob. {\it Quantification of the Spinal Cord Axon Diameter using an Extension of the PGSE Sequence}, Proc. Int. Soc. Magn. Reson. Med. 2007:2290 (Berlin 2007).
\item {\bf C. H. Ziener}, T. Kampf, W. R. Bauer, and P. M. Jakob. {\it Magnetic resonance imaging of magnetically labelled cells}, Fellows Meeting 2007 der Ernst-Schering-Foundation (Berlin 2007).
\item {\bf C. H. Ziener}. {\it Suszeptibilitätseffekte in der Kernspinresonanzbildgebung}, 11. Jahrestreffen der Deutschen Sektion der ISMRM, Vortrag G2 (Frankfurt 2008).
\item {\bf C. H. Ziener}, T.Kampf, P. M. Jakob, W. R. Bauer. {\it Diffusionseffekte im Dipolfeld bei CPMG-Sequenzen}, 12. Jahrestreffen der Deutschen Sektion der ISMRM (Basel 2009).
\item {\bf C. H. Ziener}, P. M. Jakob, W. R. Bauer. {\it Dephasierung im Myokard}, 6. Herbsttagung der Medizinischen Klinik I (Bad Brückenau 2009).
\end{enumerate}

\section*{Eingeladene Vorträge}
\begin{enumerate}
\item {\bf C. H. Ziener}. {\it From Microscopic Field Inhomogeneities to a Macroscopic MR-Signal}, Bayer Schering Pharma Symposium \glqq Keeping Track of Innovation\grqq\  anlässlich des Joint Annual Meeting ISMRM-ESMRMB (Berlin 2007).
\item {\bf C. H. Ziener}. {\it Medizin und Physik - Synergien in Forschung und Lehre}, Assistententag anlässlich des zweiten Deutschen Internistentages (Berlin 2009).
\end{enumerate}

\section*{Poster}
\begin{enumerate}
\item {\bf C. H. Ziener}, W. R. Bauer, P. M. Jakob. {\it Relaxationsverhalten magnetisch markierter Zellen}, 7. Jahrestreffen der Deutschen Sektion der ISMRM (Mainz 2004).
\item {\bf C. H. Ziener}, W. R. Bauer, and P. M. Jakob. {\it Transverse Relaxation of Cells Labeled with Magnetic Nanoparticles}, Proc. Int. Soc. Magn. Reson. Med. 2005:2611 (Miami 2005).
\item {\bf C. H. Ziener}, T. Kampf, G. Melkus, R. Kharrazian, M. Choli, W. R. Bauer, C. Faber, P. M. Jakob. {\it SSFP Signal Formed by a Lorentzian Frequency Distribution}, International Symposium on Biomedical Magnetic Resonance Imaging and Spectroscopy at Very High Fields, Poster 14 (Würzburg 2006).
\item {\bf C. H. Ziener}, T. Kampf, S. Glutsch, W. R. Bauer, P. M. Jakob, F. Bechstedt. {\it Quantum Beates in Magnetic Semiconductors}, 17th International Conference on High Magnetic Fields in Semiconductor Physics (HMF), Poster HMF$\_5\_5$ (Würzburg 2006).
\item T. Kampf, {\bf C. H. Ziener}, G. Melkus, A. Purea, M. Parczyk, W. R. Bauer, P. M. Jakob. {\it USPIO-Modelle im Vergleich}, 9. Jahrestreffen der Deutschen Sektion der ISMRM (Jena 2006).
\item T. Kampf, {\bf C. H. Ziener}, P. M. Jakob, W. R. Bauer. {\it Dependence of the frequency distribution on the orientation of the voxel}, Molekulare Bildgebung 07, Poster 1 (Kiel 2007).
\item {\bf C. H. Ziener}, T. Kampf, W. R. Bauer, P. M. Jakob. {\it Diffusion Dependent Frequency Distribution}, 9th International Conference on Magnetic Resonance Microscopy, Poster 42 (Aachen 2007).
\item T. Kampf, {\bf C. H. Ziener}, X. Helluy, P. M. Jakob, W. R. Bauer. {\it Computation of inter and intra voxel diffusion using MC-simulations in frequency and spatial domain: a comparison}, 9th International Conference on Magnetic Resonance Microscopy, Poster 34 (Aachen 2007).
\item {\bf C. Ziener}, V. Herold, G. Klug, M. Parczyk, E. Rommel, P. Jakob, W. R. Bauer. {\it Nichtinvasive in vivo Messung der regionalen Pulswellengeschwindigkeit mittels hochauflösender MRI}, 74. Jahrestagung der
Deutschen Gesellschaft für Kardiologie - Herz - und Kreislaufforschung e.V., Poster 1491 (Mannheim 2008).
\item V. Herold, M. Parczyk, {\bf C. Ziener}, G. Klug, E. Rommel, W. R. Bauer, P. Jakob {\it In vivo Magnetresonanzbildgebung zur Messung der lokalen Pulswellengeschwindigkeit an der Maus bei 17,6 Tesla}, 74. Jahrestagung der Deutschen Gesellschaft für Kardiologie - Herz - und Kreislaufforschung e.V., Poster 841 (Mannheim 2008).
\item {\bf C. H. Ziener}, V. Herold, M. Parczyk, G. Klug, T. Kampf, E. Rommel, P. Jakob, W. Bauer. {\it In-vivo-Messung der regionalen und lokalen Pulswellengeschwindigkeit in der Aorta der Maus mittels MR-Bildgebung bei 17,6 Tesla}, 114. Kongress der Deutschen Gesellschaft für Innere Medizin, Poster 245 (Wiesbaden 2008).
\item T. C. Basse-Luesebrink, T. Kampf, {\bf C. H. Ziener}, G. Klug, W. R. Bauer, P. M. Jakob, and D. Haddad. {\it Evaluation of sensitivity increase by T1 and T2 contrast agents in 19F MRI of PF15C}, Proc. Int. Soc. Magn. Reson. Med. 2008:1655 (Toronto 2008).
\item V. Herold, G. Klug, M. Parczyk, {\bf C. Ziener}, T. Weber, S. Sarkar, W. R. Bauer, E. Rommel, and P. M. Jakob. {\it In vivo measurement of local pulse-wave velocity in mice with MRI at 17.6 T}, Proc. Int. Soc. Magn. Reson. Med. 2008:907 (Toronto 2008).
\item {\bf C. H. Ziener}, T. Kampf, V. Herold, P. M. Jakob, W. R. Bauer and W. Nadler. {\it Temporal correlation function around spheres and cylinders}, 9th International Bologna Conference Magnetic Resonance in Porous Media (MRPM9), Poster 101 (Cambridge MA, USA 2008).
\item T. Kampf, {\bf C. H. Ziener}, P. M. Jakob, and W. R. Bauer. {\it Theoretical considerations on the quantification of iron oxide labeled cells in vivo}, Proc. Int. Soc. Magn. Reson. Med. 2009:3156 (Honolulu 2009).
\item T. Kampf, {\bf C. H. Ziener}, P. M. Jakob, and W. R. Bauer. {\it Iron Oxide Labeled Cell Quantification in vivo - A Theoretical Study}, Molekulare Bildgebung 09, Poster 16 (Berlin 2009).
\item {\bf C. H. Ziener}, T. Kampf, P. M. Jakob, W. R. Bauer. {\it Freier Induktionszerfall im inhomogenen Magnetfeld einer Kapillare - Analogie zum gedämpften Oszillator}, 12. Jahrestreffen der Deutschen Sektion der ISMRM (Basel 2009).
\item T. Kampf, {\bf C. H. Ziener}, P. M. Jakob, and W. R. Bauer. {\it The effect of agglomeration on the efficiency of contrast agents in a two compartment model}, 12. Jahrestreffen der Deutschen Sektion der ISMRM (Basel 2009).
\item V. Herold, M. Parczyk, T. Kampf, {\bf C. Ziener}, A. Gotschy, E. Rommel, W. R. Bauer, P. Jakob. {\it In-vivio-Messung des arteriellen Pulsdrucks der Maus mit MR-Methoden bei 17,6 Tesla}, 76. Jahrestagung der Deutschen Gesellschaft für Kardiologie - Herz - und Kreislaufforschung e.V., Poster 1430 (Mannheim 2010).
\item T. Kampf, {\bf C. H. Ziener}, P. M. Jakob, and W. R. Bauer. {\it On the effect of contrast agent internalization in a two compartment diffusion model}, Proc. Int. Soc. Magn. Reson. Med. 2010:4219 (Stockholm 2010).
\item V. Herold, {\bf C. Ziener}, E. Rommel, W. R. Bauer, P. Jakob. {\it 	In vivo measurement of 3D blood flow patterns in the murine aorta using four-dimensional magnetic resonance velocity mapping}, 77. Jahrestagung der Deutschen Gesellschaft für Kardiologie - Herz - und Kreislaufforschung e.V., Poster 702 (Mannheim 2011).
\item V. Herold, A. Gotschy, {\bf C. Ziener}, E. Rommel, W. R. Bauer, P. Jakob. {\it In vivo measurement of local pulse-wave velocity in the right common carotid artery in mice with PC-Cine-MRI at 17.6 T}, 77. Jahrestagung der Deutschen Gesellschaft für Kardiologie - Herz - und Kreislaufforschung e.V., Poster 1319 (Mannheim 2011).
\item {\bf C. Ziener}, T. Kampf, V. Herold, P. Jakob, W. R. Bauer. {\it Transversale Relaxation im Kroghschen Kapillarmodell des Myokards}, 77. Jahrestagung der Deutschen Gesellschaft für Kardiologie - Herz - und Kreislaufforschung e.V., Poster 1384 (Mannheim 2011).
\item V. Herold, A. Gotschy, {\bf C. H. Ziener}, E. Rommel, W. R. Bauer, P. M. Jakob. {\it In vivo measurement of local pulse-wave velocity in the right common carotid artery in mice with PC-Cine-MRI at 17.6 T}, Proc. Int. Soc. Magn. Reson. Med. 2011:1216 (Montreal 2011).
\end{enumerate}

\chapter*{Danksagung}
\addcontentsline{toc}{chapter}{Danksagung}
\thispagestyle{empty}
\pagestyle{empty}
Ich danke allen, die an der Entstehung dieser Arbeit mitgewirkt haben, insbesondere meinem Doktorvater Prof. Dr. Dr. Wolfgang Bauer und meinem Betreuer Prof. Dr. Peter Jakob, für die stetige Motivation und Hilfe. Den Kollegen vom Lehrstuhl für Experimentelle Physik 5 bin ich dankbar für eine angenehme Arbeitsatmosphäre und hilfreiche Diskussionen. Für viele Hinweise und Ratschläge bezüglich mathematischer Fragestellungen möchte ich Prof. Dr. Georg Reents vom Institut für Theoretische Physik und Astrophysik der Universität Würzburg danken.

Dem Berufsverband Deutscher Internisten, insbesondere Herrn Dr. Wolfgang Wesiack danke ich für die Aufnahme in das studentische Förderprogramm.

Der Arbeitsgruppe Medizinische Physik des Instituts für Diagnostische und Interventionelle Radiologie der Friedrich-Schiller-Universität Jena, insbesondere Prof. Dr. Jürgen Reichenbach, möchte ich meinen Dank aussprechen für die Zusammenarbeit und die wertvollen Diskussionen bezüglich der Experimente zur Bestätigung der analytischen Betrachtungen.

Da diese Arbeit im Rahmen des Sonderforschungsbereiches 688 "`Mechanismen und Bildgebung von Zell-Zell-Wechselwirkungen im kardiovaskulären System"' der Universität Würzburg entstand, möchte ich mich bei all seinen Mitarbeitern, insbesondere den Leitern Prof. Dr. Walter und Prof. Dr. Ertl bedanken.

Ganz besonderer Dank gilt meinen Eltern, Großeltern und meiner Freundin für die Unterstützung beim Anfertigen dieser Arbeit.

\chapter*{Lebenslauf}
\addcontentsline{toc}{chapter}{Lebenslauf}
\thispagestyle{empty}
\begin{minipage}{1\textwidth}
\vspace{0cm}
{\normalsize{\bf Persönliche Daten}}\vspace{0cm}\\
\begin{tabular}[t]{p{0.25\textwidth}p{0.7\textwidth}}
Vor- und Zuname & \underline{Christian} Herbert Ziener\\
Geburtsdatum &18.12.1978\\
Geburtsort &Weimar\\
Staatsangehörigkeit &deutsch\\
Familienstand &ledig\\
Adresse &dienstlich:\hspace{5.0cm}privat:\\
        &Deutsches Krebsforschungszentrum\\
        &Abteilung Radiologie\\
        &E010\\
        &Im Neuenheimer Feld 280\hspace{2.15cm}Zur Forstquelle 4\\
        &69120 Heidelberg\hspace{3.60cm}69126 Heidelberg\\
Telefon &06221-42\,2564\hspace{4.10cm}06221-8891798\\
E-mail &{\tt c.ziener@dkfz-heidelberg.de}\\
\end{tabular}
\\[4ex]
{\normalsize{\bf Werdegang}}\vspace{0cm}\\
\begin{tabular}[t]{p{0.25\textwidth}p{0.6\textwidth}}
09/1985 - 08/1991 & Polytechnische Oberschule \glqq Friedrich Leßner\grqq\  in Blankenhain\\
09/1991 - 08/1993 & Geschwister-Scholl-Gymnasium in Bad Berka\\
09/1993 - 06/1997 & Carl-Zeiss-Gymnasium in Jena, Spezialschule mathematisch-naturwis\-senschaftlich-technischer Richtung, Abschluss: Abitur, Durchschnittsnote: 1,1; Internat der Spezialschule in Jena\\
07/1997 - 04/1998 & Wehrdienst in Eschweiler und Gotha\\
\end{tabular}
\end{minipage}

\vspace{1.0cm}
\begin{minipage}{1\textwidth}
\begin{tabular}[t]{p{0.25\textwidth}p{0.6\textwidth}}
10/1998 - 09/2003 & Studium der Physik an der Friedrich-Schiller-Universität in Jena, Diplomarbeit am Institut für Festkörpertheorie und Theoretische Optik. Abschluss: Diplomphysiker, Durchschnittsnote: 1,0 (mit Auszeichnung), Nebenfach: Funktionalanalysis\\
03/2004 - 04/2009 & Promotionsstudium am Lehrstuhl für Experimentelle Physik 5 der Universität Würzburg. Abschluss: Dr. rer. nat., Durchschnittsnote: 1,00 (mit Auszeichnung)\\
10/2004 - 11/2010 & Studium der Humanmedizin an der Julius-Maximilians-Universität Würzburg. Abschluss: Staatsexamen, Durchschnittsnote: 1,50 (sehr gut); Approbation als Arzt am 30. 11. 2010\\
01/2005 - 12/2006 & Stipendiat der Schering-Stiftung\\
04/2007 - 11/2010 & Stipendiat des Berufsverbandes Deutscher Internisten\\
seit 01/2010 & Projektleiter des Teilprojekts B5 des Sonderforschungsbereichs 688 der Julius-Maximilians-Universität Würzburg\\
\end{tabular}

\vspace{1.0cm}
{\normalsize{\bf Berufstätigkeit}}\vspace{0cm}\\
\begin{tabular}[t]{p{0.25\textwidth}p{0.6\textwidth}}
10/2002 - 02/2003 & Übungsassistent der Vorlesung Festkörperphysik an der Universität Jena\\
03/2004 - 12/2004 & Wissenschaftlicher Mitarbeiter am Physikalischen Institut der Universität Würzburg\\
01/2007 - 12/2009 & Wissenschaftlicher Mitarbeiter im Teilprojekt B5 des Sonderforschungsbereichs 688 der Universität Würzburg\\
01/2010 - 05/2011 & Wissenschaftlicher Mitarbeiter des Integrierten Forschungs- und Behandlungszentrums Herzinsuffizienz (IFB, CHFC) der Universität Würzburg\\
seit 01/2011 & Wissenschaftlicher Mitarbeiter der Abteilung Radiologie des Deutschen Krebsforschungszentrums in Heidelberg\\
\end{tabular}
\end{minipage}

\begin{minipage}{1\textwidth}

{\normalsize{\bf Mitgliedschaften in Fachgesellschaften}}\vspace{0cm}\\
\begin{tabular}[t]{p{0.99\textwidth}p{0.1\textwidth}}
International Society for Magnetic Resonance in Medicine (ISMRM) & \\
European Society for Magnetic Resonance in Medicine and Biology (ESMRMB) & \\
Deutsche Sektion der ISMRM (DSISMRM) & \\
Berufsverband Deutscher Internisten (BDI) & \\
Hartmannbund &\\
\end{tabular}

\vspace{1.0cm}
{\normalsize{\bf Gutachter für wissenschaftliche Zeitschriften}}\vspace{0cm}\\
\begin{tabular}[t]{p{0.8\textwidth}p{0.1\textwidth}}
Magnetic Resonance in Medicine & \\
Journal of Magnetic Resonance & \\
Journal of Chemical Physics & \\
Journal of Magnetic Resonance Imaging & \\
NMR in Biomedicine & \\
Medical \& Biological Engineering \& Computing & \\
\end{tabular}

\vspace{3cm}

Würzburg, 21. Juni 2011\\
\vspace{2cm}

Christian H. Ziener
\end{minipage}


\begin{thebibliography}{200}
\addcontentsline{toc}{chapter}{Literaturverzeichnis}

\bibitem{Hombach2005}
V. Hombach, O. Grebe, R.M. Botnar. {\it Kardiovaskuläre Magnetresonanztomographie.} Schattauer-Verlag, Stuttgart, 2005.

\bibitem{Koehler03}
S. Köhler, K.-H. Hiller, M. Griswold, W.R. Bauer, A. Haase, P.M. Jakob. NMR-microscopy with TrueFISP at 11.75 T. \JMR{2003}{161}{252-257}

\bibitem{Streeter69}
D.D. Jr. Streeter, H.M. Spotnitz, D.P. Patel, J. Jr. Ross, E.H. Sonnenblick. Fiber orientation in the canine left ventricle during diastole and systole. \Circ{1969}{24}{339-347}

\bibitem{Ogawa90}
S. Ogawa, T.M. Lee, A.R. Kay, D.W. Tank. Brain magnetic resonance imaging with contrast dependent on blood oxygenation. \PNAS{1990}{87}{9868-9872}

\bibitem{Wacker03}
C.M. Wacker, A.W. Hartlep, S. Pfleger, L.R. Schad, G. Ertl, W.R. Bauer. Susceptibility-sensitive magnetic resonance imaging detects human myocardium supplied by a stenotic coronary artery without a contrast agent. \JACC{2003}{41}{834-840}

\bibitem{Bauer99}
W.R. Bauer, W. Nadler, M. Bock, L.R. Schad, C. Wacker, A. Hartlep, G. Ertl. Theory of the BOLD effect in the capillary region: an analytical approach for the determination of T2 in the capillary network of myocardium. \MRM{1999}{41}{51-62}

\bibitem{Ziener07PRE}
C.H. Ziener, T. Kampf, G. Melkus, V. Herold, T. Weber, G. Reents, P.M. Jakob, W.R. Bauer. Local frequency density of states around field inhomogeneities in magnetic resonance imaging: effects of diffusion. \PRE{2007}{76}{031915}

\bibitem{Krogh19a}
A. Krogh. The number and the distribution of capillaries in muscle with the calculation of the oxygen pressure necessary for supplying the tissue. \Jphysiol{1919}{52}{409-415}

\bibitem{Krogh19b}
A. Krogh. The supply of oxygen to the tissues and the regulation of the capillary circulation. \Jphysiol{1919}{52}{457-474}

\bibitem{Krogh70}
A. Krogh. {\it Anatomie und Physiologie der Capillaren.} Springer-Verlag, Berlin 1970.

\bibitem{Kennan94}
R.P. Kennan, J. Zhong, J.C. Gore. Intravascular susceptibility contrast mechanisms in tissues. \MRM{1994}{31}{9-21}

\bibitem{Yablonskiy94}
D.A. Yablonskiy, E.M. Haacke. Theory of NMR signal behavior in magnetically inhomogeneous tissues: the static dephasing regime. \MRM{1994}{32}{749-763}

\bibitem{Schmidt05}
R.F. Schmidt, F. Lang, G. Thews. {\it Physiologie des Menschen.} Springer-Verlag, Heidelberg, 2005.

\bibitem{Pauling36}
L. Pauling, C.D. Coryell. The Magnetic Properties and Structure of Hemoglobin, Oxyhemoglobin and Carbonmonoxyhemoglobin. \PNAS{1936}{22}{210-216}

\bibitem{Reichenbach01}
J.R. Reichenbach, E.M. Haacke. High-resolution BOLD venographic imaging: a window into brain function. \NMRBio{2001}{14}{453-467}

\bibitem{Ziener08JCP}
C.H. Ziener, T. Kampf, V. Herold, P.M. Jakob, W.R. Bauer, W. Nadler. Frequency autocorrelation function of stochastically fluctuating fields caused by specific magnetic field inhomogeneities. \JCP{2008}{129}{014507}

\bibitem{Torrey56}
H.C. Torrey. Bloch Equations with Diffusion Terms. \PR{1956}{104}{563-565}

\bibitem{Thews53}
G. Thews. Über die mathematische Behandlung physiologischer Diffusionsprozesse in zylinderförmigen Objekten. Acta biotheor. (Leiden) 1953;10:105-138.

\bibitem{Haacke99}
E.M. Haacke, R.W. Brown, M.R. Thompson, R. Venkatesan. {\it Magnetic Resonance Imaging: Physical Principles and Sequence Design.} John Wiley, New York, 1999.

\bibitem{Stoller91}
S.D. Stoller, W. Happer, F.J. Dyson. Transverse spin relaxation in inhomogeneous magnetic fields. \PRA{1991}{44}{7459-7477}

\bibitem{Kiselev98PRL}
V.G. Kiselev, S. Posse. Analytical Theory of Susceptibility Induced NMR Signal Dephasing
in a Cerebrovascular Network. \PRL{1998}{81}{5696-5699}

\bibitem{Kiselev99MRM}
V.G. Kiselev, S. Posse. Analytical Model of Susceptibility-Induced MR Signal Dephasing: Effect of Diffusion in a Microvascular Network. \MRM{1999}{41}{499-509}

\bibitem{Sukstanskii03}
A.L. Sukstanskii, D.A. Yablonskiy. Gaussian approximation in the theory of MR signal formation
in the presence of structure-specific magnetic field inhomogeneities. \JMR{2003}{163}{236-247}

\bibitem{Sukstanskii04}
A.L. Sukstanskii, D.A. Yablonskiy. Gaussian approximation in the theory of MR signal formation in the presence of structure-specific magnetic field inhomogeneities. Effects of impermeable susceptibility inclusions. \JMR{2004}{167}{56-67}

\bibitem{Jensen2000a}
J.H. Jensen, R. Chandra. Strong field behavior of the NMR signal from magnetically heterogeneous tissues. \MRM{2000}{43}{226-236}

\bibitem{Jensen2000b}
J.H. Jensen, R. Chandra. NMR relaxation in tissues with weak magnetic inhomogeneities. \MRM{2000}{44}{144-156}

\bibitem{Bauer99T2}
W.R. Bauer, W. Nadler, M. Bock, L.R. Schad, C. Wacker, A. Hartlep, G. Ertl. The relationship between T2* and T2 in myocardium. \MRM{1999}{41}{1004-1010}

\bibitem{Bauer02}
W.R. Bauer, W. Nadler. Spin dephasing in the strong collision approximation. \PRE{2002}{65}{066123}

\bibitem{Bauer99PRL}
W.R. Bauer, W. Nadler, M. Bock, L.R. Schad, C. Wacker, A. Hartlep, G. Ertl. Theory of coherent and incoherent nuclear spin dephasing in the heart. \PRL{1999}{83}{4215-4218}

\bibitem{Ziener06MRI}
C.H. Ziener, W.R. Bauer, G. Melkus, T. Weber, V. Herold, P.M. Jakob. Structure-specific magnetic field inhomogeneities and its effect on the correlation time. \MRI{2006}{24}{1341-1347}

\bibitem{Oberhettinger72}
F. Oberhettinger. {\it Tables of Bessel Transforms.} Springer-Verlag, Berlin, Heidelberg, New York, 1972.

\bibitem{Spiegel77}
M.R. Spiegel. {\it Laplace-Transformationen.} McGraw-Hill, London, 1977.

\bibitem{Spiegel76}
M.R. Spiegel. {\it Komplexe Variablen.} McGraw-Hill, London, 1976.

\bibitem{Meixner54}
J. Meixner, F.W. Schäfke. {\it Mathieusche Funktionen und Sphäroidfunktionen.} Springer, Berlin, 1954.

\bibitem{McLachlan64}
N.W. McLachlan. {\it Theory and application of Mathieu functions.} Dover, New York, 1964.

\bibitem{Strutt}
M.J.O. Strutt. {\it Lamésche- Mathieusche- und verwandte Funktionen in Physik und Technik.} Springer, Berlin, 1932.

\bibitem{Campbell}
R. Campbell. {\it Théorie générale de l'équation de Mathieu et de quelques autres équations différentielles de la mécanique.} Masson, Paris, 1955. 

\bibitem{Arscott}
F.M. Arscott. {\it Periodic differential equations.} Pergamon Press, Oxford, 1964.

\bibitem{Vega}
J.C. Gutiérrez-Vega, R.M. Rodríguez-Dagnino, M.A. Meneses-Nava, S. Chávez-Cerda. Mathieu functions, a visual approach. \AJP{2003}{71}{233-242} 

\bibitem{Ruby}
L. Ruby. Applications of the Mathieu equation. \AJP{1996}{64}{39-44}

\bibitem{Seeger97}
A. Seeger. Transverse spin relaxation of spin carriers diffusing in spatially periodic magnetic fields. \Hyperfine{1997}{105}{151-159}

\bibitem{Hochstadt}
H. Hochstadt. {\it Special Functions of Mathematical Physics.} Holt, Rinehart and Winston, New York, 1961.

\bibitem{Abramowitz72}
M. Abramowitz, I.A. Stegun. {\it Handbook of Mathematical Functions with Formulas, Graphs, and Mathematical Tables.} Dover, New York, 1972.

\bibitem{Chaos}
L. Chaos-Cador, E. Ley-Koo. Mathieu functions revisited: matrix evaluation and generating functions. \RMF{2002}{48}{67-75}

\bibitem{Ikebe}
Y. Ikebe, N. Asai, Y. Miyazaki, D. Cai. The eigenvalue problem for infinite complex symmetric tridiagonal matrices with application. \LAA{1996}{241-243}{599-618}

\bibitem{Mechel}
F.P. Mechel. {\it Mathieu Functions.} S. Hirzel Verlag, Stuttgart, 1997.

\bibitem{Hunter81}
C. Hunter, B. Guerrieri. Eigenvalues of Mathieu's Equation and Their Branch Points. \SAM{1981}{64}{113-141}

\bibitem{Olver}
F.W.J. Olver, D.W. Lozier, R.F. Boisvert, C.W. Clark. {\it NIST Handbook of Mathematical Functions.} University Press, Cambridge, 2010.

\bibitem{Blanch69}
G. Blanch, D.S. Clemm. {\it Mathieu's Equation for Complex Parameters: Tables of Characteristic Values. (Box 2)}, Aerospace Research Laboratories, Washington D.C., 1969.

\bibitem{Watson95}
G.N. Watson. {\it A Treatise on the Theory of Bessel Functions.} University Press, Cambridge, 1995.

\bibitem{Spiegel76a}
M.R. Spiegel. {\it Fourier-Analysis.} McGraw-Hill, London, 1976.

\bibitem{Gottlieb}
H.P.W. Gottlieb. Eigenvalues of the Laplacian with Neumann boundary conditions. \JAMSSB{1985}{26}{293-309}

\bibitem{Lommel75}
E. Lommel. Ueber eine mit den Bessel'schen Functionen verwandte Function. \MA{1875}{9}{425-444}

\bibitem{Magnus}
W. Magnus, F. Oberhettinger, R.P. Soni. {\it Formulas and Theorems for the Special Functions of Mathematical Physics.} Springer, Berlin, 1966.

\bibitem{Bottomley84}
P.A. Bottomley. Selective volume method for performing localized NMR spectroscopy. United States Patent No. 4 480 228, 1984.

\bibitem{Bottomley87}
P.A. Bottomley. Spatial Localization in NMR Spectroscopy in Vivo. \ANYAS{1987}{508}{333-348}

\bibitem{Gruetter93}
R. Gruetter. Automatic, localized in vivo adjustment of all first- and second-order shim coils. \MRM{1993}{29}{804-811}

\bibitem{Schachner}
H. Schachner. {\it Kleine Theorie zur Kernspintomographie: Eine Einführung für Medizinphysiker, Medizintechniker und Studierende dieser Fachrichtungen.} Lehmanns-Media-Verlag, Berlin, 2005.

\bibitem{Reese95}
T.G. Reese, R.M. Weisskoff, R.N. Smith, B.R. Rosen, R.E. Dinsmore, V.J. Wedeen. Imaging myocardial fiber architecture in vivo with magnetic resonance. \MRM{1995}{34}{786-791}

\bibitem{Ziener05MAGMA}
C.H. Ziener, W.R. Bauer, P.M. Jakob. Frequency distribution and signal formation around a vessel. \MAGMA{2005}{18}{225-230}

\end{thebibliography}
\end{document}